\documentclass[a4paper,11pt]{article}
\pdfoutput=2
\usepackage{jheppub} 
\usepackage[T1]{fontenc} 
\usepackage[utf8]{inputenc}
\usepackage[table,x11names]{xcolor}

\usepackage{slashed}
\usepackage{booktabs}
\usepackage{arydshln}
\usepackage{pdflscape}
\usepackage{amssymb}
\usepackage{tikz}
\usepackage{subfigure}
\usepackage{multirow}

\DeclareUnicodeCharacter{2217}{}
\DeclareUnicodeCharacter{2212}{}

\usepackage{hyperref}
\hypersetup{
  colorlinks=true,
  linkcolor=red,
  urlcolor=blue,
  citecolor=red, 
  pdfauthor={Name}
}

\usepackage{lineno}

\newcommand{\be}{\begin{equation}}
\newcommand{\ee}{\end{equation}}
\newcommand{\tev}{{\rm TeV}}

\newcommand{\fr}{{\sc \small FeynRules}}
\newcommand{\micromegas}{{\sc\small MicrOMEGAs}}
\newcommand{\mg}{{\sc\small MG5\_aMC}}

\def\RD{\ifmmode R_{D^{(*)}} \else $R_{D^{(*)}}$ \fi}
\def\RK{\ifmmode R_{K^{(*)}} \else $R_{K^{(*)}}$ \fi}
\def\mLQ{\ifmmode M_{S_1} \else $M_{S_1}$ \fi}
\def\mxOne{\ifmmode M_{\chi_1} \else $M_{\chi_1}$ \fi}
\def\mxZero{\ifmmode M_{\chi_0} \else $M_{\chi_0}$ \fi}
\def\lag{{\cal L}}
\def\sss{\scriptscriptstyle}
\def\Ll{L_{\sss L}}
\def\eR{\ell_{\sss R}}

\def\QLbar{\bar Q_{\sss L}}
\def\uR{u_{\sss R}}
\def\uRbar{{\bar u}_{\sss R}}
\def\ydm{y_{\sss \chi}}
\def\lamL{\ifmmode \lambda_L \else $\lambda_L$ \fi}
\def\lamR{\ifmmode \lambda_R \else $\lambda_R$ \fi}
\newcommand{\MET}{$\vec{p}_{\textrm{miss}}$}
\def\DX{\ifmmode \Delta_{X} \else $Delta_{X}$ \fi}
\def\bsp#1\esp{\begin{split}#1\end{split}}

\newcommand{\beq}{\begin{equation}}
\newcommand{\eeq}{\end{equation}}

\newcommand{\diff}{\mathrm{d}}
\newcommand{\eq}{\mathrm{eq}}

\def\ltl{\ifmmode \widetilde{\lambda}_L \else $\widetilde{\lambda}_L$ \fi }
\def\ltr{\ifmmode \widetilde{\lambda}_R \else $\widetilde{\lambda}_R$ \fi }

\preprint{
 KIAS-Q21013, MSUHEP-21-031, IFIC/21-46, \begin{flushright}  TTK-21-44,  P3H-21-091
\end{flushright}
}

\title{
Leptoquark manoeuvres in the dark: a simultaneous solution of the dark matter problem and the $R_{D^{(*)}}$ anomalies
}
\author[a]{Genevi\`eve B\'elanger}
\author[b]{\!\!, Aoife Bharucha} 
\author[c]{\!\!, Benjamin Fuks} 
\author[d]{\!\!, Andreas Goudelis} 
\author[e,f]{\!\!, Jan~Heisig} 
\author[g,h]{\!\!, Adil Jueid}
\author[i]{\!\!, Andre Lessa} 
\author[j]{\!\!, Kirtimaan A. Mohan}
\author[k]{\!\!, Giacomo Polesello} 
\author[l]{\!\!, Priscilla Pani}
\author[m]{\!\!, Alexander Pukhov} 
\author[n]{\!\!, Dipan Sengupta}
\author[o]{\!and\! Jos\'e~Zurita} 

\affiliation[a]{LAPTh, Univ. Grenoble Alpes, USMB, CNRS, 9 Chemin de Bellevue, F-74940 Annecy, France}
\affiliation[b]{Aix Marseille Univ, Universit\'e de Toulon, CNRS, CPT, Marseille, France}
\affiliation[c]{Laboratoire de Physique Th\'eorique et Hautes Energies (LPTHE), UMR 7589, Sorbonne Universit\'e et CNRS, 4 place Jussieu, 75252 Paris Cedex 05, France}
\affiliation[d]{Laboratoire de Physique de Clermont (UMR 6533), CNRS/IN2P3, Univ.\ Clermont Auvergne, 4 Av.\ Blaise Pascal, F-63178 Aubi\`ere Cedex, France}
\affiliation[e]{Institute for Theoretical Particle Physics and Cosmology, RWTH Aachen University, Sommerfeldstr. 16, D-52056 Aachen, Germany}
\affiliation[f]{Centre for Cosmology, Particle Physics and Phenomenology (CP3), Universit\'e catholique de Louvain, Chemin du Cyclotron 2, B-1348 Louvain-la-Neuve, Belgium}
\affiliation[g]{Quantum Universe Center, Korea Institute for Advanced Study, Seoul 02455, Republic of Korea}
\affiliation[h]{School of Physics, Konkuk University, 05029, Seoul, Republic of Korea}
\affiliation[i]{Centro de Ci$\hat{e}$ncias Naturais e Humanas, Universidade Federal do ABC, Santo Andr\'e, 09210-580 SP, Brazil}
\affiliation[j]{Department of Physics and Astronomy, 567 Wilson Road,  East Lansing, Michigan-48824, USA}
\affiliation[k]{INFN, Sezione di Pavia, Via Bassi 6, 27100 Pavia, Italy}
\affiliation[l]{Deutsches Elektronen Synchrotron, DESY, 15738 Zeuthen, Germany}
\affiliation[m]{Skobeltsyn Institute of Nuclear Physics, Moscow State University, Moscow 119992, Russia}
\affiliation[n]{Department of Physics and Astronomy, 9500 Gilman Drive, University of California, San Diego, USA}
\affiliation[o]{Instituto de F\'{\i}sica Corpuscular, CSIC-Universitat de Val\`encia,
 E-46980 Paterna, Valencia, Spain}
 
\emailAdd{belanger@lapth.cnrs.fr}
\emailAdd{aoife.bharucha@cpt.univ-mrs.fr}
\emailAdd{fuks@lpthe.jussieu.fr}
\emailAdd{andreas.goudelis@clermont.in2p3.fr}
\emailAdd{heisig@physik.rwth-aachen.de}
\emailAdd{adiljueid@kias.re.kr}
\emailAdd{andre.lessa@ufabc.edu.br}
\emailAdd{kamohan@msu.edu}
\emailAdd{giacomo.polesello@cern.ch}
\emailAdd{priscilla.pani@cern.ch}
\emailAdd{pukhov@lapp.in2p3.fr}
\emailAdd{disengupta@physics.ucsd.edu}
\emailAdd{jzurita@ific.uv.es}

\begin{document}

\abstract{The measured branching fractions of $B$-mesons into leptonic final states derived by the LHCb, Belle and BaBar collaborations hint towards the breakdown of lepton flavour universality. In this work we take at face value the so-called $R_{D^{(*)}}$ observables that are defined as the ratios of neutral $B$-meson charged-current decays into a $D^{(*)}$-meson, a charged lepton and a neutrino final state in the tau and light lepton channels. A well-studied and simple solution to this charged current anomaly is to introduce a scalar leptoquark $S_1$ that couples to the second and third generation of fermions. We investigate how  $S_1$ can also serve as a mediator between the Standard Model and a dark sector. We study this scenario in detail and estimate the constraints arising from collider searches for leptoquarks, collider searches for missing energy signals, direct detection experiments and the dark matter relic abundance. We stress that the production of a pair of leptoquarks that decays into different final states ({\it i.e.}\ the commonly called ``mixed'' channels) provides critical information for identifying the underlying dynamics, and we exemplify this by studying the $t \tau b \nu$ and the resonant $S_1$ plus missing energy channels. We find that direct detection data provides non-negligible constraints on the leptoquark coupling to the dark sector, which in turn affects the relic abundance. We also show that the correct relic abundance can not only arise via standard freeze-out, but also through conversion-driven freeze-out. We illustrate the rich phenomenology of the model with a few selected benchmark points, providing a broad stroke of the interesting connection between lepton flavour universality violation and dark matter.
}

\keywords{Dark Matter, Hadron-Hadron Collisions, Scalar Leptoquarks, Lepton Flavour Universality Violation.}

\maketitle

\section{Introduction}
\label{sec:intro}
Strong evidence for Lepton Flavour Universality Violation (LFUV) has been established by the LHCb, Belle, and BaBar collaborations in their measurements of the \RK~\cite{Aaij:2017vbb,Aaij:2019wad,Aaij:2021vac,Belle:2015qfa,BaBar:2013mob} and \RD~\cite{Aaij:2015yra,Aaij:2017uff,Aaij:2017deq} observables. Two classes of new physics models can accommodate these results: scenarios featuring either leptoquarks or an extra $Z'$ boson (see {\it e.g.}~\cite{Blanke:2019pek} and references therein). None of these particles, however, address on their own some additional  shortcomings of the Standard Model of particle physics such as electroweak naturalness, neutrino masses and the dark matter (DM) puzzle, just to mention a few.

In view of this situation, it is interesting to entertain the idea of using the same models that address the LFUV anomalies to simultaneously explain one of the aforementioned flaws of the Standard Model (SM). In this work, we focus on solving the dark matter problem (for previous work connecting the B--flavour anomalies and dark matter, see {\it e.g.} \cite{Belanger:2015nma,Carvunis:2020exc,Arcadi:2021cwg,Guadagnoli:2020tlx,Baker:2021llj}) with a  scalar leptoquark with the gauge quantum numbers of a right-handed SM quark, often referred to as $S_1$ in the literature~\cite{Davidson:1993qk,Dorsner:2013tla,Bauer:2015knc,Becirevic:2016oho,Cai:2017wry,Angelescu:2018tyl,Azatov:2018kzb,Aydemir:2019ynb,Dorsner:2019itg,Crivellin:2019qnh}.

Considering a scalar leptoquark as a mediator to the dark sector has only been attempted previously, to the best of our knowledge, in refs.~\cite{Queiroz:2014pra,Baker:2015qna,Cline:2017aed,Azatov:2018kzb,DEramo:2020sqv,Choi:2018stw}. Coupling a leptoquark to a dark sector requires the introduction of at least two additional particles, $\chi_0$, our dark matter candidate, and $\chi_1$. As the latter must carry colour charge, it can be looked for at colliders in final states with transverse missing energy (MET) plus SM particles. Direct searches for leptoquarks (LQ) would also in principle constrain the parameter space in a way in which $m_{\rm LQ} \gtrsim 1~\rm{TeV}$ for ${\cal O} (1)$ couplings~\cite{ATLAS:2020sxq,CMS:2020wzx,ATLAS:2021aui}. However those constraints are relaxed in our scenario, given that novel LQ decays in the dark sector dilute the ``visible'' branching fractions ({\it i.e.}~those associated with SM final states). This extends the opportunities to test this setup at the LHC in the near future, and also to consider novel, currently unexplored final states. In addition, direct detection experiments, due to their high-precision, can provide important constraints on the coupling of the leptoquark to the dark sector, which can also impact the regions of the parameter space consistent with the measured relic abundance. Finally, in order to establish a link between the flavour anomalies and the dark sector, it is of paramount importance to have in place searches for final states with both visible and invisible LQ decays. We thus pay particular attention to the existing CMS search for a resonant leptoquark plus missing energy signal~\cite{Sirunyan:2018xtm}, which would allow us to directly probe the $\RD$-DM connection (RDM).

This article, which heavily expands upon preliminary results presented in~\cite{Brooijmans:2020yij}, is organised as follows. In section~\ref{sec:model} we introduce our notation, conventions and the model under consideration. A simple setup with just two couplings can explain the $\RD$ anomalies, and we briefly discuss the most salient phenomenological features of our construction. In section~\ref{sec:lhc} we detail the collider constraints originating from searches for missing energy, leptoquark and LQ+MET signals. We emphasise the role of a few overlooked searches for leptoquark pair production, stressing the relevance of studies where each leptoquark decays in a different final state: the \emph{mixed} searches. In section~\ref{sec:dm} we study the constraints coming from direct detection and the relic abundance. In particular, we examine dark-matter genesis beyond the standard freeze-out paradigm, considering solutions in the conversion-driven freeze-out regime (CDFO)~\cite{Garny:2017rxs}. We show that current XENON data forbids a coupling to the dark sector $\gtrsim {\cal O}(1)$, while DARWIN has the potential to strongly constrain the parameter space where thermal freeze-out is the pathway to dark matter production in the early universe. The bounds on the dark coupling favour the CDFO regime: solutions with very small couplings to the dark sector are a natural feature of this mechanism. Finally, in section~\ref{sec:bench} we define a few benchmark scenarios that illustrate the large amount of phenomenological possibilities, showing that a wide-open search program is necessary, as the strongest hint can come from either direct detection or collider searches in any sub-category. We reserve our conclusions for section~\ref{sec:conclu}.

\section{Theoretical setup}
\label{sec:model}
In this section we first introduce the particle content and the Lagrangian of our model in section~\ref{sec:lagrangian}, and we then explain in section~\ref{sec:anomalies} how this setup is able to solve the $\RD$ anomalies. The knowledge of the particle content and of the regions of the parameter space that yield a solution to the $R_{D^{(*)}}$ anomalies allows us to anticipate the most salient phenomenological features of our setup, that we depict in section~\ref{sec:pheno}. Appendix~\ref{app:model} is dedicated to technical details about the implementation of the model in numerical tools. Moreover, as a byproduct of our analysis, we discuss in appendix~\ref{app:tchannel} the impact of  $t$-channel SM lepton exchanges contributions to leptoquark pair production, an effect that is currently ignored in the vast majority of the leptoquark phenomenological and experimental studies and that could be important~\cite{Borschensky:2020hot,Borschensky:2021hbo}.

\subsection{Particle content and interactions}
\label{sec:lagrangian}
Enlarging the Standard Model particle content with a single scalar leptoquark $S_1$ suffices to explain the $R_D$ anomalies, as detailed in section~\ref{sec:anomalies}. Here, we follow a common notation that is inspired by the generic classification of ref.~\cite{Buchmuller:1986zs} and that is widespread in the literature. In this setup, $S_1$ stands for a weak singlet state carrying the same quantum numbers as the SM right-handed down-type quarks, except that $\chi_1$ does not carry lepton number while $S_1$ does.  Since our goal is to employ $S_1$ as a mediator to a dark sector (where dark matter resides) through tree-level interactions, and as the dark matter candidate is colourless and electrically neutral,
we are forced to introduce two \emph{dark} particles, which we pick for simplicity as fermions. The first one is a Majorana fermion, $\chi_0$, that consists of our dark matter candidate and that is singlet under the Standard Model gauge group. The second particle, $\chi_1$, is taken to be a Dirac fermion. Both $\chi_0$ and $\chi_1$ are then assumed to be odd under a $\mathcal{Z}_2$-type symmetry, whereas all other particles are chosen to be $\mathcal{Z}_2$-even.

The most general renormalisable Lagrangian describing the dynamics of the three considered new fields reads
\be\bsp
 \lag = &\ \lag_{\rm SM} + \lag_{\rm kin}
   + \bigg[
    {\bf \lambda_{\sss R}}\ \uRbar^c\ \eR^{\phantom{c}} \ S_1^\dag
  + {\bf \lambda_{\sss L}}\ \QLbar^c \!\cdot\! \Ll^{\phantom{c}} \ S_1^\dag
  + \ydm \bar\chi_1\chi_0 S_1
   + {\rm H.c.} \bigg]\ .
\esp\label{eq:RDM:lag}\ee
In this expression, the matrices ${\bf \lambda_{\sss R,L}}$ are understood to carry quark and lepton flavour indices (in that order), and the dot appearing in the second term in the square bracket indicates an $SU(2)$ invariant product of two fields lying in its fundamental representation, {\it i.e.} $\QLbar^c \!\cdot\! \Ll^{\phantom{c}} \equiv \QLbar^c \epsilon \Ll^{\phantom{c}}$ where $\epsilon=i\sigma_2$ and $\sigma_2$ is the second Pauli matrix. The kinetic and mass terms of the new fields are encoded in $\lag_{\rm kin}$, while the Standard Model Lagrangian is denoted by $\lag_{\rm SM}$. The $Q_L$  and $\Ll$ fields stand for the $SU(2)_L$ doublets of left-handed quarks and leptons, whereas the $\uR$ and $\eR$ fields stand for the $SU(2)_L$ singlets of up-type quarks and charged leptons. As we have chosen $\chi_0$ to be a SM singlet, then $\chi_1$ and $S_1$ must have the same representation under the SM gauge group, which in this particular case corresponds to $({\bf 3}, {\bf 1})_{-1/3}$.

For the analysis performed in this study, the Lagrangian~\eqref{eq:RDM:lag} has been implemented in \fr~\cite{Alloul:2013bka,Christensen:2009jx}, and therefore connected to various high-energy physics tools. Relevant details are collected in appendix~\ref{app:model}. While the matrices ${\bf \lambda_{\sss R}}$ and ${\bf \lambda_{\sss L}}$ appearing in this Lagrangian introduce a large number of new free parameters, we define in the next subsection a minimal set of parameters that can account for an explanation to the $\RD$ anomalies (all unnecessary entries in the coupling matrices being set to zero).

\subsection{Leptoquark solutions to \texorpdfstring{$R_D$}{} anomalies}
\label{sec:anomalies}
Considering $(\lambda_{\sss L})_{33} \equiv \lamL$ and $(\lambda_{\sss R})_{23} \equiv -\lamR$ as the only non-zero entries of the leptoquark couplings to the SM sector provides a minimal framework to explain the $\RD$ anomalies, the minus sign allowing us to match the conventions of ref.~\cite{Gherardi:2020qhc} and earlier studies\footnote{Even though we do not introduce direct couplings to the first generation, they’re necessarily CKM--induced and therefore we expect $b \to u$ transitions which are, however, not particularly constraining.}. The class of scenarios obtained by exchanging the {\it left} and {\it right} labels ({\it i.e.} by taking $(\lambda_{\sss L})_{23}$ and $(\lambda_{\sss R})_{33}$ as the only non-zero entries of the ${\bf \lambda_{\sss L,R}}$ matrices) is in contrast disfavoured. Fixing $(\lambda_{\sss L})_{23} \ne 0$ would indeed give an unacceptable contribution to $B \to X_s \nu \bar{\nu}$, although this could be circumvented through destructive interference with other leptoquark states~\cite{Crivellin:2017zlb,Buttazzo:2017ixm,Marzocca:2018wcf,Crivellin:2019dwb}, and having $(\lambda_{\sss R})_{33} \ne 0$ would explicitly require right-handed neutrinos as done {\it e.g.} in \cite{Azatov:2018kzb}. A global fit with $\lamL,\lamR$ as free parameters has been carried out, for example, in refs.~\cite{Buttazzo:2017ixm,Marzocca:2018wcf,Azatov:2018kzb,Angelescu:2018tyl}, and was recently updated in ref.~\cite{Gherardi:2020qhc}. In such a fit, the $S_1$ contributions to the $\RD$ ratios are computed at one-loop accuracy. Moreover, besides solely performing a fit to provide an explanation for the $\RD$ anomalies, other observables such as the $\text{BR}(B_c^+ \to \tau^+ \nu)$ branching ratio, lepton flavour universality tests in $\tau$ decays, and reproducing the high-$p_T$ tail of $p p \to \tau \tau$ at the LHC\footnote{After the completion of this work we become aware that a new update of the constraints on $\lambda_{L,R}$ from the high-$p_T$ tail in $pp\to\ell\ell$ has been carried in \cite{Angelescu:2021lln} using full dataset of $140~{\rm fb}^{-1}$. However, these new constraints do not change the main results of this paper.}, are also considered. Similarly, we make sure in this work that not only the $\RD$ anomalies are accommodated, but that predictions for the above-mentioned observables do not challenge observations.

 \begin{figure}[t]
  \centering
  \includegraphics[width=0.495\textwidth]{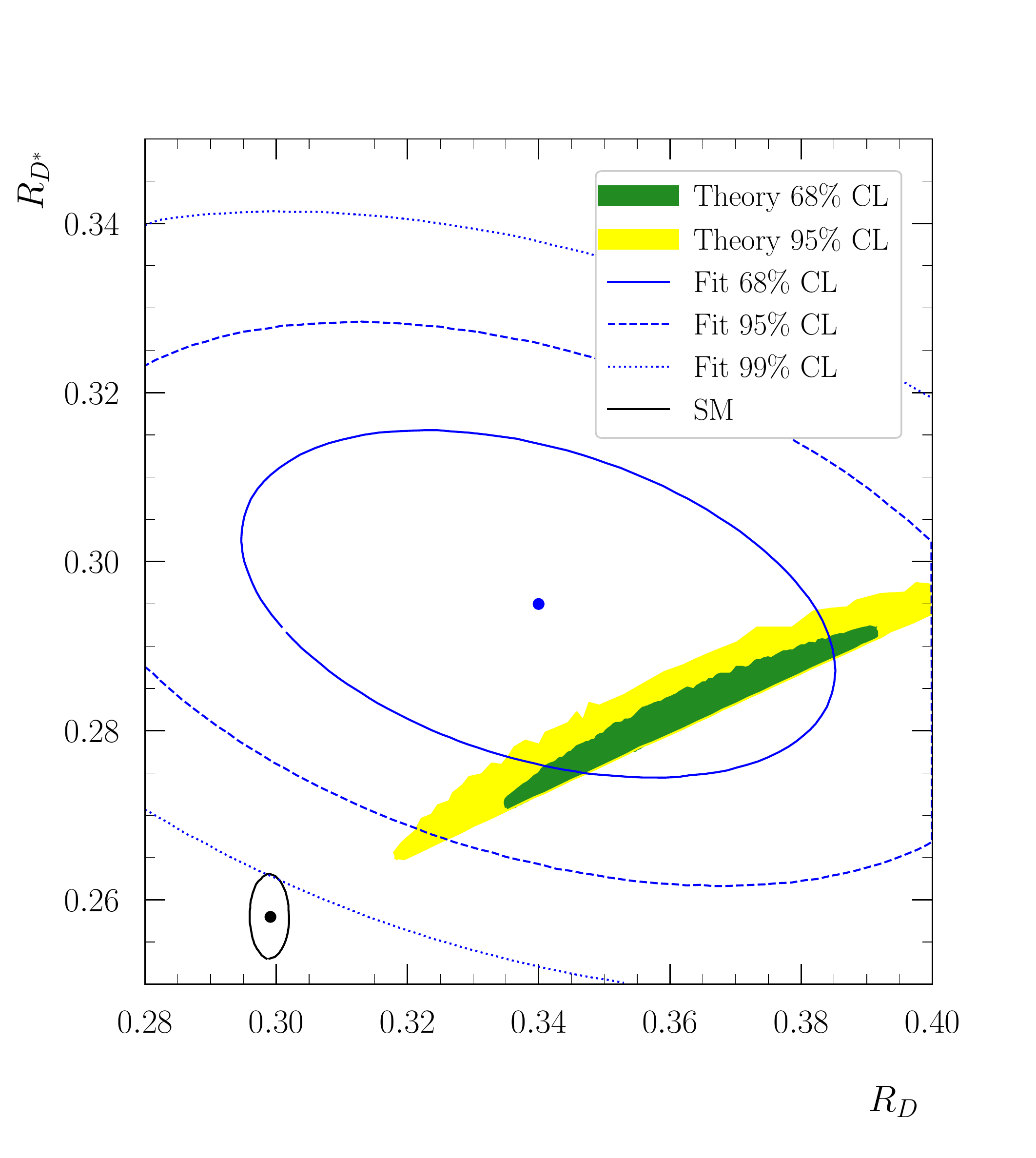}\hfill
  \includegraphics[width=0.495\textwidth]{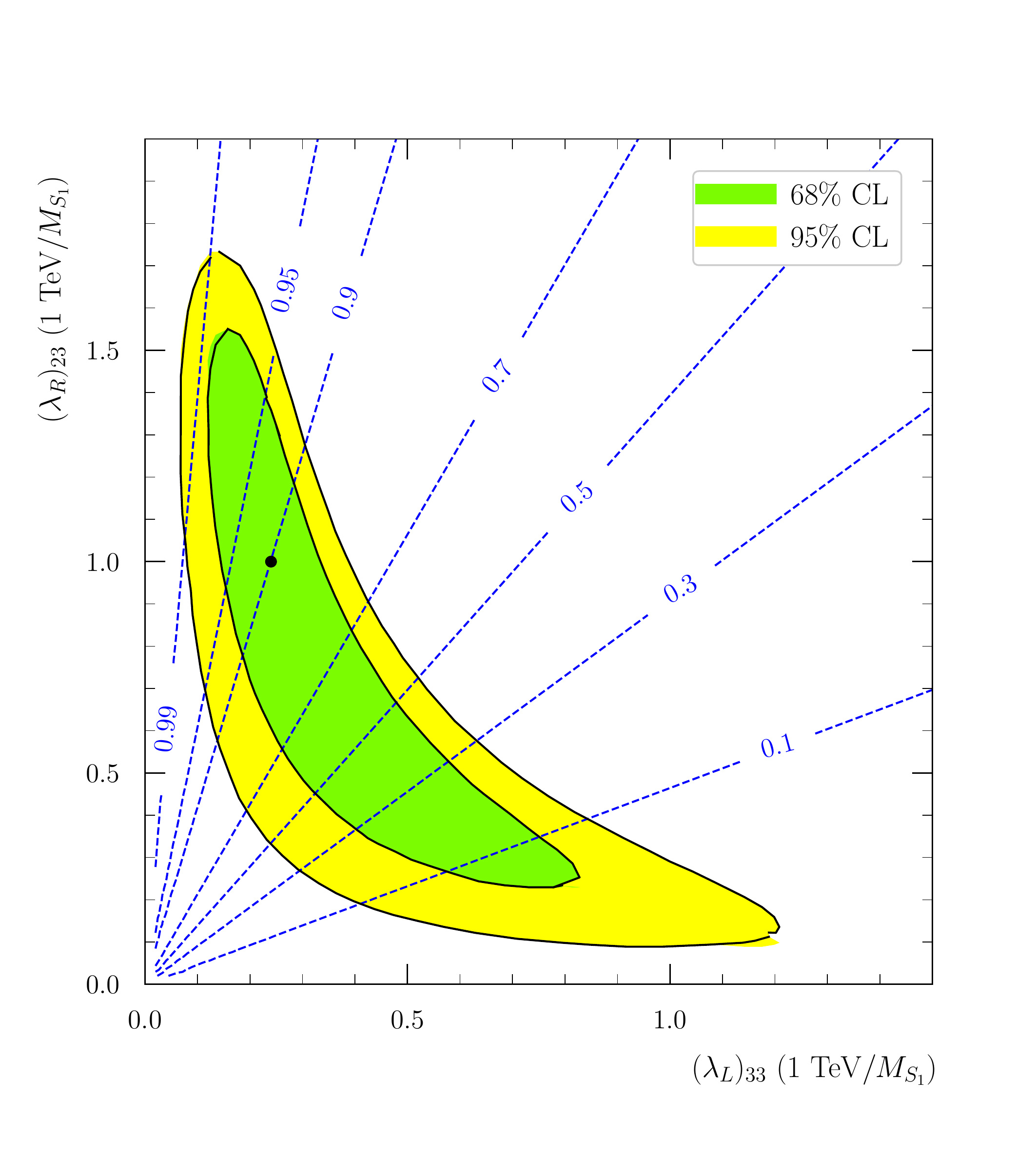}
 \caption{{\it  Left}:
 Flavour fit to the charged current anomalies. The blue lines represent the current status in the measurements of the two $\RD$ observables, compared with the SM predictions (gray point and contour), using the Spring 2019 update from HFLAV~\cite{HFLAV:2016hnz}. The green and yellow dots correspond to the predicted values of $R_{D}$--$R_{D^*}$ in our model. Our results are adapted from figure 1 (right) of ~\cite{Gherardi:2020qhc}. {\it Right}: Viable parameter space region as originating from the fit of flavour and precision observables as function of $\lambda_{L,R} / \mLQ$. Blue dashed lines are iso-lines of the branching ratio ${\rm BR}(S_1 \to c_R \tau_R)$ when setting $\ydm = 0$. The black dot represents the best-fit point.}
\label{fig:RDM:FlaFit}
\end{figure}

The outcome of our fitting procedure is given in figure~\ref{fig:RDM:FlaFit}, which is obtained by adapting figure~1 of ref.~\cite{Gherardi:2020qhc}. The left panel of figure~\ref{fig:RDM:FlaFit} presents the results of our scan in the plane of the $\RD$ ratios, while its right panel shows them in the plane of $\ltl$ and $\ltr$ where we have defined  
\begin{eqnarray}
\ltl = \lambda_L (\tev / \mLQ), \quad \ltr= \lambda_R (\tev / \mLQ),
\end{eqnarray}
with $\mLQ$ being the $S_1$ mass. The shape of the favoured regions of the parameter space stems from two considerations: while the $\RD$ ratios are proportional to the product $\ltl \ltr$, other considered observables scale with $\ltl^2$ (lepton-flavour universality in $\tau$ decays) or with $\ltr^2$ ($p p \to \tau \tau$). 

From now on, we introduce the following notation. A given choice of ${\ltl,\ltr}$ values is called a {\bf benchmark slope} (BS), and is identified with an Arabic numeral. In addition, a given benchmark slope supplemented by additional parameters (often $\mLQ$) is called a {\bf benchmark scenario} (BS, in a slight abuse of notation) and is identified with the slope numeral and a Latin character. The precise set of additional parameters that enter here depends on the specific context. For instance, when dealing with visible LQ searches at colliders, the dark sector parameters only play a minor role as their effect is solely to modify the visible $S_1$ branching ratios. In contrast, when dealing with MET searches at colliders, $\mLQ$ is irrelevant and the main relevant parameters are the dark sector masses.

To select our benchmark slopes, we use as a guide the ratio $\lamL / \lamR$, or equivalently $\ltl/\ltr$, and we restrict ourselves to the 68\% confidence level (CL) contour obtained in our fit (green area of figure~\ref{fig:RDM:FlaFit}). For $\lamL > \lamR$ we pick $(\ltl,\ltr) = (0.7,0.3) \equiv $ BS1, with the largest ratio value in the contour being $(0.83,0.25)$. For the opposite case we settle for the best fit point $(0.24,1.0) \equiv$ BS2. The lowest coupling ratio is achieved for $(0.16,1.55)$. For such large values of $\ltr$, it is important to keep in mind that lepton $t$--channel exchange diagrams and the validity of the narrow width approximation for large $S_1$ masses could be relevant for leptoquark production and decay at colliders. We analyse this issue in appendix~\ref{app:tchannel}. 

From these two benchmark slopes we construct the benchmark scenarios BS1a/BS2a for $\mLQ = 1.25$ TeV, BS1b/BS2b for $\mLQ = 1.5$ TeV and BS1c/BS2c for $\mLQ = 1.7$ TeV. These points are detailed in table~\ref{tab:RDM:benchmark_points}. For reference purposes we also include the leptoquark branching fractions into visible states when the dark sector coupling $y_\chi$ is set to zero.  The chosen benchmark scenarios exhibit the wide range of possibilities for the $S_1\to c \tau$ decay channel (with branching ratios ranging from about 10\% to roughly 90\%), while the decays into a $b$ plus neutrino or a top plus neutrino system can feature branching ratios varying in the range $0.1\%-50\%$. However, being an off-shell interference with the SM amplitude, the contributions to the $\RD$ ratios depend only on the $\ltl$ and  $\ltr$ parameters and not on the specific leptoquark branching ratios. Hence opening up a dark world for $S_1$ ($y_\chi \ne 0$ and $\mLQ > \mxOne + \mxZero$ where $\mxOne$ and $\mxZero$ are the masses of the $\chi_1$ and $\chi_0$ states respectively) only impacts the constraints coming from direct LQ searches, which are discussed in detail in section~\ref{sec:lhc}, and not the potential explanations for the flavour anomalies. 

\begin{table}[t]
\renewcommand{\arraystretch}{1.4}
\resizebox{\textwidth}{!}{
\begin{tabular}{cccccccc}
 Name &$\mLQ$ [GeV] & $\lamL$ & $\lamR$ & ${\rm BR}(S_1 \to b \nu)$ & ${\rm BR}(S_1 \to t \tau)$ & ${\rm BR}(S_1 \to c \tau)$ & $\Gamma_{S_1}$ [GeV] \\
\hline
BS1a &1250      & 0.875     & 0.375  & 0.466   & 0.448   & 0.086   & $40.9$   \\
BS2a & 1250     & 0.3       & 1.25   & 0.053   & 0.050   & 0.897   & $43.24$  \\   
BS1b &1500      & 1.05     & 0.45  & 0.463   & 0.451   & 0.086   & $70.98$   \\
BS2b & 1500     &   0.36     & 1.5   & 0.052   & 0.050   & 0.898   & $74.78$  \\
BS1c & 1700     & 1.19       & 0.51  & 0.462   & 0.452   & 0.085   & $103.60$ \\
BS2c & 1700     & 0.408      & 1.7   & 0.052   & 0.051   & 0.897   &  $108.88$
\end{tabular}
}
\caption{Benchmark scenarios providing an explanation for the flavour anomalies. We have assumed negligible LQ decays into the dark sector ($\ydm = 0$).
}
\label{tab:RDM:benchmark_points}
\end{table}

\subsection{Phenomenological features}
\label{sec:pheno}

\begin{figure}[t]
	\centering
	\includegraphics[width=0.75\textwidth,trim={0 12.5cm 0 0},clip]{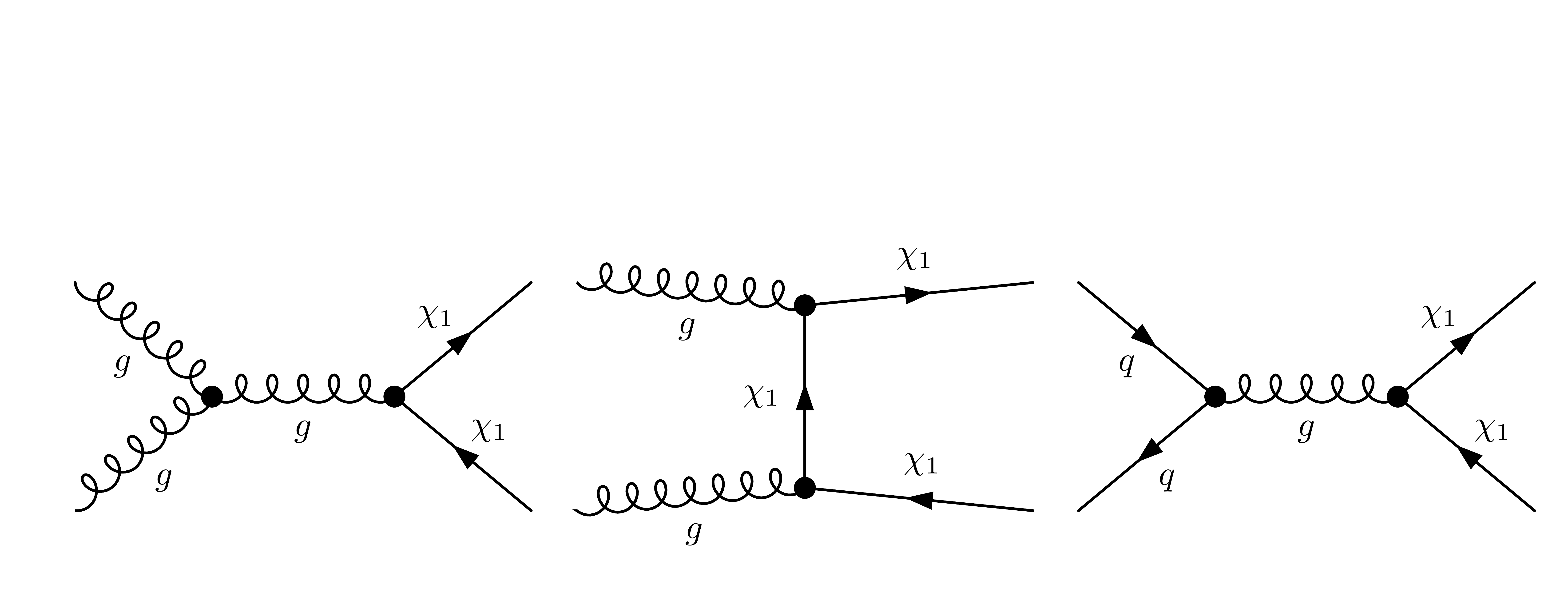}\\ \vspace{.6cm}
	\includegraphics[width=0.87\linewidth,trim={0 25cm 0 0},clip]{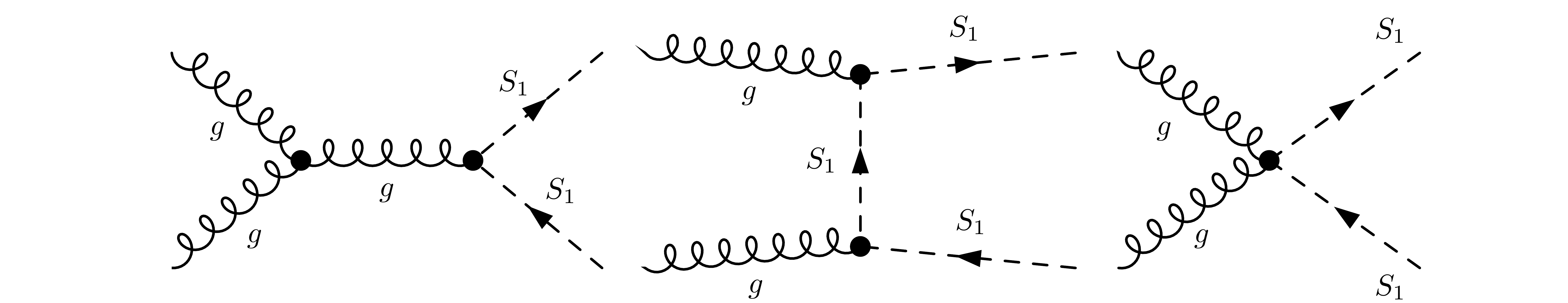}\\
	\includegraphics[width=0.87\linewidth,trim={0 2.5cm 0 20cm},clip]{figures/fig_lqlq_production.pdf}\\
	\vspace{.25cm}\hspace{-1.4cm}
	\includegraphics[width=0.6\textwidth,trim={7.5cm 12.5cm 0 7cm},clip]{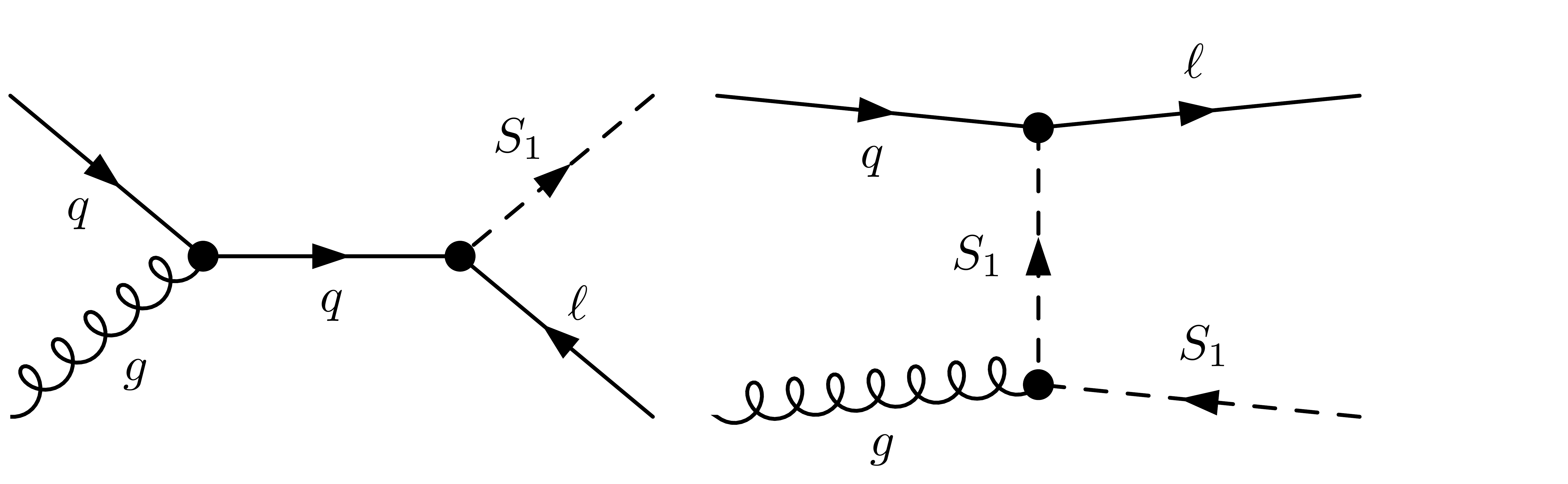}
	\caption{Representative Feynman diagrams for the hadron collider processes $p p \to \overline{\chi}_1 \chi_1$ (top row), $p p \to \overline{S}_1 S_1$ (second and third rows) and $p p \to S_1 \tau^{\pm}$ (bottom row).}
		\label{fig:collider_diagrams}
\end{figure}

Before diving into a full and detailed study of the dark matter and collider phenomenology of our setup, it is worth taking a pause to make a quick tour of its main collider phenomenological features. The signatures of the considered model at the LHC originate from  three dominant new physics production processes. They consist of $\chi_1$ pair production, $S_1$ pair production and the associated production of a leptoquark and a tau-lepton. The corresponding tree-level Feynman diagrams are depicted in figure~\ref{fig:collider_diagrams}. From the top row of the figure, we can see that the process  $p p \to \overline{\chi}_1 \chi_1$ only depends on a single new free parameter, $\mxOne$, as the three diagrams only involve $g\overline{\chi}_1\chi_1$ vertices whose dynamics is fully dictated by the quantum numbers of $\chi_1$ and the SM strong coupling constant. Similarly, $\overline{S}_1 S_1$ production is also driven by one new physics parameter, the leptoquark mass $\mLQ$, and the SM strong coupling (see the first four diagrams of the second and third rows of figure~\ref{fig:collider_diagrams}). The fifth diagram relevant for leptoquark pair production scales in contrast with $\lamR^2$. This is in principle a worrisome feature, as we have discussed that a possible solution to the $\RD$ anomalies might prefer a $\lamR$ coupling value of ${\cal O} (1)$. We assess the numerical impact of this contribution for the scenarios considered in this work in appendix~\ref{app:tchannel}. It turns out that for our practical purposes of studying the $S_1$ leptoquark phenomenology at the (HL-)LHC, this $t$-channel lepton exchange contribution can be safely neglected. Finally, in the bottom row of figure~\ref{fig:collider_diagrams}, we show diagrams relevant for the associated production of a leptoquark $S_1$ with a SM lepton. Such a process depends both on the leptoquark Yukawa coupling $\lamR$ (as any diagram involving the $\lamL$ coupling is relatively suppressed by virtue of small third-generation quark densities), as well as on the leptoquark mass. In addition, it is less phase-space suppressed than the pair-production mode, and could thus become important for larger $\mLQ$ values.

\begin{figure}[t]
	\centering
	\includegraphics[width=0.48\textwidth]{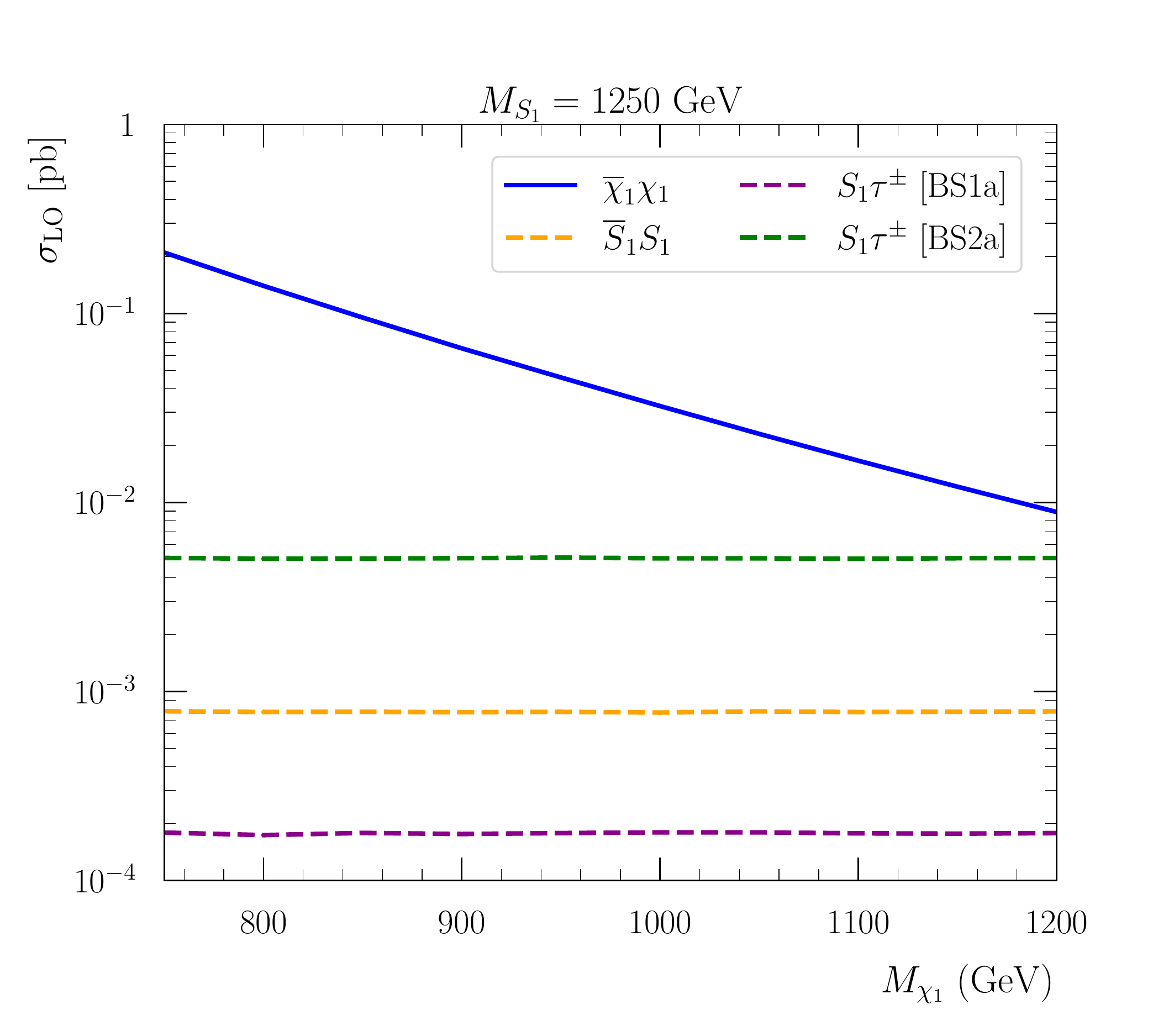}
	\includegraphics[width=0.48\textwidth]{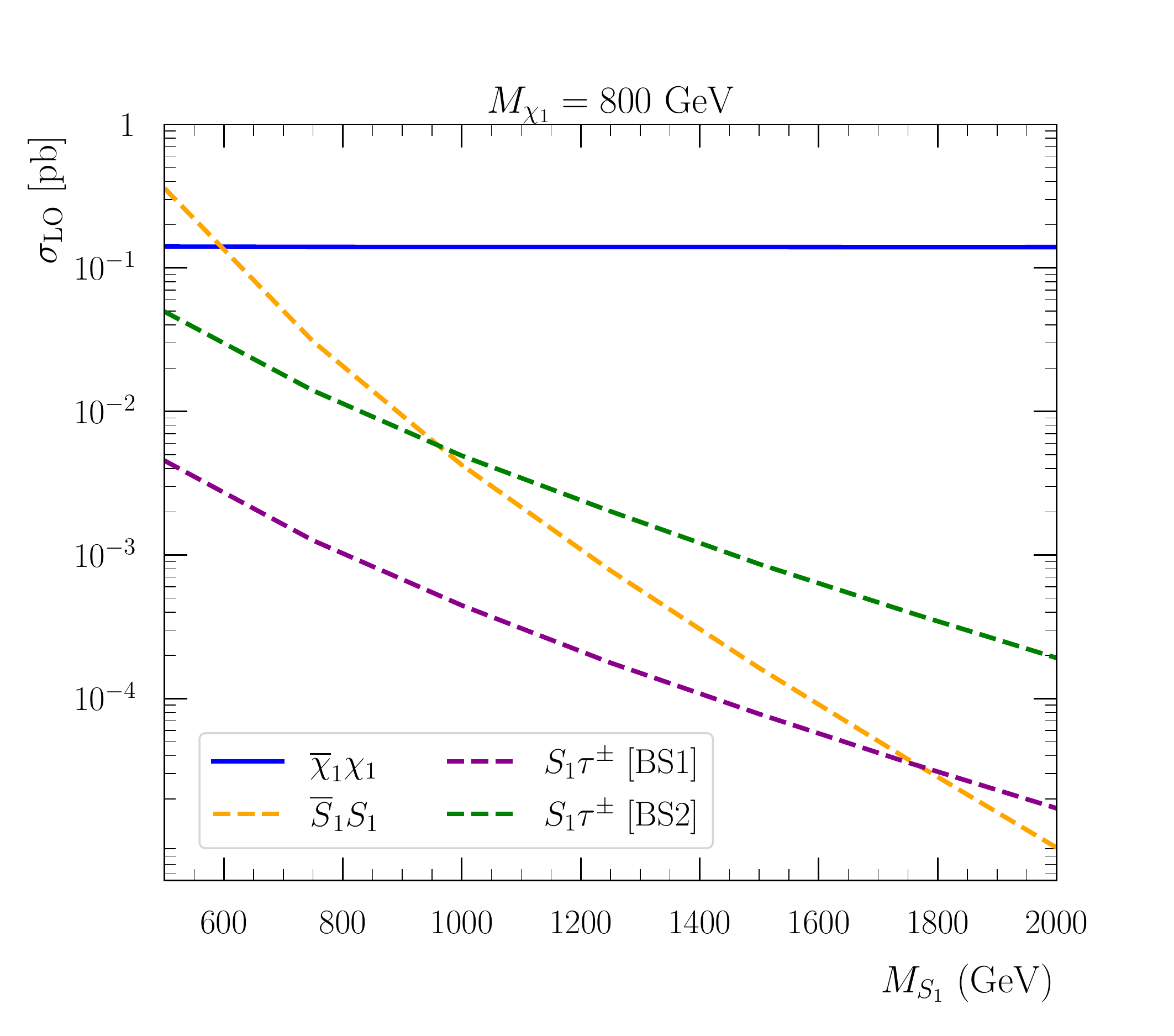}
	\caption{Cross sections $\sigma_{\rm LO}$ for $S_1$ and $\chi_1$ production at the LHC as a function of  $\mxOne$ (left panel) and $\mLQ$ (right panel). We consider the production of a pair of $\chi_1$ states (solid blue, LO), of a pair of $S_1$ states (dashed orange, LO) and the associated production of $S_1$ with a tau lepton (LO). The latter cross section is shown for the benchmark points BS1a (dashed purple) and BS2a (dashed green). In the right panel we keep $\ltl=\lambda_L \tev / \mLQ$ and $\ltr = \lambda_R \tev / \mLQ$ fixed to the corresponding BS1 and BS2 values.}
		\label{fig:RDM:LHCxsec}
\end{figure}

From the \fr\ implementation of our model, we generate its UFO version~\cite{Degrande:2011ua} so that \textsc{MadGraph5\_aMC@NLO} (\mg)~\cite{Alwall:2014hca} could be employed for cross section computations and parton-level event generation. For all collider results included in this paper, we convolve leading-order (LO) matrix elements with the \texttt{NNPDF30\_lo\_as\_0118}~\cite{Ball:2014uwa} set of parton distribution functions (PDF), that we handle through {\sc LHAPDF}~\cite{Buckley:2014ana}. Since predictions at next-to-leading-order (NLO) in QCD are in the same ballpark as the corresponding LO rates~\cite{Kramer:1997hh,Kramer:2004df,Mandal:2015lca,Borschensky:2020hot,Borschensky:2021hbo}, higher-order corrections are not expected to largely modify the LHC constraints that we derive on the model in this work. We therefore employ LO cross sections throughout this study. We present in figure~\ref{fig:RDM:LHCxsec} cross sections as a function of $\mxOne$ (left panel) and $\mLQ$ (right panel) for the three considered new physics production processes. Among them, only single leptoquark production with an associated lepton explicitly depends on the leptoquark Yukawa couplings. Hence we present it for our two benchmark scenarios BS1b and BS2b in the left panel. On the right panel we fix instead $\mxOne = 800$ GeV and scan over $\mLQ$. In this case we keep the ratio of Yukawa couplings fixed as the leptoquark mass varies, and we choose this ratio to the values corresponding to the BS1 and BS2 benchmarks.

It comes as no surprise that the pair production of $\chi_1$ is the process with the largest cross section, as both $S_1$ and $\chi_1$ have the same colour charges but $\chi_1$ is a fermion. Moreover, we also assume that $S_1 \to \chi_1 \chi_0$ occurs on-shell, so that $\chi_1$ is lighter than $S_1$. If we abandon this hypothesis the take-home message is exactly the same: the largest event rate corresponds to the QCD-induced pair-production of the lightest new physics coloured particle in the spectrum. This inverse regime in which the $S_1$ state is lighter than the dark matter candidate is illustrated in the left part of the right panel of figure~\ref{fig:RDM:LHCxsec}, in which $\mLQ<\mxOne$. On the other hand, the $S_1 \tau$ process becomes relevant for large values of $\lambda_R$: the cross section intercepts that of $S_1$ pair production at about 950 GeV for BS2, as the $S_1$ pair production cross section is relatively suppressed by phase space. Hence, if we restrict ourselves to inclusive cross sections larger than 0.1 fb (otherwise the number of signal events at the high-luminosity phase of the LHC would be fairly limited, rendering the process a challenging one to observe), this effectively means that the LHC sensitivity to $S_1$ leptoquarks in our model is limited to  $\mLQ\lesssim 1.6$~TeV\footnote{This statement is based only on the low signal count. A fair assessment would require a detailed examination of the signal and the SM backgrounds on a case-by-case basis.}. In the case of BS1 where $\lambda_L$ and $\lambda_R$ are comparable, we should in principle also consider the associated production of $S_1$ with a bottom or a top quark. These processes turn out to be PDF-suppressed, and thus give a negligible contribution to the total new physics cross section at the LHC. They could however be important when considering future hadron colliders with larger centre-of-mass energies~\cite{Benedikt:2018csr}.

\begin{figure}
	\centering
	\includegraphics[width=0.48\textwidth]{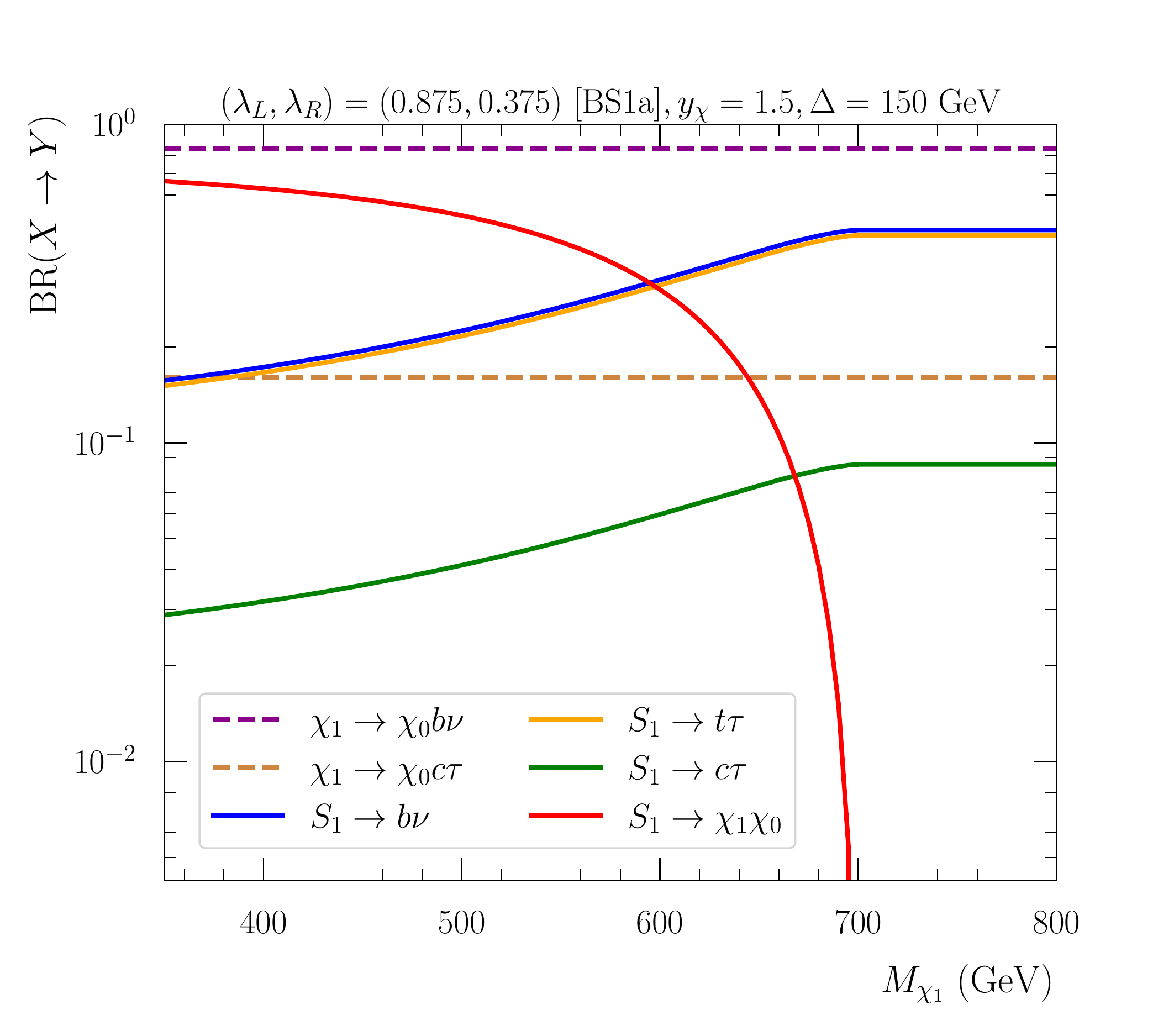}
	\includegraphics[width=0.48\textwidth]{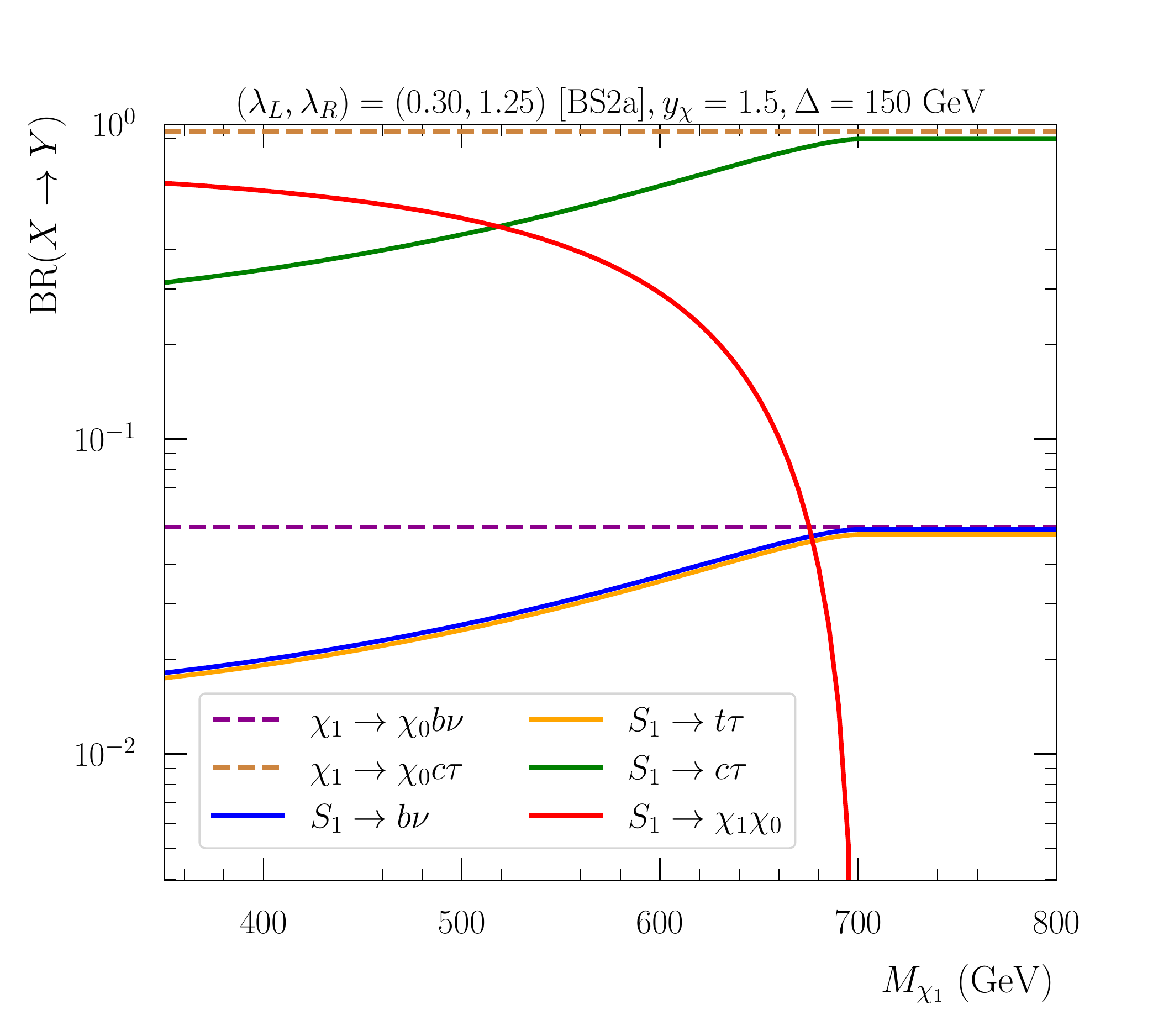}
	\caption{Branching ratios for the $S_1$ and $\chi_1$ states as a function of the $\chi_1$ mass. The mass difference between $\chi_1$ and $\chi_0$ is kept at 150~GeV, $y_{\chi} = 1.5$, and the other parameters correspond to the BS1a (left) and BS2a (right) points defined in table~\ref{tab:RDM:benchmark_points}.}
		\label{fig:RDM:BRs}
\end{figure}

The collider phenomenology of the model also strongly depends on the decay rates of the $S_1$ and $\chi_1$ states, which we present in figure~\ref{fig:RDM:BRs} for the BS1a (left panel) and BS2a (right panel) benchmark scenarios. For illustration purposes, we have set $\Delta= \mxOne - \mxZero = 150$~GeV and $\ydm=1.5$, such a large value of $\ydm$ being currently allowed by cosmology (see the detailed dark matter analysis of section~\ref{sec:dm}). In practice, the former choice forbids the $\chi_1 \to \chi_0 t \tau$ decay to happen, while the latter one aims to illustrate the impact of opening a dark decay channel for $S_1$. In all cases we have checked that the width over mass ratio of $S_1$ is below 10\%, so that the narrow-width approximation holds and all decays are prompt. 
In the case of the BS1a scenario $\chi_1$ decays mostly into the $\chi_0 b \nu$ final state  due to $\lamL > \lamR$. For the same reason, $S_1$ decays dominantly into $S_1 \to t \tau$ and $S_1 \to b \nu$ with almost comparable branching fractions, unless the dark channel $S_1 \to \chi_1 \chi_0$ is open. Moreover, the $S_1 \to c \tau$ decay rate is of about 10\% (see table~\ref{tab:RDM:benchmark_points}). For the BS2a scenario, the situation is reversed: both $\chi_1$ and $S_1$ decay predominantly through the $\lamL$ coupling into $\chi_0 c \tau$ and $c \tau$ systems respectively, with branching ratios well above 90 \%. 

We can summarise the collider phenomenology of the model in a few words as follows. The largest production rate at hadron colliders leads to a signature comprising a significant amount of missing energy and SM particles, and corresponds to $\overline{\chi}_1 \chi_1$ production and decay. Leptoquark pair and single production then subdominantly contribute to the full new physics signal, and yield a variety of signatures. These depend on the relative $S_1$ branching ratio into the dark mode ($S_1\to \chi_1 \chi_0)$ and into visible modes ($S_1\to c \tau$, $t \tau$ and  $b \nu$). In addition, a \emph{mixed} decay of the leptoquark pair into one invisible and one visible branch could contribute as well. We examine these options in detail in the next section.

\section{LHC constraints}
\label{sec:lhc}
Our model contains three new fields whose signatures can be tested by several searches at the LHC. These can be split into two main categories: searches for new physics in the missing transverse energy channel (targeting invisible final state particles, {\it i.e.}\ mostly $\chi_1$ decays in our case), and searches for leptoquark pair-production and visible decays (targeting $S_1$ decays to SM particles). A third category, inspired by the Coannihilation Codex~\cite{Baker:2015qna}, combines both these categories and is illustrated by a CMS search for leptoquarks plus dark matter~\cite{Sirunyan:2018xtm}. This search is important not only because it can provide relevant constraints on the viable regions of the parameter space, but also because it is the only search that allows us to unravel the link between the dark world and the visible world of the leptoquark decays. It is therefore useful in the characterisation phase of a newly discovered signal. 

\subsection{Missing energy searches}
\label{subsec:met}
In the considered class of models, $\chi_1$ pair production is the new physics process with the largest cross section. In general, the $\chi_1 \to \chi_0 l q$ decay subsequently gives rise to a final state involving a significant amount of MET. The corresponding LHC signature would thus contain leptons, jets and missing energy. Such a class of signatures is thoroughly searched for, so that tight constraints exist on the model parameter space. The only possibility to relax them all is to consider low values for the dark sector mass gap $\Delta = \mxOne-\mxZero$. In this case, $\chi_1$ decays proceed via $\chi_1 \to S_1^{(*)}\chi_0 \to q \ell\chi_0$. For $m_b < \Delta < m_t$ the dominant channel turns out to be $q\ell = b \nu$ ($c \tau$) for BS1 (BS2) scenarios, whereas for a more compressed situation in which $\Delta < m_b$, $\chi_1\to c \tau \chi_0$ occurs with a branching ratio of 100\% irrespectively of the specific $\lamL$ and $\lamR$ values. In this case $\chi_1$ becomes long-lived due a compressed spectrum in concomitance with a highly off-shell $S_1$ mediating the decay. In the rest of this work, we restrict ourselves to the prompt regime, which implies that $\Delta > 5$ GeV. Consequently, the proper lifetime $c \tau_{\chi_1} \lesssim 1$ mm in all the parameter space region under consideration.

\begin{table}
\begin{center}
\renewcommand{\arraystretch}{1.2}
\setlength\tabcolsep{17.5pt}
\begin{tabular}{lllll}
Search                & arXiv      & ${\cal L}~[{\rm fb}^{-1}]$ & BS1          & BS2          \\ \hline
CMS $b/c$ + MET       & 1707.07274~\cite{CMS:2017kil} & 35.9        & $\checkmark$ & X            \\
ATLAS $b\bar{b}$+MET  & 2101.12527~\cite{ATLAS:2021yij} & 139         & $\checkmark$ & X            \\
CMS $\ell_{\rm soft}$ + MET & 1801.01846~\cite{CMS:2018kag} & 35.9 & X            & $\checkmark$ \\
ATLAS mono-jet         & 2102.10874~\cite{ATLAS:2021kxv} & 139         & $\checkmark$ & $\checkmark$ \\
ATLAS $\tau^+ \tau^-$+MET & 1911.06660~\cite{Aad:2019byo} & 139         & X            & $\checkmark$ \\
ATLAS multi-jet        & 2010.14293~\cite{ATLAS:2020syg} & 139        & X            & $\checkmark$ 
\end{tabular}
\end{center}
\caption{List of MET searches at the LHC considered in this work. We indicate whether they can target the BS1 and BS2 scenarios.}
\label{tab:RDM:met_searches}
\end{table}

\begin{figure}
\centering
\includegraphics[width=0.75\textwidth]{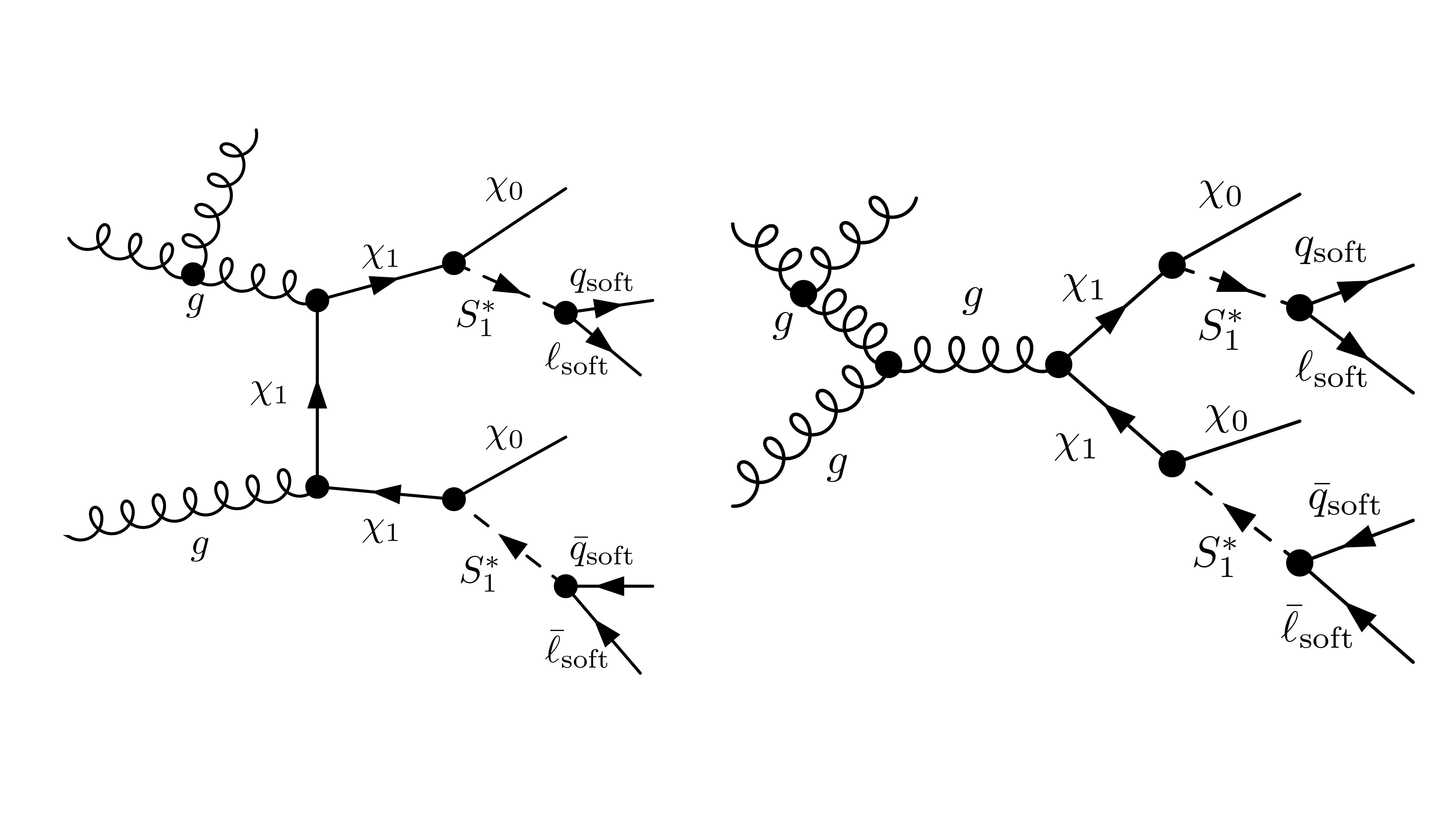}
\vspace{-1cm}
\caption{Feynman diagram representing mono-jet production (plus soft leptons and quarks) in the model considered.}
\label{fig:RDM:MonoJet}
\end{figure}

In order to assess the LHC constraints on the model that originate from missing-energy searches, we reinterpret the results of the searches collected in table~\ref{tab:RDM:met_searches}, in which we also specify whether the searches are appropriate to target $\chi_1\to b \nu \chi_0$ (BS1) and $\chi_1\to c \tau \chi_0$ (BS2) decays. The targeted signatures include the production of a significant amount of missing energy in association with a $b\bar b$ pair~\cite{ATLAS:2021yij}, one jet ({\it i.e.} the mono-jet channel) or more (the multi-jet plus MET channel)~\cite{ATLAS:2020syg,ATLAS:2021kxv}, heavy-flavour jets~\cite{CMS:2017kil}, a $\tau^+ \tau^-$ pair~\cite{Aad:2019byo} and soft leptons~\cite{CMS:2018kag}. Those missing energy searches lose sensitivity when the particle spectrum becomes compressed. For such a spectrum configuration, the traditional approach is to boost the system against additional SM objects, which gives rise to the so-called mono-X signals. In this case, while the SM decay products in the $\chi_1 \to \ell q \chi_0$ decay might not be hard enough to be triggered on, they might pass the $p_T$ reconstruction thresholds (see  figure~\ref{fig:RDM:MonoJet} for an illustration in the mono-jet case). The presence of these additional soft objects in the final state can then potentially enhance the sensitivity of the LHC searches, as exemplified in~\cite{Schwaller:2013baa} for electroweakinos searches with soft-leptons and pursued by the experimental collaborations in, for instance, refs.~\cite{ATLAS:2017vat,CMS:2018kag}. The more compressed the spectrum becomes, the higher the probability is that the soft decay products fail to pass the reconstruction thresholds, in which case only the additional radiation is reconstructed. In this limit one can set, for a given $\Delta$ value, an unavoidable model-dependent lower bound on the dark sector masses. The reinterpretation of experimental LHC studies in the highly compressed region can nonetheless be tricky, and requires a careful validation. For this reason, we only consider the multi-jet ATLAS~\cite{ATLAS:2020syg} study for values of $\Delta \gtrsim 30$~GeV. 

\begin{figure}
  \centering
  \includegraphics[width=0.48\textwidth]{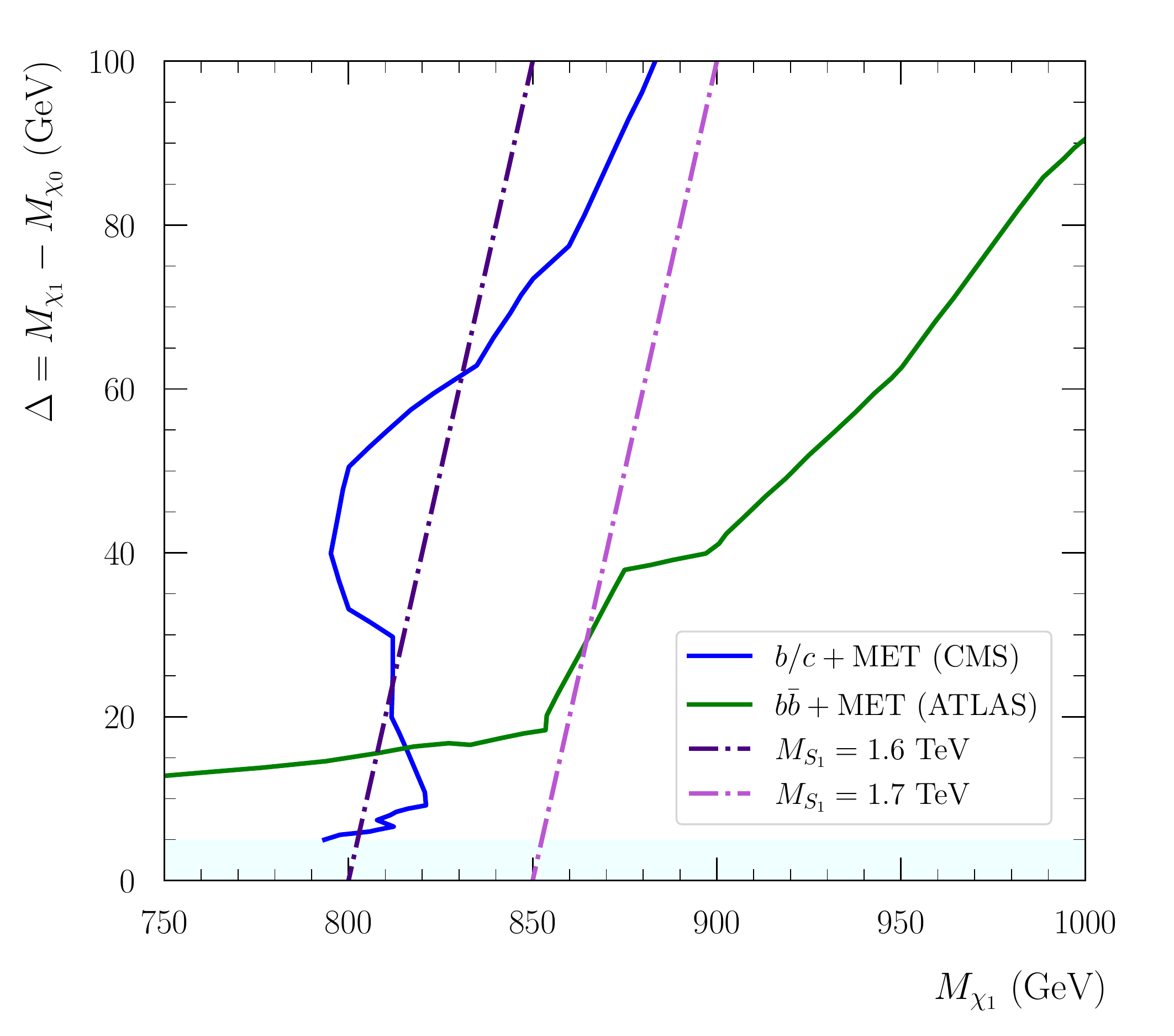}
  \includegraphics[width=0.48\textwidth]{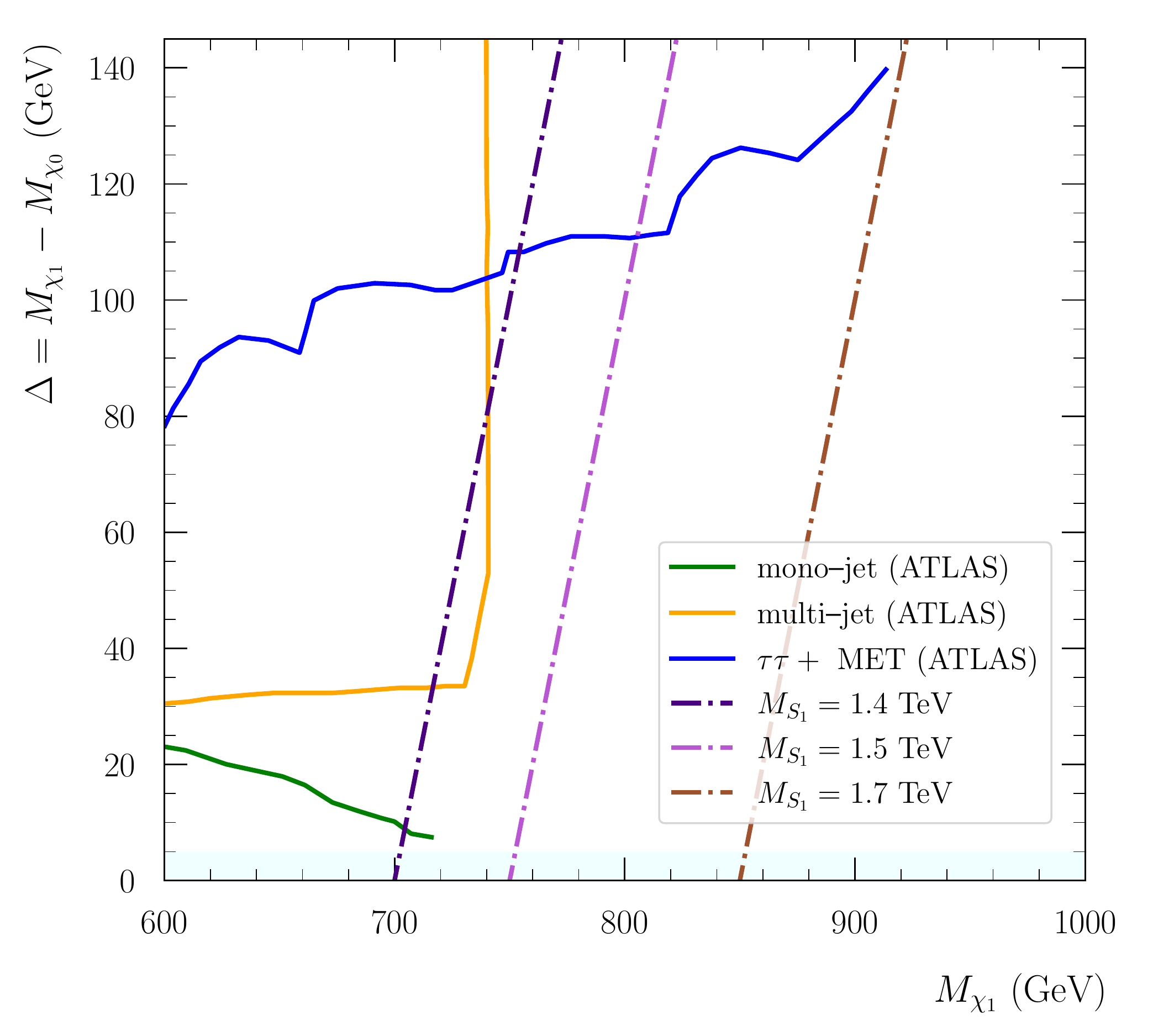}
  \caption{Collider constraints on our model, presented in the $(\mxOne, \Delta)$ plane for the BS1 (left panel) and BS2 (right panel) scenarios. We only show exclusions for the studies that are the most constraining ones in a given region of the parameter space, and the dot-dashed isolines correspond to $\mLQ = \mxOne + \mxZero= 1.6$~TeV (black) and 1.7~TeV (purple) in the left panel, and to $\mLQ = \mxOne + \mxZero=1.4$~TeV (black), 1.5~TeV (purple) and 1.7~TeV (brown) in the right panel. The region with $\Delta<5$~GeV is not considered in this study, as $\chi_1$ becomes long-lived and the collider phenomenology is markedly different. In the left panel we show constraints originating from the reinterpretation of the $b\bar{b}$ plus MET searches from ATLAS (green) and CMS (blue). In the right panel, we show mono-jet (green), multi-jet plus MET (orange), soft lepton plus MET (red) and $\tau^+ \tau^-$ + MET (blue) exclusions.}
\label{fig:RDM:met_bounds}
\end{figure}

All the analyses under consideration rely on the full LHC run~2 dataset of 139 fb$^{-1}$, except for~\cite{CMS:2018kag} and~\cite{CMS:2017kil} which only use 35.9 fb$^{-1}$ of data. The details of the reinterpretation of each of these studies are left for Appendix~\ref{app:met_searches}, and we present the resulting constraints in the $(\mxOne, \Delta)$ plane in figure~\ref{fig:RDM:met_bounds} for the benchmark points BS1 (left panel) and BS2 (right panel). Our results are obtained by generating hard scattering events as in section~\ref{sec:pheno}, that we then match with parton showering and hadronisation as modelled in {\sc Pythia 8.2}~\cite{Sjostrand:2014zea}. Detector effects, event reconstruction and the computation of exclusions for all considered analyses are next carried out through the reinterpretation frameworks of {\sc MadAnalysis}~5~\cite{Conte:2018vmg}, {\sc CheckMATE~2}~\cite{Dercks:2016npn} and {\sc SModelS}~\cite{Ambrogi:2018ujg}, the former two programs depending on {\sc Delphes}~3~\cite{deFavereau:2013fsa} for the simulation of the detector response. In order to help visualising the lower bound on $\mLQ$ set by the considered searches, we overlay in figure~\ref{fig:RDM:met_bounds}, in dot-dashed style, lines of constant $\mxOne + \mxZero$ values. We use values of 1.6~TeV (black) and 1.7~TeV (purple) for the left panel of figure~\ref{fig:RDM:met_bounds} (BS1), and we employ values of 1.4~TeV (black), 1.5~TeV (purple) and 1.7~TeV (brown) for its right panel (BS2).
 
For the BS1 scenarios, $b\bar{b}$+MET searches strongly constrain the viable regions of the parameter space, and they imply that $\mxOne \gtrsim$ 800 GeV. Mono-jet searches are not competitive with such a high limit, that additionally leaves very little room for leptoquark masses smaller than 1.6~TeV. We indeed find that either a large spectrum compression has to be enforced, or $\mLQ \gtrsim 1.7$ TeV. For such a mass of $\mLQ=1.7$~TeV, the LO $p p \to S_1 S_1$ cross section at the 14 TeV LHC is approximately of 53 ab, hence yielding about 1500 signal events at the HL-LHC. The accompanying leptoquark signals (discussed in section~\ref{subsec:lqs}) may thus be out of the HL-LHC reach. Since we are interested in the interplay among the different searches, we do not consider heavier $S_1$ masses in the benchmark setups studied in section~\ref{sec:bench}.

In contrast, the allowed parameter space region for BS2 scenarios (right panel) is larger. The tight constraints set by the $b\bar{b}$ and $b/c$ + MET searches for scenarios BS1 do not apply anymore, as those searches implement an explicit tau veto that effectively reduce their sensitivity to BS2 setups. The signal regions are indeed only populated by events featuring tau leptons that fail identification. All other considered searches therefore yield important constraints on the model. The mono-jet and soft lepton searches leave open an interesting parameter space region in which $\mLQ>1.5$ TeV, $\Delta \in [5, 20]$ GeV and $\mxOne \in [680, 750]$ GeV. Lower leptoquark masses are still allowed, although $\Delta$ values between 20 and 50~GeV are quite restricted by the soft leptons plus MET search and $\Delta$ values larger than 80~GeV are quite constrained by $\tau^+ \tau^-$ + MET searches. In the most extreme situations, scenarios with $\mLQ=1.4$ TeV are still viable, provided that $\Delta \in [80-100]$ GeV. For heavier leptoquarks, $\Delta$ values up to more than 100 GeV are allowed, the exact limit being set by $\tau^+ \tau^-$ + MET searches. BS2 setups are therefore promising (yet challenging) phenomenological scenarios, due to the dominance of the $c \tau$ final state in both the $\chi_1$ and $S_1$ decays.

To summarise this subsection, we have found that, as anticipated, the dark sector must exhibit some degree of compression to be viable relative to the new physics LHC search program. This in turn sets a lower limit on the $S_1$ mass in order for the decays into the dark sector to be kinematically open. If decays through a $\lamL$ coupling dominate (BS1), then we are forced to have $\mLQ \gtrsim 1.6$ TeV. If in contrast $\lamR$ dominates (BS2), then the range $\mLQ \in [1.4, 1.5]$ TeV is open as long as $\Delta \in [5, 30]$ GeV. In addition, for $\mLQ \gtrsim 1.4$~TeV multi-jet constraints barely apply, thus enlarging the viable mass gaps to $\Delta \in [50, 100]$~GeV. In this case, however, some parameter configurations are excluded by the soft-lepton plus MET search. We now dive in the next subsection into direct searches for leptoquarks, knowing that they have to be heavier than 1.4--1.5 TeV.

\subsection{Leptoquark searches}
\label{subsec:lqs}
\subsubsection{Searches for leptoquark pair-production under consideration}
In this section we discuss constraints that can be set on our model from ATLAS and CMS searches for leptoquark pair-production and decay into SM final states. The subset of these searches dedicated to third generation leptoquarks rely on the same leptoquark model as in this work, so in most cases the reinterpretation of the results is 
straightforward. The only caveat is that in the simplified model considered by the LHC collaborations, it is assumed that ${\rm BR}(S_1 \to t \tau^-) = 1- {\rm BR}(S_1 \to b \nu)$. This is equivalent, in our language, to consider all leptoquark couplings to be zero, except for $\lamL$. As we are interested in exploring a more generic case where these two branching fractions do not add up to unity, certain care must be taken for a proper reinterpretation\footnote{We note that the sensitivity potential of the LHC and the HL-LHC on scalar leptoquarks through the non-resonant $pp\to \ell q$ and the resonant pair production have been carried in {\it e.g.}\cite{Chandak:2019iwj,Bhaskar:2021gsy,Iguro:2020keo,Endo:2021lhi}.}.

\begin{table}\renewcommand{\arraystretch}{1.2}
\begin{center}\resizebox{\textwidth}{!}{\begin{tabular}{c|ccc}
Decays   & \multicolumn{1}{c}{$t \tau$} & \multicolumn{1}{c}{$b \nu$} & \multicolumn{1}{c}{$c \tau$}\\ \hline
$t \tau$ & \cellcolor{LightBlue1}ATLAS-CONF-2020-029~\cite{ATLAS:2020sxq} & \cellcolor{LightBlue1}ATLAS-CONF-2021-008~\cite{ATLAS:2021aui} & \cellcolor{LightPink1} \\
$b \nu$ & $-$ & \cellcolor{LightBlue1}2101.12527~\cite{ATLAS:2021yij} & \cellcolor{LightPink1}\\
$c \tau$ & $-$ & $-$ & \begin{tabular}[c]{@{}l@{}}\cellcolor{LightGoldenrod1}{Rescaling of 1803.08103~\cite{Aaboud:2019bye}}\end{tabular}
\end{tabular}}
\end{center}
\caption{SM final states originating from leptoquark pair production and decay, provided together with their coverage by the LHC searches considered in this work. The first (second) leptoquark decay is indicated by the row (column) of the table, and the colour code shows the relevance of the signatures for the BS1 (blue) and BS2 (yellow) benchmarks. Signatures appearing in red are currently not covered experimentally.}
\label{tab:LQ_pairs}
\end{table}

We collect in table~\ref{tab:LQ_pairs} the different possibilities for the (SM) final state originating from $S_1 S_1$ production and decay, together with their relevance for the considered benchmarks BS1 and BS2. Moreover, we indicate for each channel the corresponding LHC searches whose results are reinterpreted within this work. As visible from the table, we recast the results of several leptoquark searches that focus on the $t \tau t \tau$~\cite{ATLAS:2020sxq}, $b \nu b \nu$~\cite{ATLAS:2021yij} and $b \nu t \tau$~\cite{ATLAS:2021aui} signatures. The second of these searches has already been used in section~\ref{subsec:met} to constrain $\chi_1$ pair production and decay through its $b\bar{b}$ + MET signature. The last of these searches, dedicated to a \emph{mixed} decay of the leptoquark pair, was pushed forward in a prequel of this work~\cite{Brooijmans:2020yij}\footnote{The kinematic reconstruction of this channel was studied in reference~\cite{Gripaios:2010hv}.}, and has been since then integrated in the LHC new physics search program~\cite{ATLAS:2021yij,CMS:2020wzx}. In the following, we recast only the ATLAS analysis and not the CMS one, as it has a larger sensitivity in the parameter space region in which both branching ratios BR($S_1 \to b \nu$) and BR($S_1 \to t \tau$) are large.

The table also shows that while there is no dedicated $pp\to S_1 S_1\to c \tau c \tau$ search yet, we can extract bounds in this channel. We consider the results of the existing $pp\to S_1 S_1\to b \tau b \tau$ search~\cite{Aaboud:2019bye}, and we rescale the excluded rates by a factor $\omega$ defined as the square of the $c$-tagging over $b$-tagging average efficiencies~\cite{ATLAS:2016gsw}
\be 
\omega = \bigg[\frac{8.3\%}{70\%}\bigg]^2 \sim 1/71.
\ee
On the contrary, even if they are not relevant for the chosen benchmarks, the mixed channels $pp\to S_1 S_1 \to c \tau b \nu$ and $pp\to S_1 S_1\to c \tau t \tau$ are currently not covered experimentally. While we could design a new benchmark slope BS3 for which such channels could play a very important role, we refrain from doing so as we would end up in a situation with poor LHC constraints originating from direct searches for leptoquarks. This would indeed undermine our motivation of building scenarios emphasising the interplay between different sorts of searches for dark matter and the $\RD$ anomalies. We instead embolden the experimental collaborations to scrutinise any not-yet-probed channel ($pp\to S_1 S_1 \to c \tau b \nu$, $c \tau t \tau$ and $c\tau c\tau$) through dedicated searches, so that we could obtain more accurate bounds on the benchmarks considered and be able to obtain relevant bounds on new, not yet considered, scenarios.

\subsubsection{Reinterpreting LHC leptoquark search results in the mixed $b\nu t\tau$ channel}
\label{sec:lqmixed}

In the next subsection, we will present the coverage of the considered searches for leptoquark pair production and decay. In the latter, the experimental results are often reported in terms of branching ratio exclusions as a function of the leptoquark mass. We adopt the same convention here. For the $b \nu b \nu$ and $t \tau t \tau$ final states this information can be directly extracted from the published results. For the mixed search, we decide to employ the geometric mean of the $b \nu$ and $t \tau$ branching ratios, $\sqrt{{\rm BR} (S_1 \to b \nu) {\rm BR}(S_1 \to t \tau)}$, instead of choosing a particular branching ratio. Contrary to the naive expectation, the sensitivity does not scale with the product of the branching ratios, which makes the reinterpretation of the search not as straightforward as for the two previous cases, where both $S_1$ decay into the same final state.

Our reinterpretation procedure will make use of the acceptances, efficiencies and upper limits reported by the ATLAS collaboration in the auxiliary material of~\cite{ATLAS:2021jyv}. We start by noting that ATLAS implicitly assumes that only the $b \nu$ and $t \tau$ channel are open, hence $x={\rm BR}(S_1 \to t \tau) = 1- {\rm BR}(S_1 \to b \nu)$. The acceptance $A$ and efficiency $E$ are reported as a function of $\mLQ$ and $x$\footnote{While ATLAS presents acceptances and efficiencies for the di-tau and single-tau signal regions, the former are suppressed by more than an order of magnitude with respect to the latter in the parameter space of interest, and hence can be neglected.}, these functions encoding the probability that a given $b \nu t \tau$ partonic final state appears, after reconstruction, in a given signal region. The relevant ingredients are the reconstruction and misidentification of $b$-jets and $\tau$ leptons, which can not be simply obtained by factorising each decay. Nonetheless, these functions are invariant if both the $t\tau$ and $b \nu$ branching fractions are scaled by the same amount. In other words, the values of $A$ and $E$ depend only on
\be\label{eq:xval}
x^{\prime} = \frac{ {\rm BR} (S_1 \to t \tau)}{ {\rm BR} (S_1 \to t \tau) +  {\rm BR} (S_1 \to b \nu)} \, ,
\ee
and not on the actual values of the two branching ratios. In the special case considered by the ATLAS collaboration, where no additional decays are present, the denominator of eq.~\eqref{eq:xval} is equal to unity, hence $x^{\prime}=x$. The experimental dilution factors can be read-off from the ATLAS tables at $A(x') E(x')$. Therefore the number of expected events for a given leptoquark mass and $x^{\prime}$ is proportional to
\be\label{eq:n95_mixed}
N \propto \sigma(\mLQ)~{\rm BR}(S_1 \to t \tau)~{\rm BR}(S_1 \to b \nu) ~A(\mLQ,x^{\prime})~E(\mLQ,x^{\prime}) \, ,
\ee
where $\sigma$ is the cross section for $p p \to S_1 S_1$. For the sake of simplicity we assume in what follows both branching ratios to be equal, hence i) $A$ and $E$ are evaluated at $x^{\prime}=0.5$, and ii) the geometric mean is equal to  ${\rm BR} (S_1 \to t \tau)$. Moreover, we must assume that additional decay channels that have zero acceptance and/or efficiency. Extending the method for a general case  with arbitrary branching fractions into $b \nu$ and $t \tau$ is straightforward.

For $x=0.5$ the ATLAS collaboration reports an upper limit of  $m_{95} \lesssim  1250$ GeV. Exploiting the fact that the excluded number of signal events does not change, with the help of eq.~\eqref{eq:n95_mixed} we express the maximum allowed branching fraction for a given mass $m$,
\be
{\rm BR} (S_1 \to t \tau)^{95} (m) = 0.5 \Biggl(\frac{\sigma(m_{95}) A(m_{95}, 0.5) E(m_{95},0.5)}{\sigma(m) A(m, 0.5) E(m,0.5)} \Biggr)^{1/2} \, ,
\ee
which allows us to derive a exclusion curve in the $\mLQ$-branching ratio plane. In our final results we consider the effect of finite top quark masses ($x \neq 0.5$) and for $\sigma$ we employ LO production cross sections.

\subsubsection{LHC sensitivity to our model via searches for leptoquark pair production}

\begin{figure}
 \centering
 \includegraphics[width=.65\textwidth]{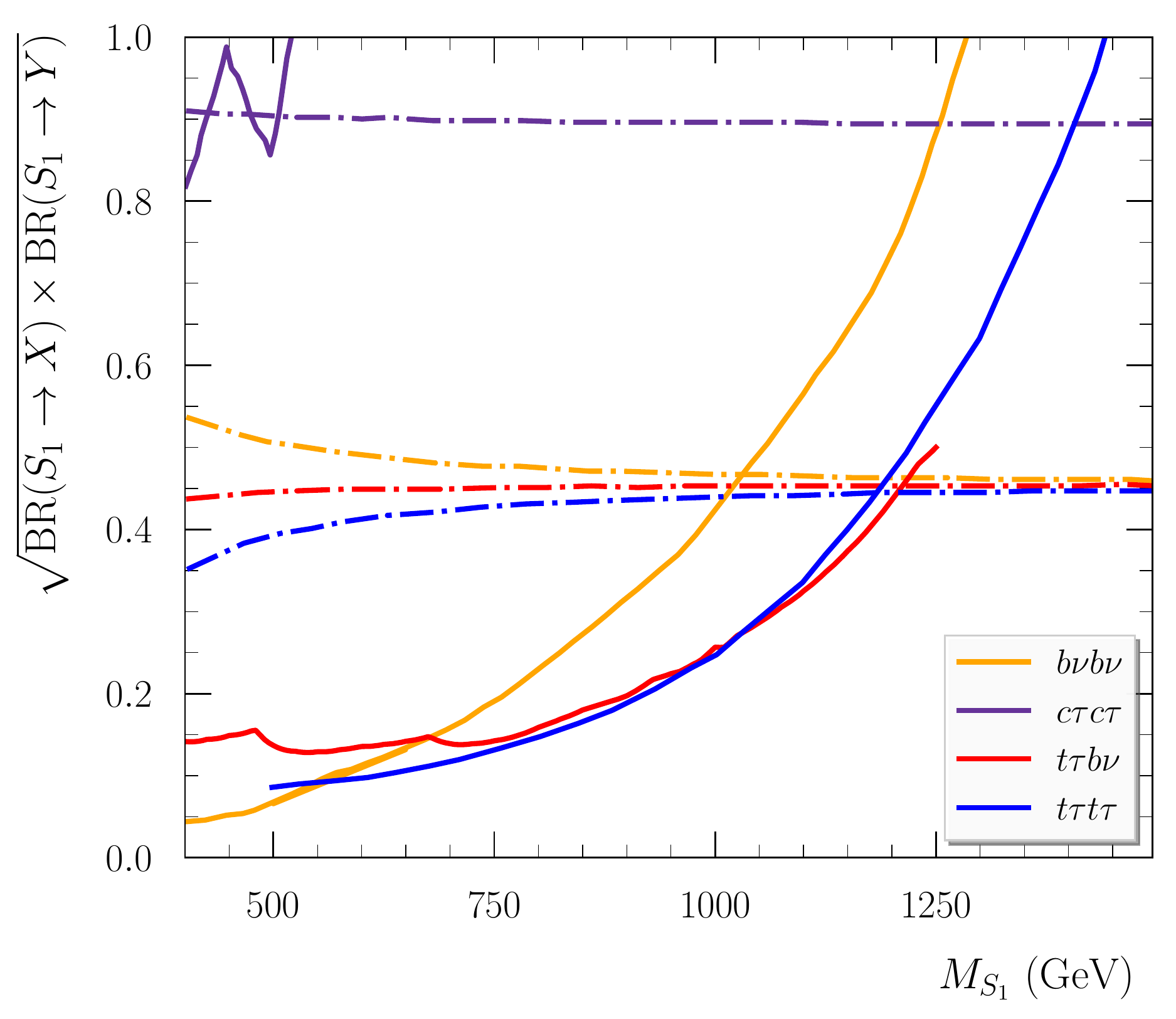}
 \caption{Sensitivity of the considered LHC searches for leptoquark pair production and decay into third generation fermions. We present as solid lines the branching ratio reach as a function of the leptoquark mass for the $t \tau t \tau$ (blue), $b \nu b \nu$ (orange), and $c \tau c \tau$ (purple) channels, as well as for the mixed mode in the $t \tau b \nu$ final state (red). We show through dashed lines the branching ratios corresponding to our benchmark points BS1 (for the $c \tau c \tau$ channel) and BS2 (for the three other channels). For the mixed decay the dashed line represents the geometric mean of ${\rm BR}(S_1 \to t \tau)$ and ${\rm BR}(S_1 \to b \nu)$.}
\label{fig:RDM:lq3_summary}
\end{figure}

In figure~\ref{fig:RDM:lq3_summary} we present the sensitivity of the considered leptoquark searches through coloured solid contours in the $\mLQ$ versus BR($S_1\to X$) plane. We overlap to those lines dashed lines representing the $S_1$ branching ratios as a function of $\mLQ$ for the BS1 and BS2 scenarios. In the BS1 case, we consider decays involving third generation fermions, while for the BS2 case, we consider the $c \tau$ branching ratio as a $y$-axis variable. The intersection of a solid and dashed line of a specific colour then provides the maximum lower bound on $\mLQ$ for these scenarios originating from the corresponding search, when we assume $\ydm=0$ ({\it i.e.}\ dark leptoquark decays being kinematically forbidden).

In the case of the mixed channel we use the geometric mean of the two relevant branching ratios as a $y$-axis quantity, as discussed in section~\ref{sec:lqmixed}. We see that, as obtained by the ATLAS collaboration, the mixed search has a slightly larger sensitivity than the other two channels for $x \approx 0.5$.

For BS1 scenarios, the leptoquark search in the mixed decay channel leads to $\mLQ > 1.21$ TeV while for BS2 scenarios, the bounds from the rescaled $c \tau c\tau$ search turn out to be of about 500 GeV. With a 5\% leptoquark branching fraction into $b \nu$ and $t \tau$ final states, the other searches are found to set lower limits (of about 450 GeV in the extreme case of the $b \nu b\nu$ search), while the mixed search has a rate 100 times lower than in the BS1 scenario and can thus be ignored. Since the $c \tau c \tau$ search can be improved by employing dedicated $c$-tagging algorithms, we should take the associated bounds with a grain of salt. Even with a lot of potential improvements, it seems hard to reach a mass limit of the order of 1~TeV.

To conclude, the constraints arising from direct leptoquark searches do not compete with the ones stemming from missing energy searches via $\chi_1$ pair-production for both BS1 and BS2 scenarios, of course provided that leptoquark dark decays are open. However, for BS1 scenarios, the bounds that we find are not very different. We could thus imagine a future situation in which both the leptoquark searches via decays into third generation fermions and the missing energy searches into the $b\bar{b}+ {\rm MET}$ channel would see seemingly uncorrelated excesses. We further discuss this outcome and potential benchmark points in section~\ref{sec:bench}.

\subsubsection{LHC sensitivity to our model via leptoquark single production and decay}

Before embarking with the study of the BS1 and BS2 phenomenology of leptoquark pair production and decay in a mixed visible/dark decay channel in the next section, we analyse the bounds that could stem from single leptoquark production in association with a lepton. Due to the large PDF suppression associated with an initial $b$-quark, this is only considered in the BS2 case (that involves initial states containing a charm quark). As for the $c \tau c \tau$ search channel, the lack of a dedicated $c \tau \tau$ search forces us to rescale the results of existing $b \tau \tau$ searches. In this case, we rely on a CMS analysis of 36 fb$^{-1}$ of LHC data~\cite{Sirunyan:2018jdk}, as the corresponding full run-2 search~\cite{CMS:2020wzx} does not focus on the $b \tau \tau$ final state, but instead on the $t \tau \nu b$ and $t \tau \nu$ ones. We proceed in an analogous manner as done for the $c \tau c \tau$ case, using however this time only one $b$-tagging to $c$-tagging rescaling factor with the average efficiencies relevant for the analysis~\cite{Sirunyan:2018jdk}. Our bounds are thus derived with a suppression factor $\omega$ given by
\be \omega = \bigg[\frac{63\%}{12\%}\bigg]\sim 5.2.\ee

\begin{figure}
 \centering
 \includegraphics[width=.75\textwidth]{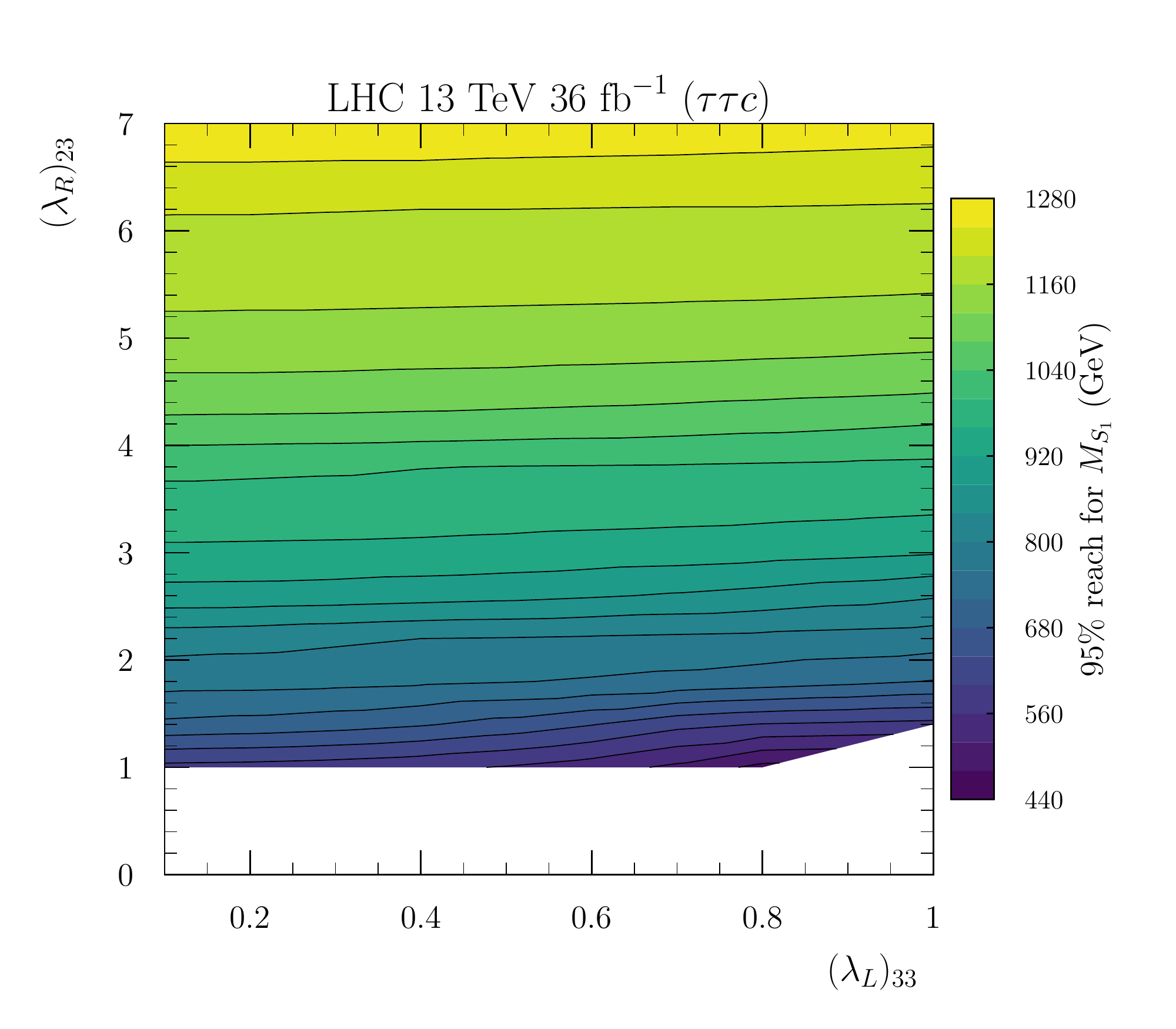}
\caption{Exclusion bounds on our model derived from the $c \tau \tau$ search~\cite{Sirunyan:2018jdk}, presented in the $(\lamL, \lamR)$ plane.}
\label{fig:RDM:lq3_summary_bis}
\end{figure}

The results of our scaling are presented in figure~\ref{fig:RDM:lq3_summary_bis}, in the $(\lamR, \lamL)$ plane. We conclude that this search is only relevant for the BS2 case, as expected, but that even in this case we still obtain quite mild bounds that do not compete with those originating from the missing energy searches. In particular, for $\lambda=1.5$, we obtain a lower limit on the leptoquark mass in the ballpark of 700~GeV, which lies further away from our 1.5~TeV benchmark value.

\subsection{Resonant leptoquark plus missing energy search}
\label{sec:lqmet}

\begin{figure}
\centering
\includegraphics[width=0.65\textwidth]{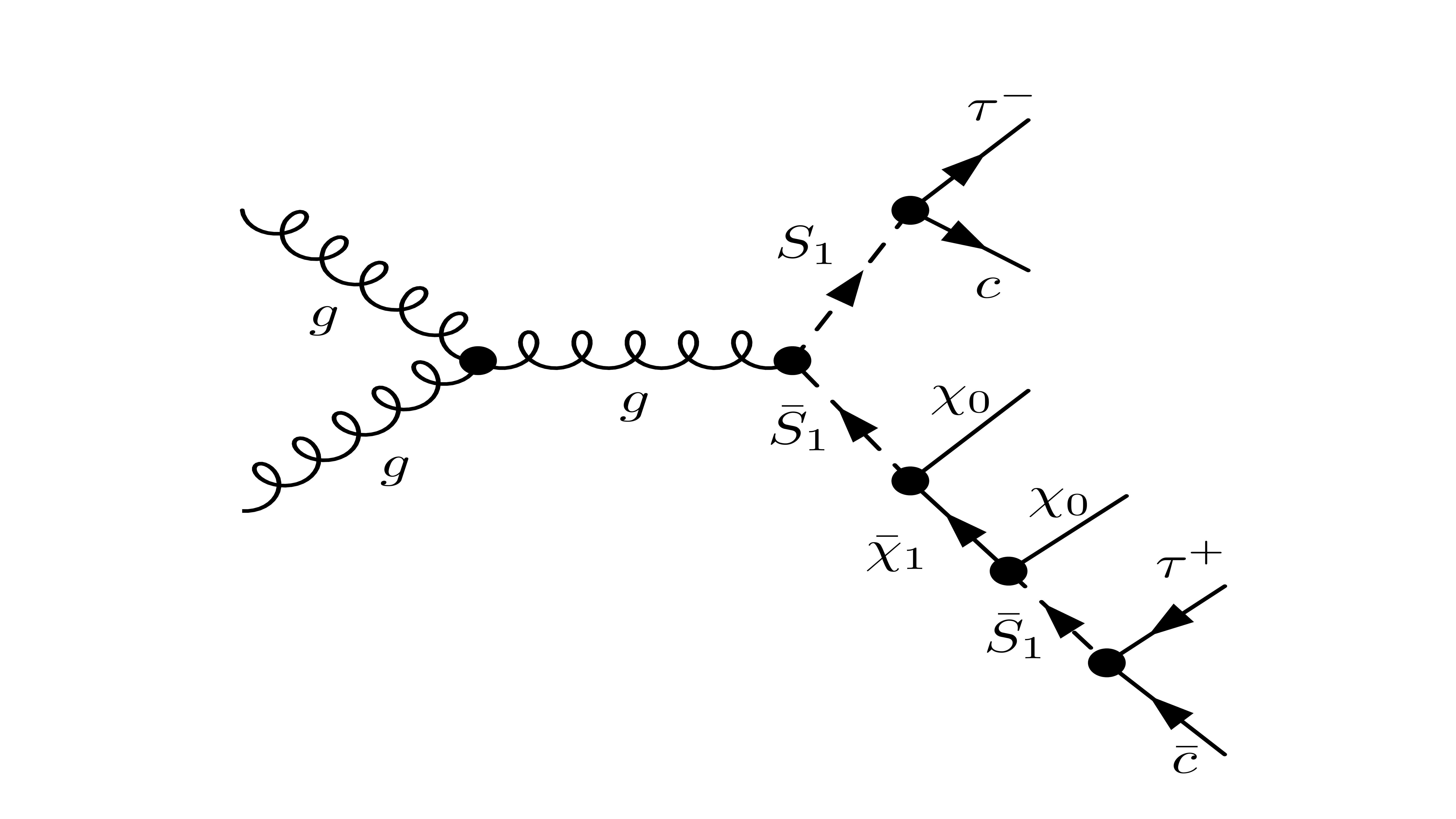}
\caption{Example of Feynman diagram illustrating the QCD-driven leptoquark pair production and decay and leading to the $c \tau$ plus MET signature.}
\label{fig:FD-LQMET}
\end{figure}

The model under consideration can also be constrained by reinterpreting the results of the CMS analysis~\cite{Sirunyan:2018xtm} specifically searching, in 77.4~fb$^{-1}$ of LHC data, for signatures of dark matter that originate from the decay of a heavy leptoquark. In this analysis, the signal is assumed to arise from the production of a pair of heavy leptoquarks which decay differently. One leptoquark is assumed to decay into a quark of the second generation (a charm or a strange quark depending on the quantum numbers of the LQ) and a muon, while the second one decays into a $\chi_1 \chi_0$ pair with $\chi_1$ subsequently decaying into two second-generation fermions  in association with dark matter. The resulting process, for which a representative Feynman diagram is displayed in figure~\ref{fig:FD-LQMET}, is therefore $ p p \to S_1 S_1 \to c \mu \chi_1 \to c \mu\ \chi_0 c \mu$. Consequently, the searched for signal consists of a significant amount of missing energy, jets and a high-$p_T$ muon.

Our interest would lie in a similar search targeting a final state featuring tau leptons instead of muons. As discussed in section~\ref{sec:anomalies}, a leptoquark explanation for the $D^{(*)}$ anomalies indeed prefers vanishing couplings of the scalar leptoquarks to second generation fermions, \emph{i.e.} $(\lambda_R)_{22} = (\lambda_L)_{22} = 0$, and non-vanishing couplings to tau leptons and second generation quarks. However, such a dedicated search does not exist. While a detailed study by the experimental collaborations including also electrons and tau leptons would be very important, we can only in the meantime make use of the only existing analysis~\cite{Sirunyan:2018xtm}.

We expect that the above search would not constrain significantly the model configurations investigated in this work. All the muons in the process should indeed arise from leptonically-decaying tau leptons, and should thus be softer. The signal selection efficiency is therefore in principle quite different from the one in ref.~\cite{Sirunyan:2018xtm}. To test this assumption, we have implemented this search in the {\sc MadAnalysis}~5 framework~\cite{Conte:2012fm,Conte:2014zja,Dumont:2014tja,Conte:2018vmg} and validated it~\cite{Fuks:2020xxz, Fuks:2021wpe} by reproducing the detailed cutflow tables kindly provided by the CMS collaboration\footnote{The source code and the validation material can be found on the \textsc{MadAnalysis} 5 dataverse \cite{ICOXG9:2020}.}. For three leptoquark masses of $\mLQ = 500$~GeV, 1~TeV and 1.5~TeV, the cumulative efficiency after the full selection is of around 1\%, as shown in table~\ref{tab:efficiency-LQmuc} (see appendix~\ref{app:lqmet_HL-LHC}, that contains extra details about our recast). Such an efficiency is as expected extremely small as compared with the case where the leptoquarks decay predominantly into muons.

\begin{figure}[!t]
\centering
\includegraphics[width=0.495\linewidth]{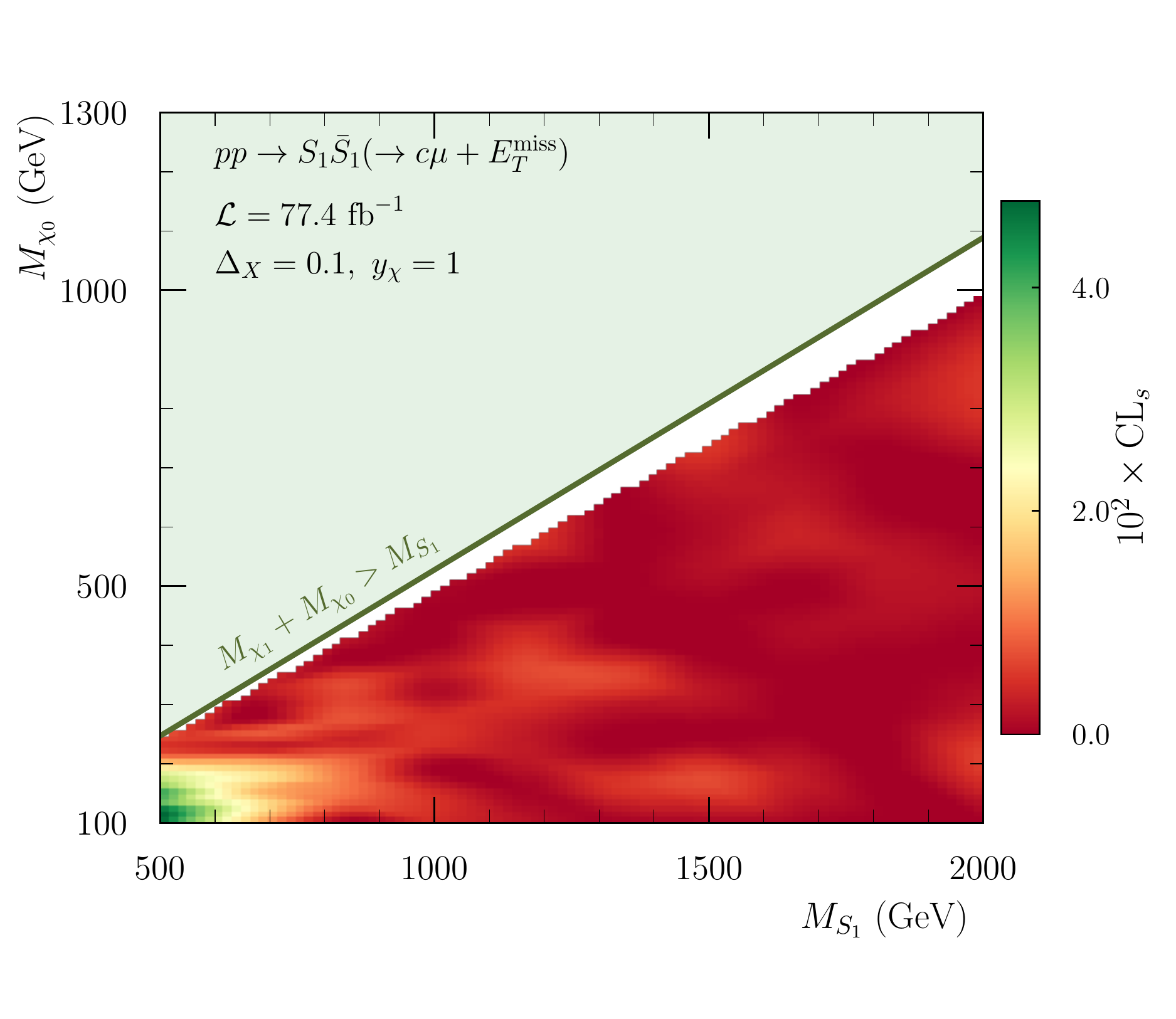}
\hfill
\includegraphics[width=0.495\linewidth]{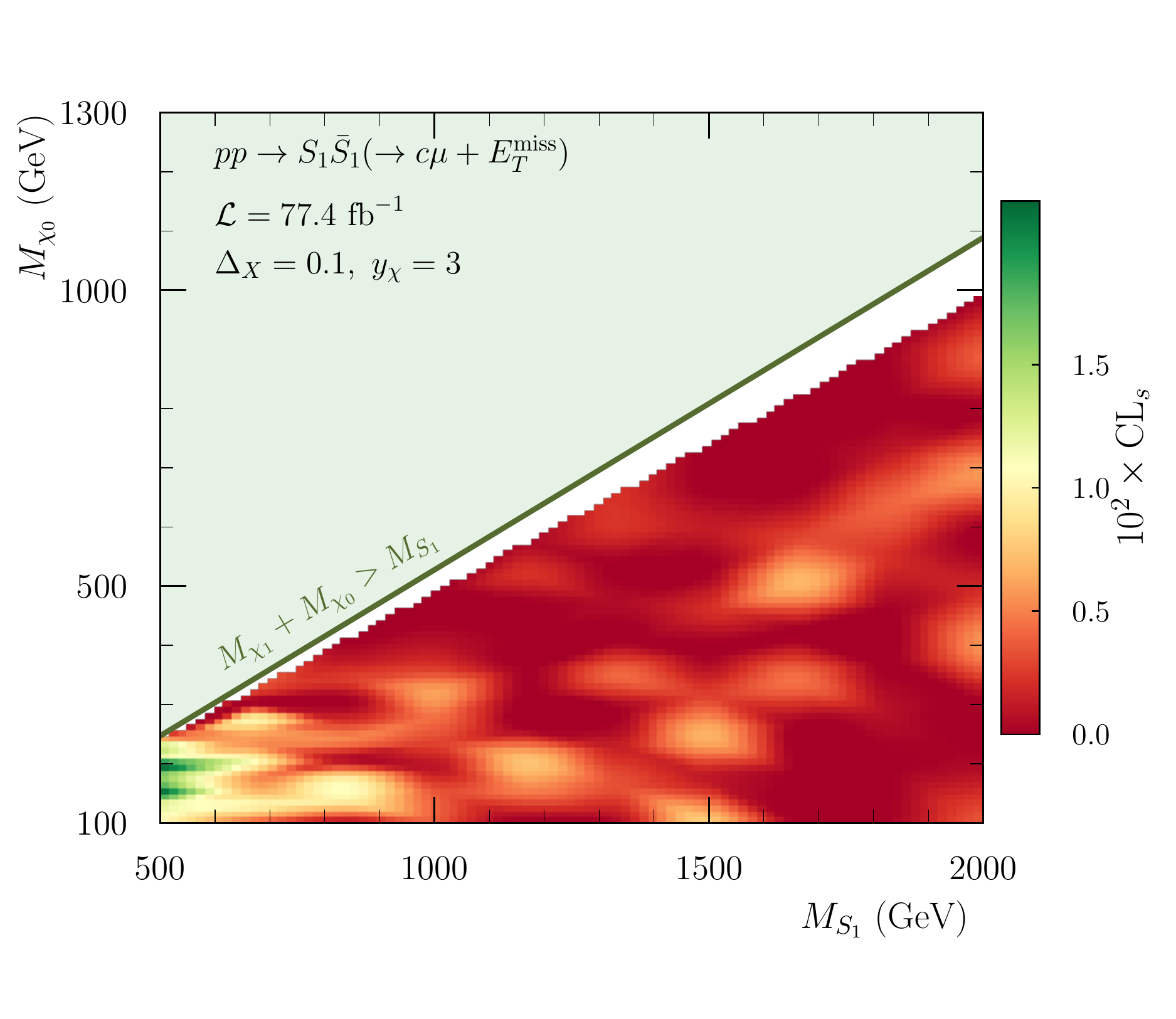}
\vfill
\includegraphics[width=0.495\linewidth]{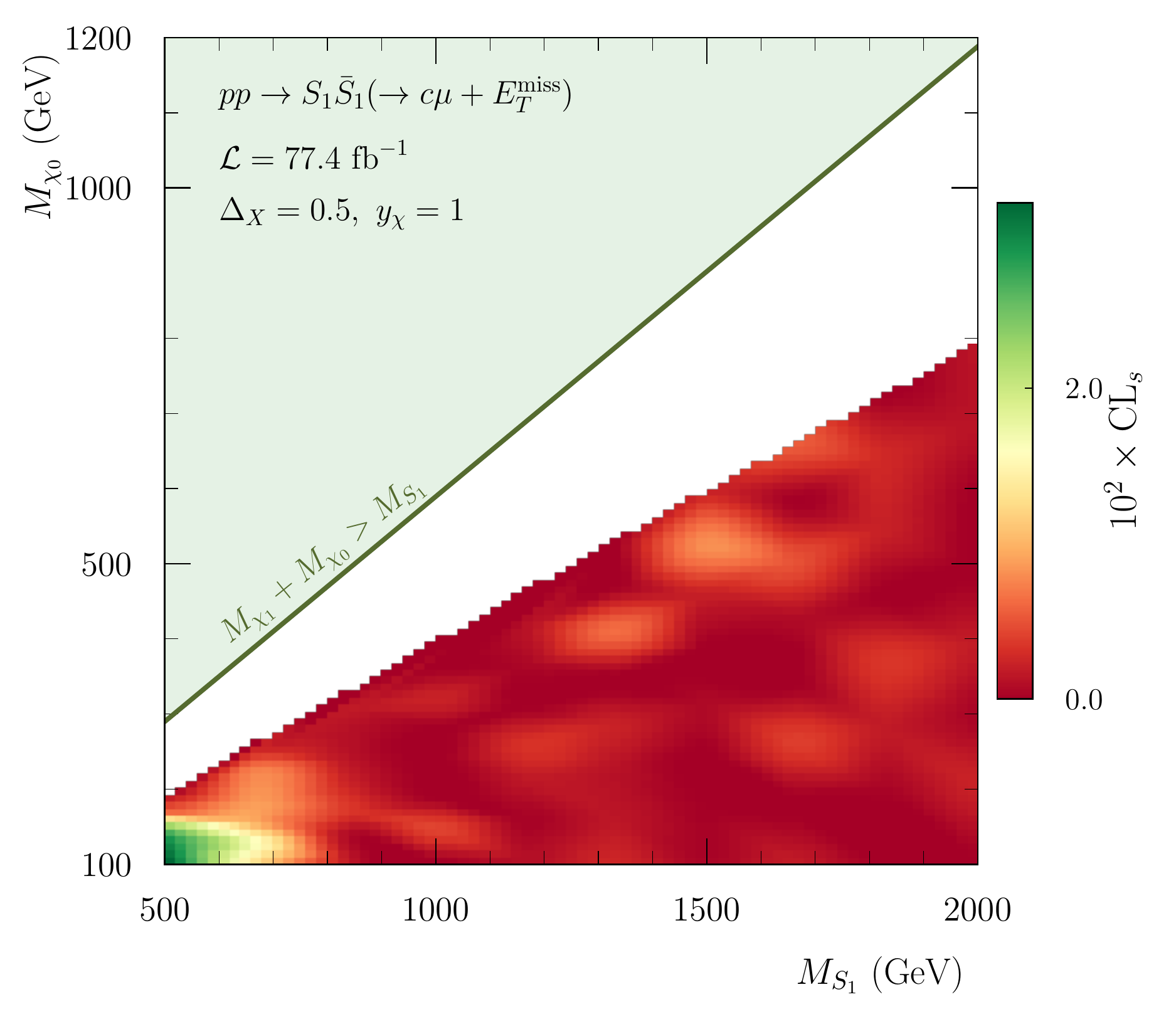}
\hfill
\includegraphics[width=0.495\linewidth]{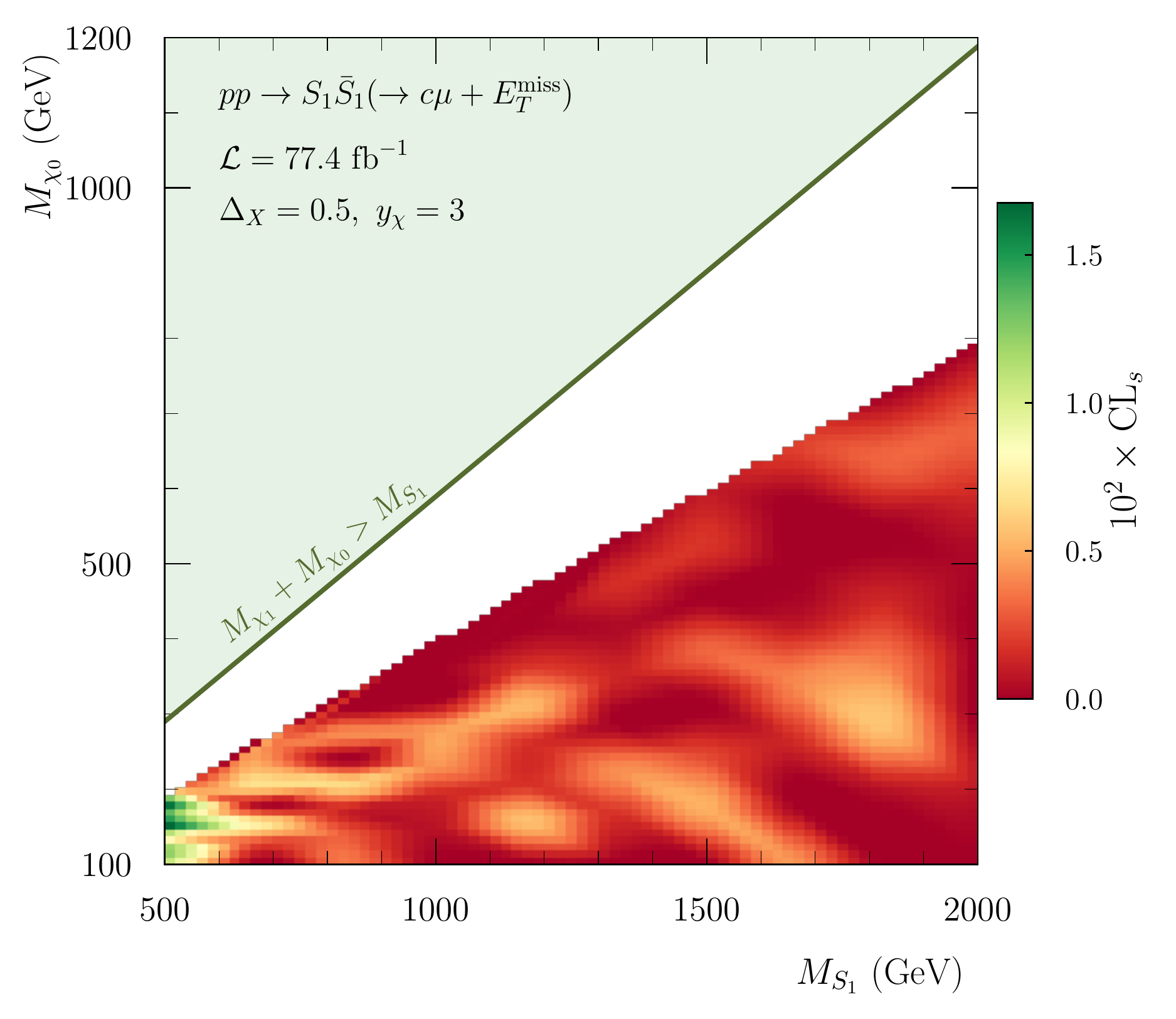}
\caption{Exclusion contour in the $(\mLQ, M_{\chi_0})$ mass plane for $\ydm = 1$ (left) and $\ydm = 3$ (right), and for $\DX = 0.1$ (top) and 0.5 (bottom). These bounds are obtained from the CMS leptoquark plus MET search~\cite{Sirunyan:2018xtm}. The dark green line in all the panels defines the kinematical boundary $\mLQ < M_{\chi_0} + M_{\chi_1}$, above which the $S_1 \to \chi_1 \chi_0$ decay is forbidden.}
\label{fig:CLs-resonantLHC}
\end{figure}

To get illustrative exclusion contours in our parameter space, we design new  benchmark scenarios belonging to the BS2 slope (as the CMS study targets $S_1 \to c \tau$ decays). By introducing the relative dark mass splitting 
\begin{eqnarray}
\DX = \frac{M_{\chi_1} - M_{\chi_0}}{M_{\chi_0}},
\end{eqnarray}
we define four sets of benchmark points by fixing the $\DX$ and $\ydm$ parameters,
\be\bsp
{\rm BP1}:& \qquad \DX = 0.1,\ \ydm = 1, \\
{\rm BP2}:& \qquad \DX = 0.1,\ \ydm = 3, \\
{\rm BP3}:& \qquad \DX = 0.5,\ \ydm = 1, \\
{\rm BP4}:& \qquad \DX = 0.5,\ \ydm = 3.
\esp\ee
We then scan over the leptoquark mass $\mLQ$ and the dark matter mass $\mxZero$ in the range $[500, 2000]~$GeV. The results are  shown in figure~\ref{fig:CLs-resonantLHC}. Almost no point is excluded at the $95\%$ confidence level, and the situation is similar at the HL-LHC (see appendix~\ref{app:lqmet_HL-LHC} for an extrapolation of the reach following the guidelines of~\cite{Araz:2019otb}). The investigated scenarios are therefore outside the reach of the LHC, at least in the channel under consideration. Nonetheless, we remind that a full characterisation of the model would require this channel (or at least the similar one where taus are produced in the final state, instead of muons) to establish the connection between dark matter and the $\RD$ anomalies.

\subsection{Conclusive statements about all considered LHC searches}
To conclude this section, we have found that in our benchmark scenarios the suite of missing energy searches should be the first ones to catch a glimpse of new physics. Next, depending on the value of $\mLQ$, one could expect an additional excess over the SM expectation in some leptoquark studies. These two seemingly unrelated excesses should finally be connected by means of mixed leptoquark plus MET searches, whose current incarnations are not sufficient to probe the considered scenarios, even at the HL-LHC reach. Estimating the reach of this search for the next generation of colliders would moreover be an important task, which is outside the scope of the current work. We however provide preliminary steps in that direction in appendix~\ref{app:lqmet_HL-LHC}, for the interested reader.

\section{Dark matter constraints}
\label{sec:dm}
In this section we study the dark matter (DM) phenomenology of our model. We start by considering direct detection rates in section~\ref{sec:directdetection}: while loop-suppressed, current spin-independent cross section ($\sigma_{\rm SI}$) measurements could set bounds on the scattering rate, and an explicit check is thus in order. We next focus in section~\ref{sec:relicdensity} on the calculation of the relic density, where two different mechanisms are involved. On the one hand, we study the standard thermal freeze-out (co)-annihilation case with a leptoquark mediator as done in~\cite{Baker:2015qna}. On the other hand, we consider the novel conversion-driven freeze-out (CDFO) mechanism~\cite{Garny:2017rxs}, also known as co-scattering~\cite{DAgnolo:2017dbv}. In CDFO, DM self-annihilation is negligible throughout the freeze-out process in the early Universe. Its chemical decoupling is instead initiated by the inefficiency of the conversion rates between the DM and the strongly interacting Dirac fermion $\chi_1$, which is driving the annihilation process. This scenario requires a small coupling between the SM and the dark sector ({\it i.e.}\ $\ydm$ in our setup), which suppresses the $S_1 \to \chi_1 \chi_0$ decay rate. This additionally reduces the potential impact of the mixed leptoquark+MET searches discussed in section~\ref{sec:lqmet} (and appendix~\ref{app:lqmet_HL-LHC}).

Although the model considered could lead to interesting monochromatic photon signals in DM indirect detection experiments if $\mLQ < \mxZero$ and for sizeable $\ydm$, this regime lies outside of the scope of this work. Indirect detection is thus not discussed in what follows. In such an $\mLQ < \mxZero$ regime, dark matter annihilation into a leptoquark pair $S_1 S_1$ followed by $S_1$ decays into fermions could moreover also play a role for the computation of the relic density.

\subsection{Direct detection}
\label{sec:directdetection}
\begin{figure}
\center
\includegraphics[width=0.8\textwidth,trim={0 13.5cm 0 0},clip]{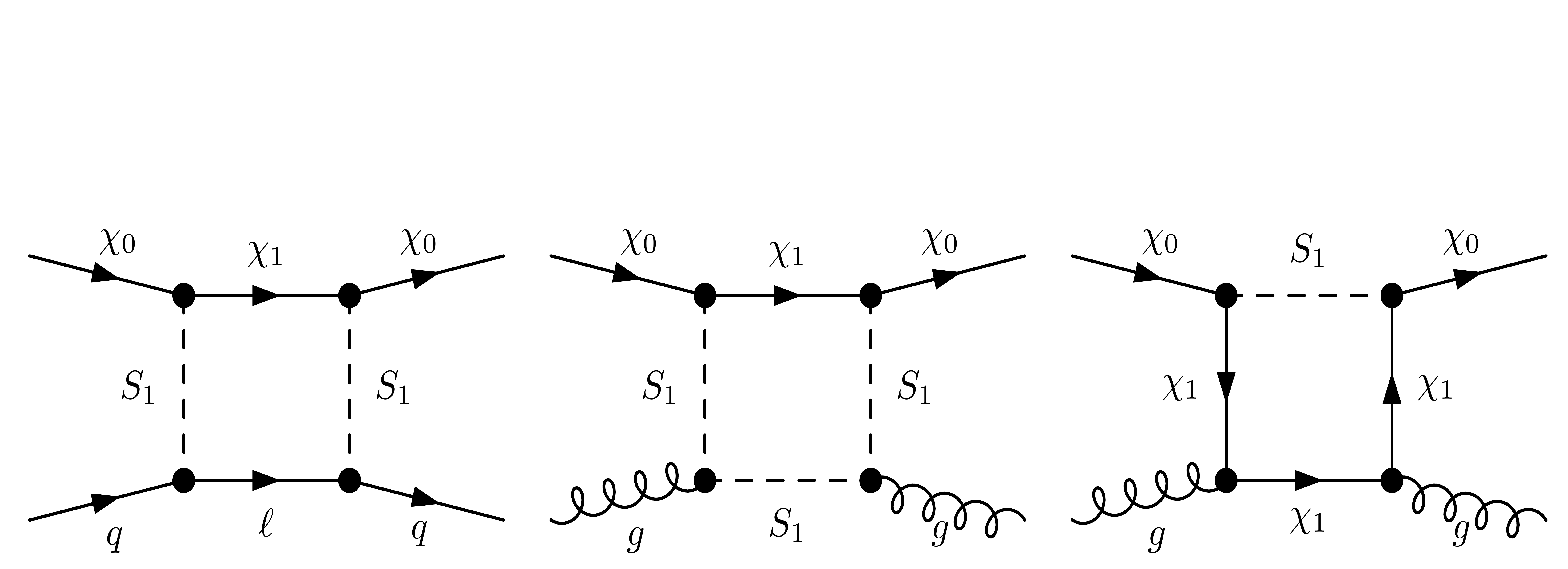}
\caption{Representative LO Feynman diagrams relevant for DM direct detection in our model.}
\label{fig:RDM:dd_1loop}
\end{figure}

The scattering of the DM state $\chi_0$ off nucleons proceeds via one-loop diagrams like the ones shown in figure~\ref{fig:RDM:dd_1loop}. The corresponding cross section is thus both loop- and mass-suppressed, the latter suppression being associated with the presence of heavy $\chi_1$ and $S_1$ particles running into the loops. Even so, the sensitivity of DM direct detection experiments has reached a level of precision such that they may be able to probe such rare loop-induced processes~\cite{Klasen:2013btp}. We therefore confront predictions of our model with existing constraints from Xenon1T~\cite{XENON:2018voc}, as well as with the projected sensitivity of the proposed DARWIN experiment~\cite{DARWIN:2016hyl}.

\begin{figure}
    \centering
    \includegraphics[width=0.495\textwidth]{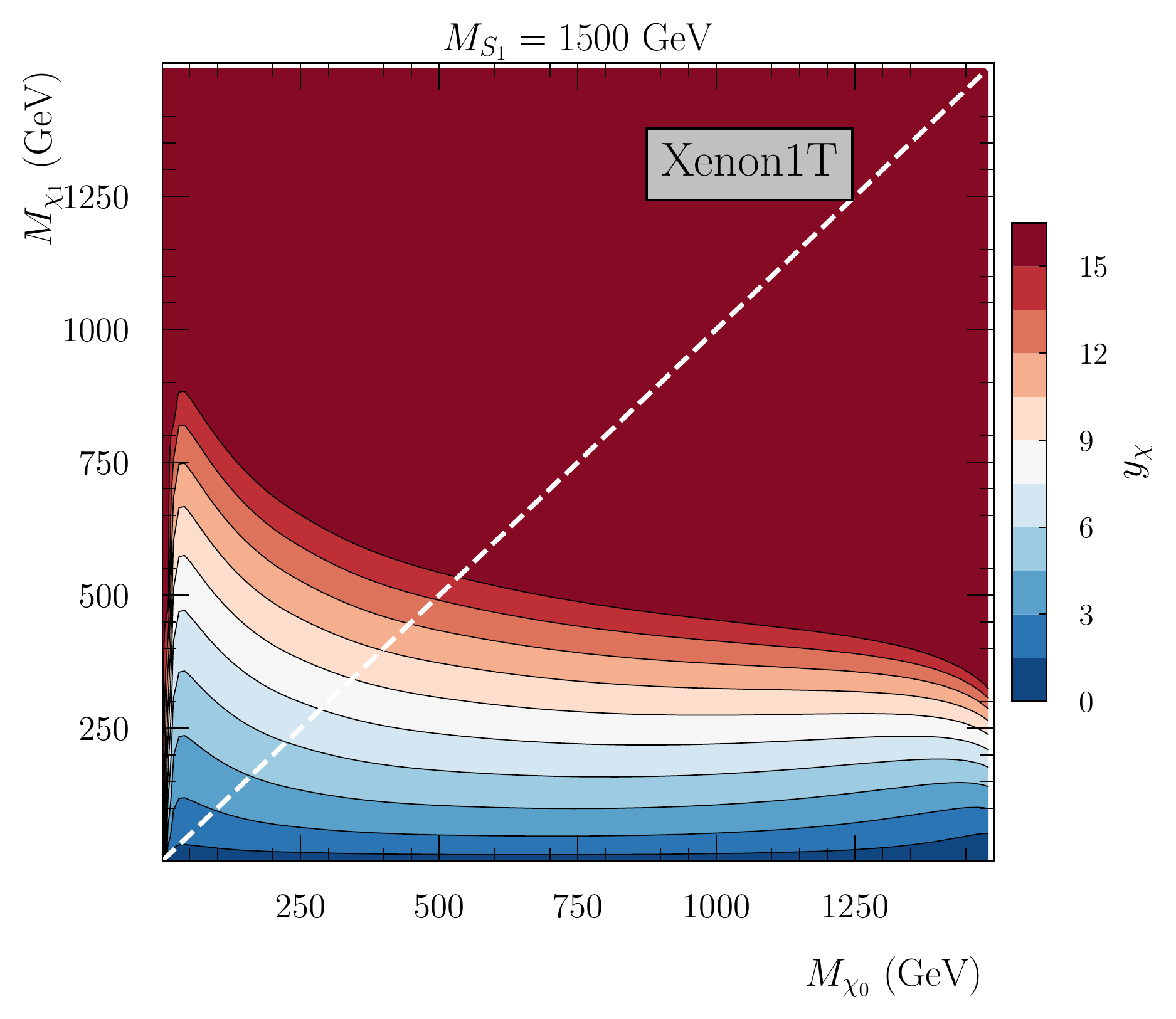}\hfill
    \includegraphics[width=0.495\textwidth]{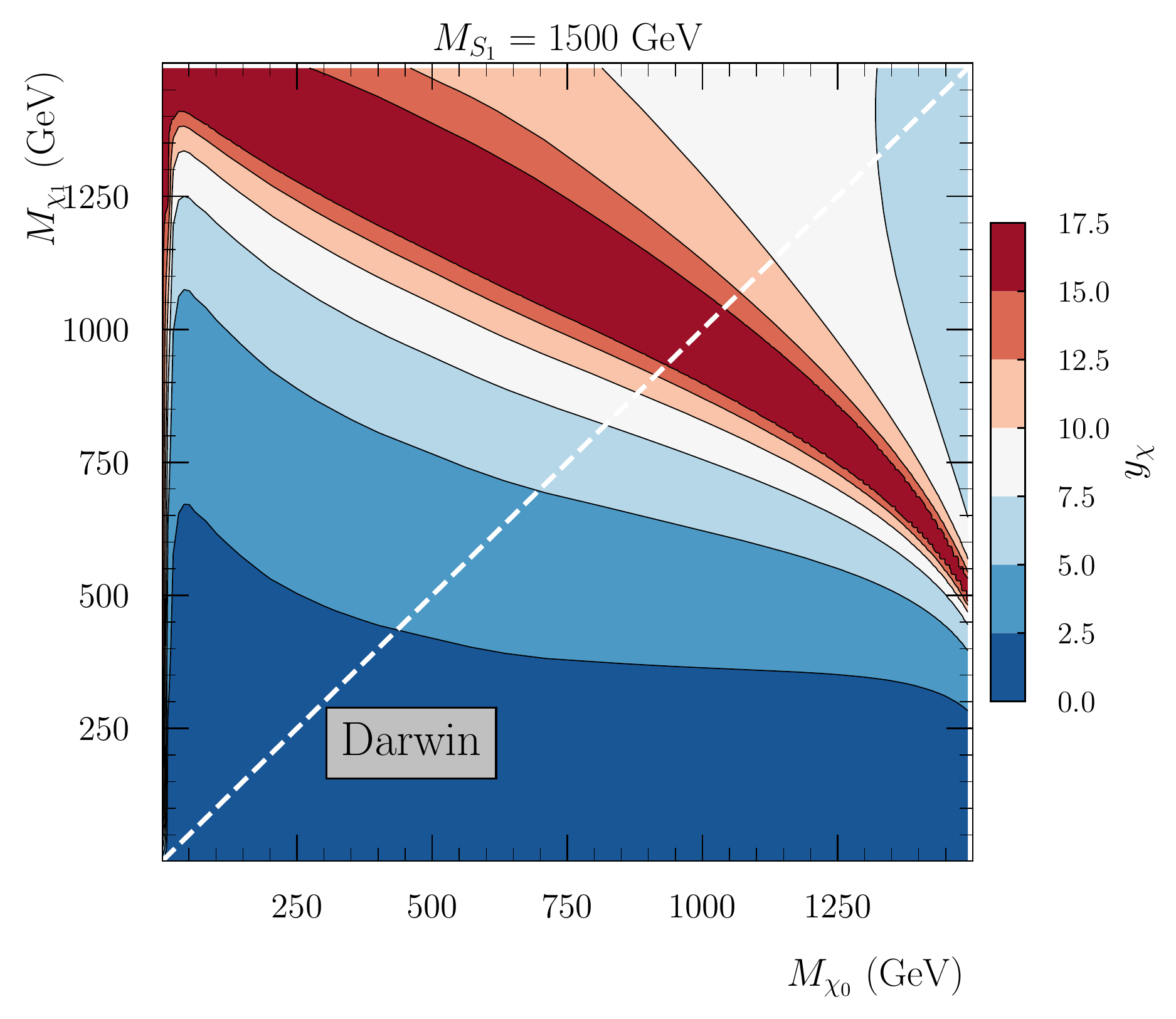}
    \vfill
    \includegraphics[width=0.495\textwidth]{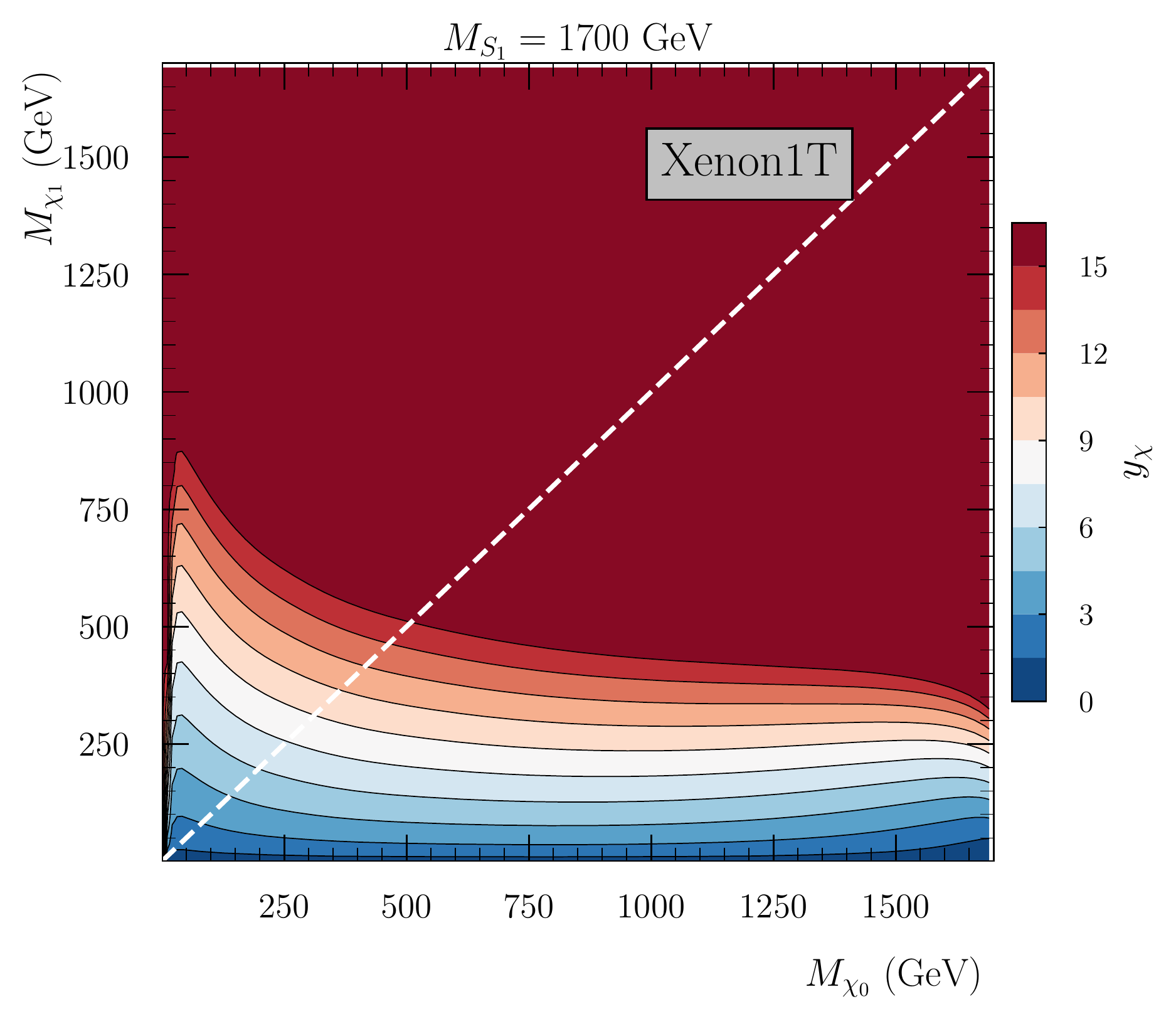}
    \hfill
    \includegraphics[width=0.495\textwidth]{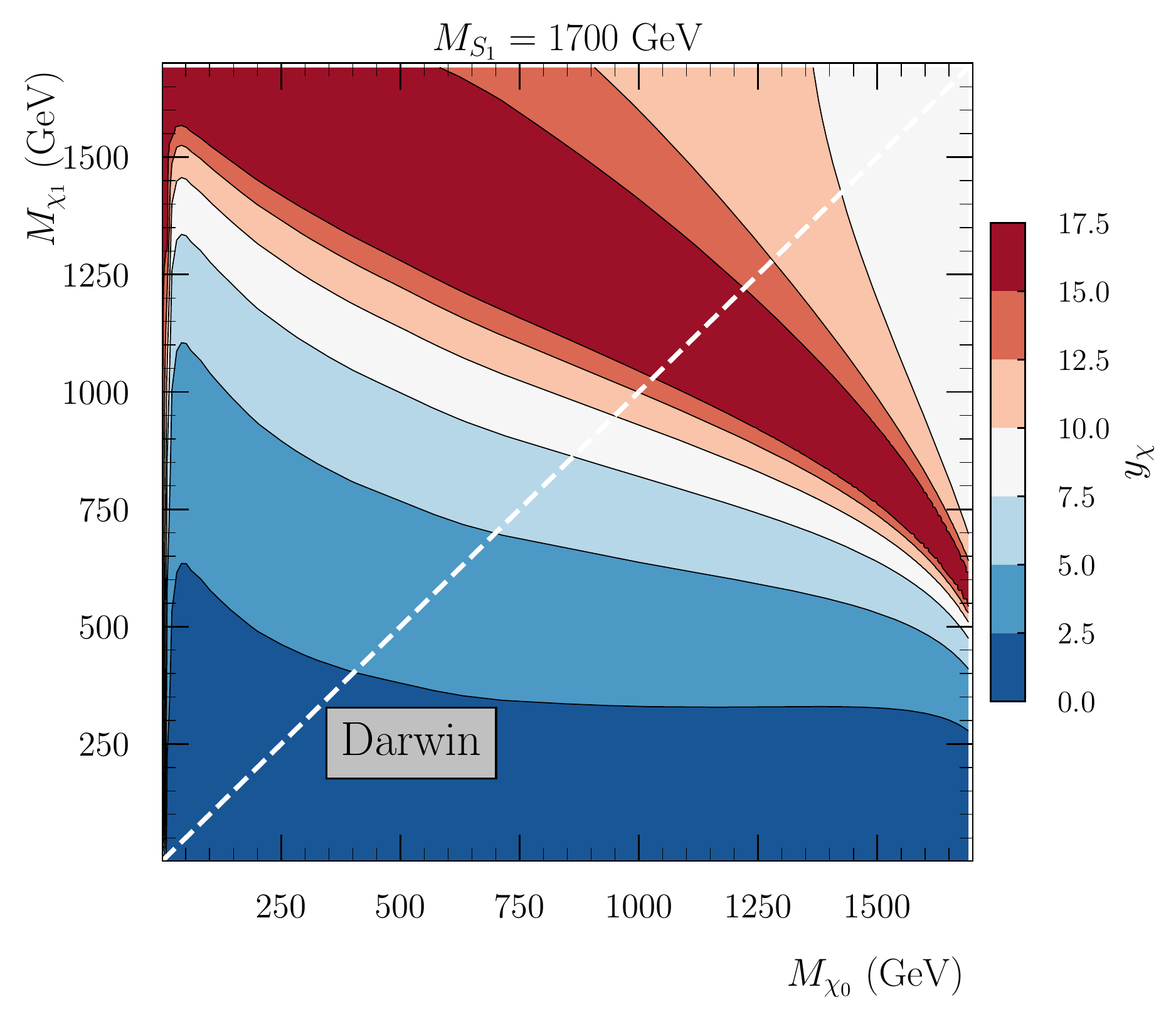}
    \caption{Contours of excluded values of $y_\chi$ originating from the spin-independent direct detection cross section constraints in Xenon1T~\cite{XENON:2018voc} (left) and from the expectation of the DARWIN experiment~\cite{DARWIN:2016hyl} (right). These contours are projected on the plane spanned by $\mxZero$ and $\mxOne$, and we assume $\mLQ = 1500~{\rm GeV}$ (upper panels) and $\mLQ = 1700~{\rm GeV}$ (lower panels). The white dashed line corresponds to the kinematical boundary ($\mxOne = \mxZero$) below which $\chi_0$ is not a suitable dark-matter candidate.}
    \label{fig:RDM:dd_contours}
\end{figure}

In order to compute the spin-independent dark matter-nucleon scattering cross section predicted by our model we perform a complete one-loop matching of the relevant Wilson coefficients taking into account all possible diagrams and interference effects. This is detailed in appendix~\ref{app:dd_calculation}. While such a calculation was carried out originally in~\cite{Drees:1993bu}, it has been revisited recently (albeit for different models than those considered here) in~\cite{Garny:2018icg,Mohan:2019zrk}. As we only consider $\chi_0$ scattering off gluons (see the appendix), $\sigma_{\rm SI}$ only depends on the three new masses and the dark coupling $\ydm$, and it is thus independent of the $\lamL,\lamR$ couplings. We present in figure~\ref{fig:RDM:dd_contours} contours of maximum allowed $\ydm$ value in the $(\mxZero, \mxOne)$ plane, for a fixed leptoquark mass of $1.5$ TeV. We find that for the parameter space regions allowed by the missing energy searches at the LHC (see figure~\ref{fig:RDM:met_bounds}), current direct detection searches by Xenon1T are not sensitive to model parameters compatible with the perturbative regime. Their naive evaluation indeed constrains $y_\chi$ to be of  ${\cal O} (10)$ (or even larger), where the validity of our computation is highly questionable. On the other hand, the expected sensitivity of DARWIN may reduce the viable range of $y_\chi$ to less than $3-5$, depending on the values of $M_{\chi_1}$ and $M_{\chi_0}$.

\subsection{Relic density}
\label{sec:relicdensity}

\begin{figure}
    \centering
   \includegraphics[width=0.8\textwidth,trim={2cm 19.0cm 1cm 5.5cm},clip]{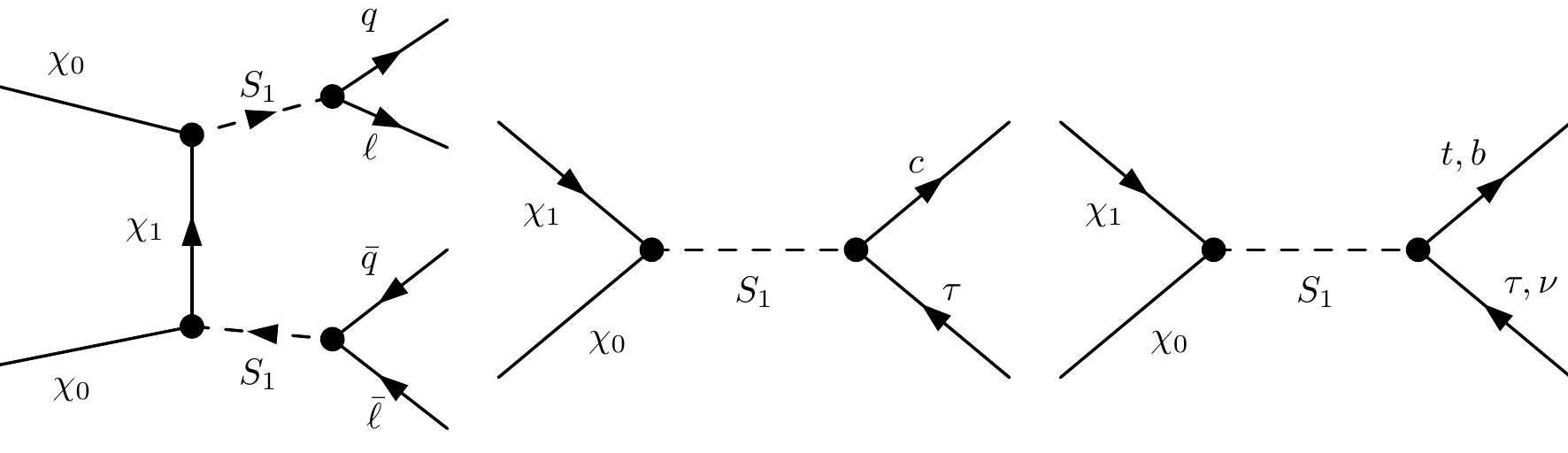}\vspace{.05cm}
   \includegraphics[width=0.8\textwidth,trim={1.5cm 13.0cm 1cm 12.5cm},clip]{figures/fig_relic_diagrams.pdf}\vspace{.3cm}
   \includegraphics[width=0.8\textwidth,trim={2cm 6.5cm 1cm 19.0cm},clip]{figures/fig_relic_diagrams.pdf}
    \caption{Representative Feynman diagrams of processes contributing to the DM relic density. We show contributions from $\chi_0$ self-annihilations (first diagram of the top row), $\chi_0 \chi_1$ co-annihilations (two last diagrams of the top row), $\chi_1$ self-annihilations into quarks and gluons (second row) and those additional diagrams for CDFO through $\chi_1 q \to \chi_0 \ell$, $\chi_1 \ell \to \chi_0 q$, and non-prompt $\chi_1$ decays (bottom row).}
    \label{fig:RDM:relic:diagrams}
\end{figure}

Our model features two regimes for the generation of the relic density of dark matter: one of them is associated with the standard freeze-out mechanism, and the other one with CDFO. Representative Feynman diagrams are shown in figure~\ref{fig:RDM:relic:diagrams}. The first row of the figure shows diagrams relevant for sizeable DM couplings, while the second and the third display those relevant in the CDFO regime. The latter mechanism had not been included in public relic density calculators, so that we implemented it in \micromegas\ version 5.3.7~\cite{Belanger:2021xxxy, Belanger:2014vza}, which we have used for all relic density computations performed in this work. Related technical details are supplied in appendix~\ref{app:cdfo_micro}.

\begin{figure}
    \centering
    \includegraphics[width=0.49\textwidth]{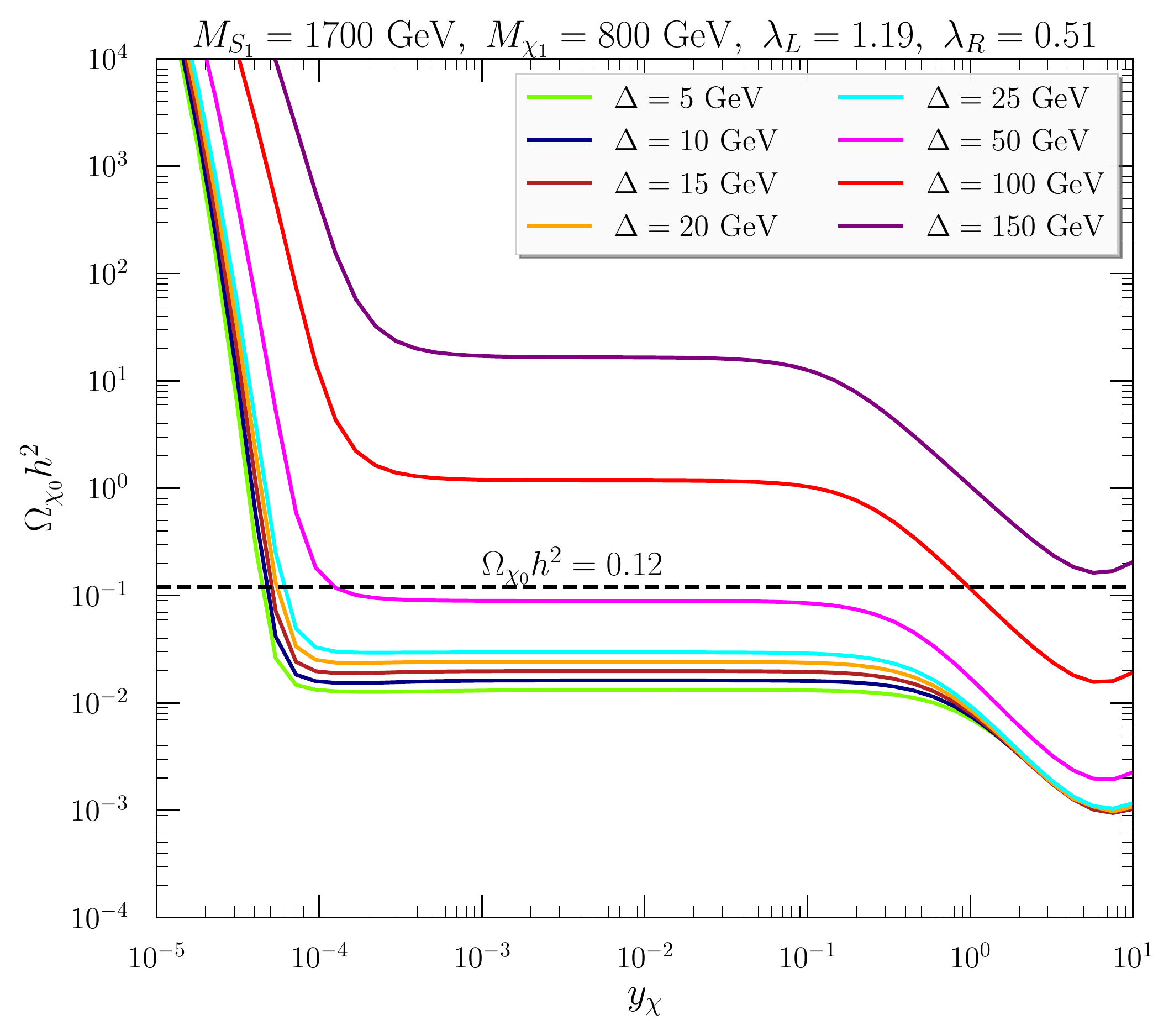}\hfill
    \includegraphics[width=0.49\textwidth]{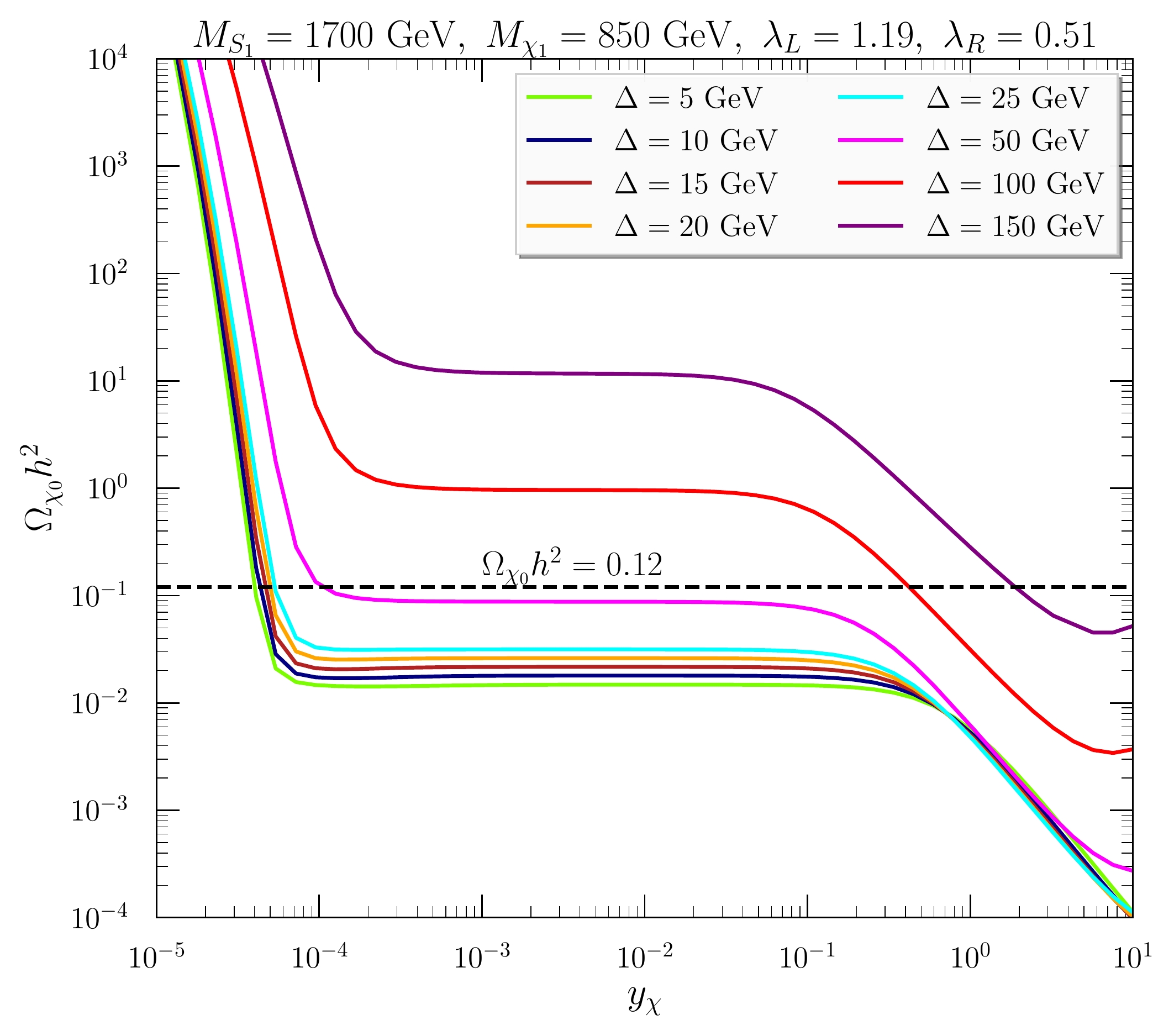}\vfill
    \includegraphics[width=0.49\textwidth]{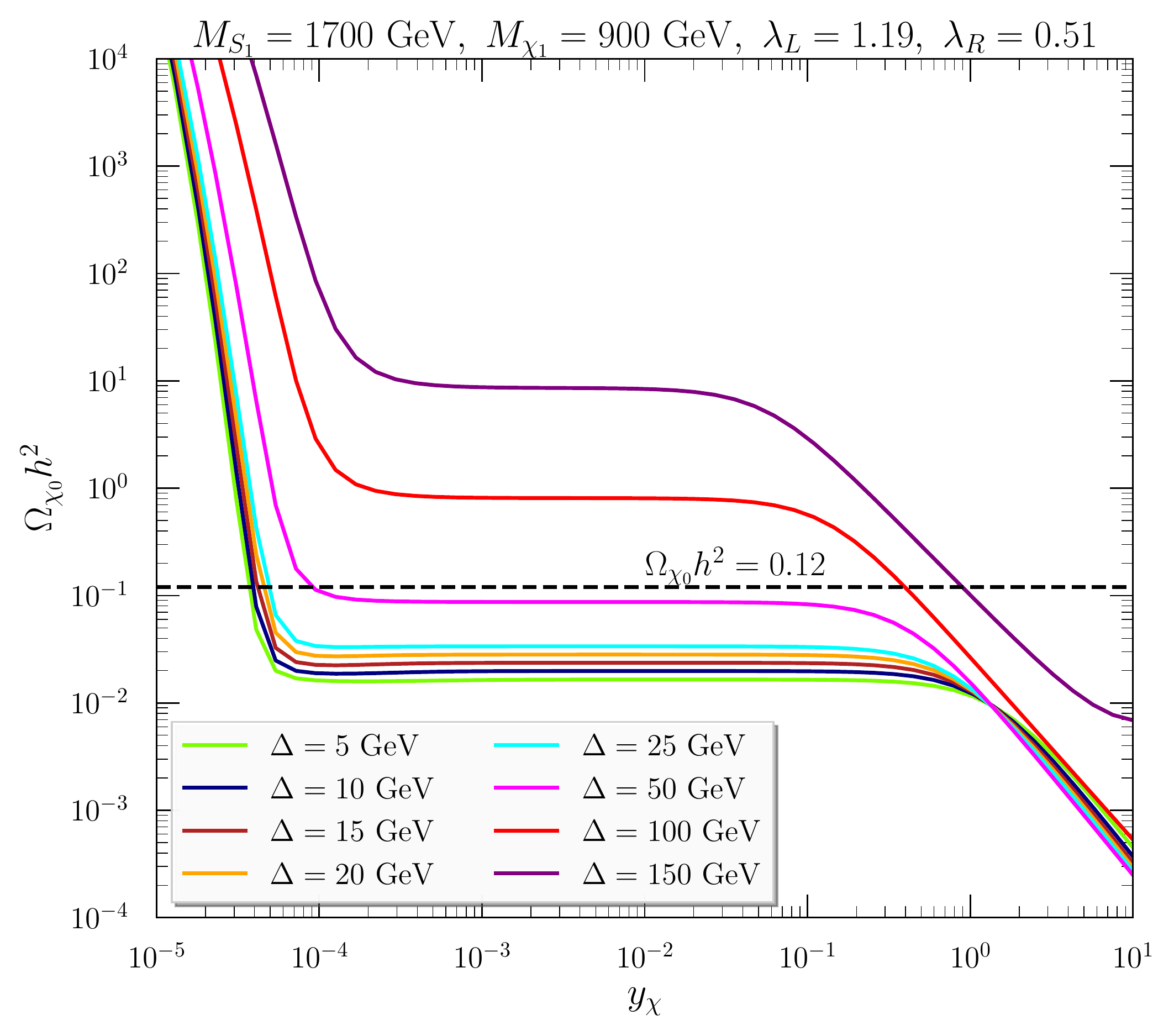}\hfill
    \includegraphics[width=0.49\textwidth]{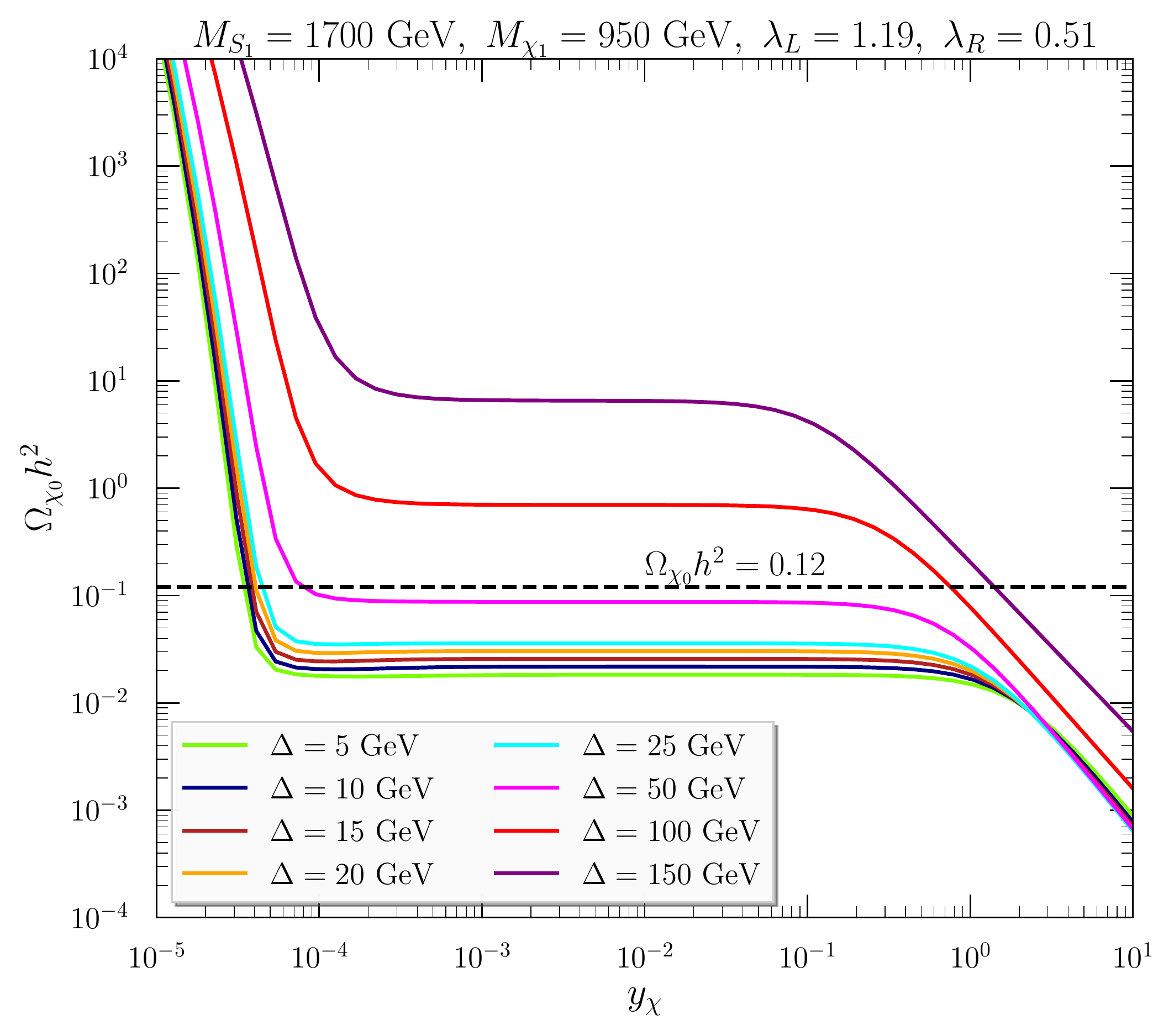}
    \caption{The DM relic abundance as a function of $y_\chi$ in BS1 scenarios for $\mLQ = 1700~{\rm GeV}, \lambda_L = 1.19$ and $\lambda_R = 0.51$. We consider four options for the $\chi_1$ mass consistent with constraints from missing energy searches at the LHC: $M_{\chi_1} = 800~{\rm GeV}$ (upper left), $850~{\rm GeV}$ (upper right), $900~{\rm GeV}$ (lower left) and $950~{\rm GeV}$ (lower right). The curves in green, navy, sienna, orange, cyan, magenta, red and purple correspond to values of $\Delta$ of 5, 10, 15, 20, 25, 50, 100 and 150~GeV respectively.\label{fig:Omh2:yDM:BS1}}
\end{figure}

\begin{figure}
    \centering
    \includegraphics[width=0.49\textwidth]{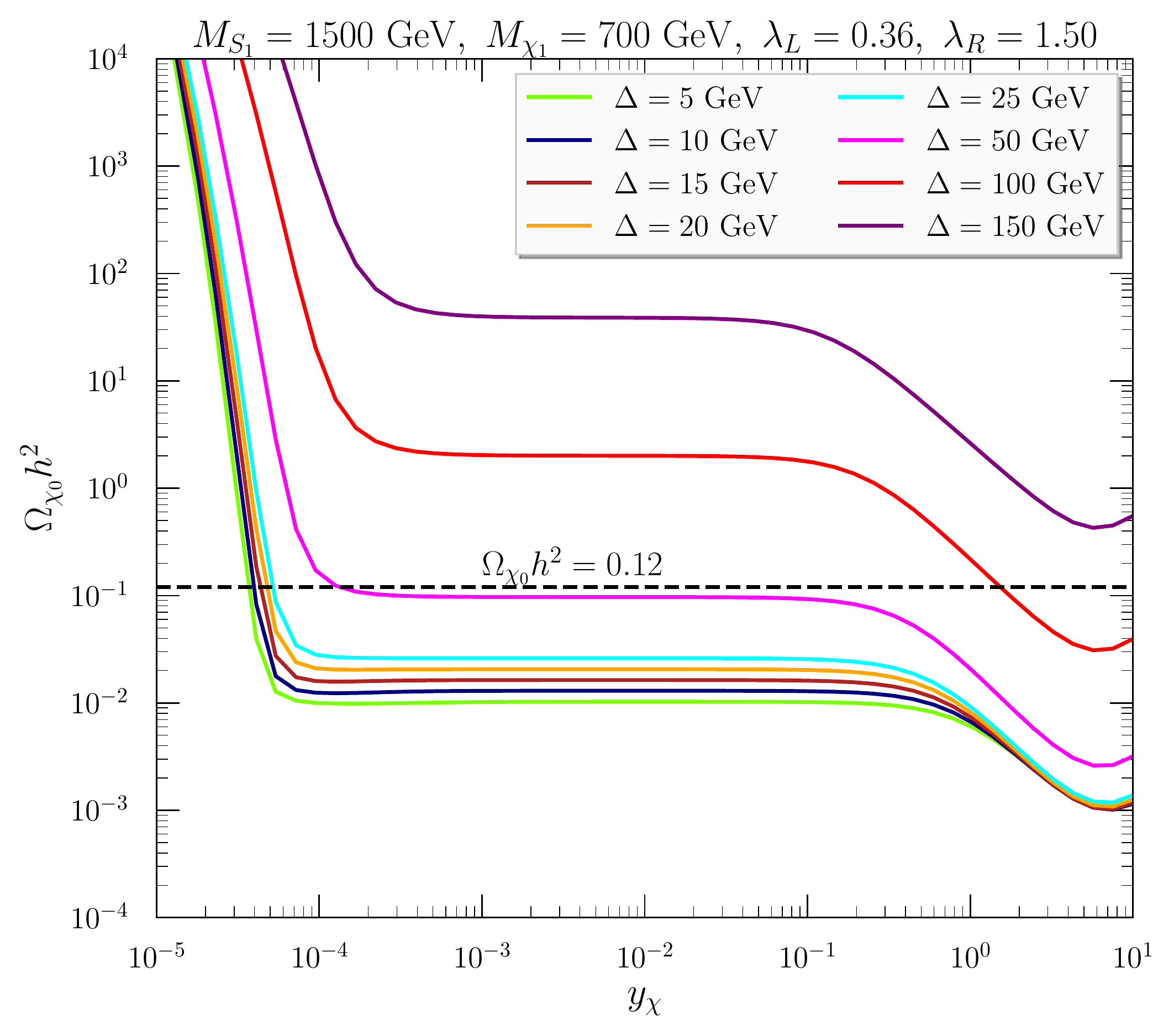}\hfill
    \includegraphics[width=0.49\textwidth]{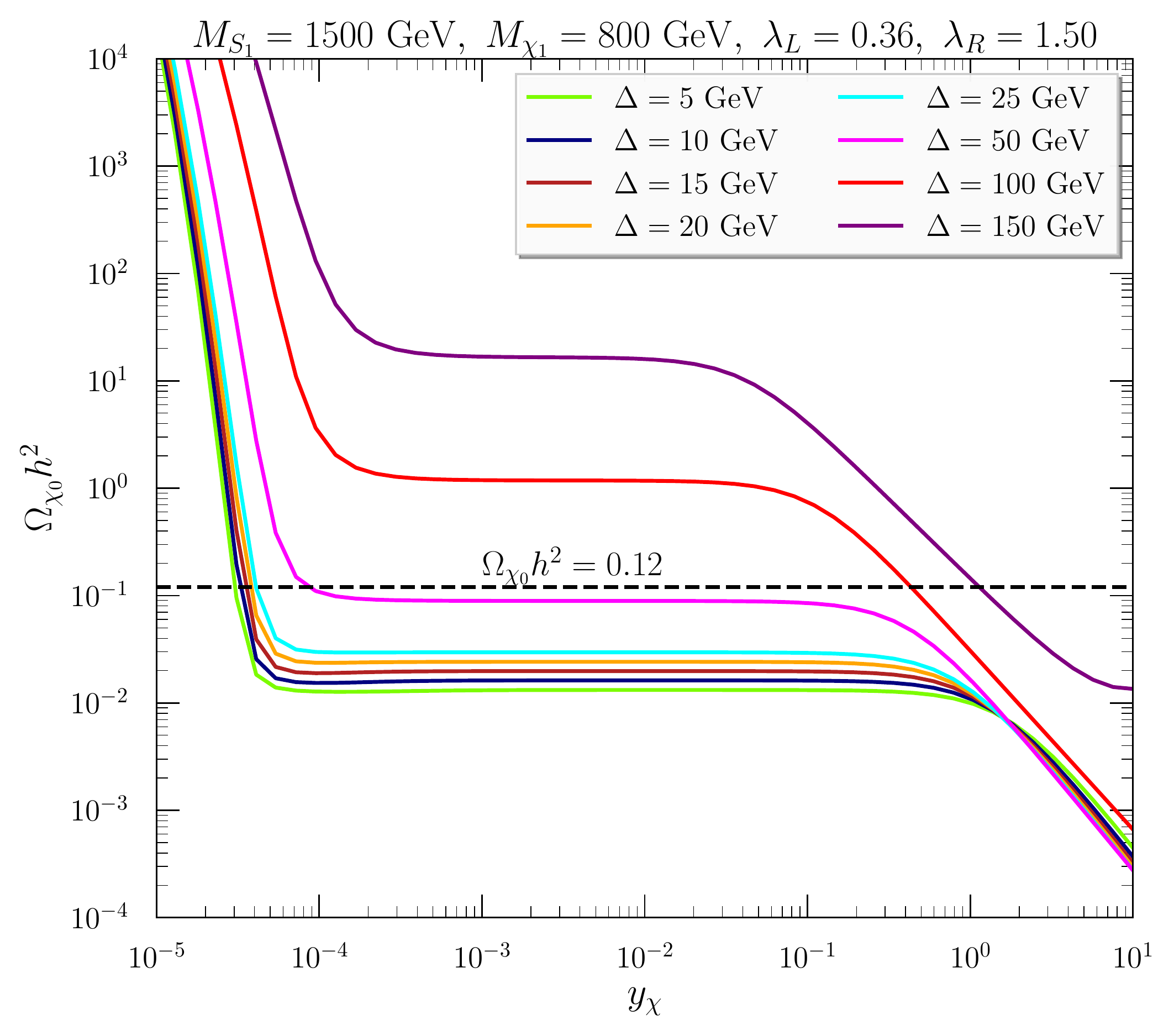}
    \includegraphics[width=0.49\textwidth]{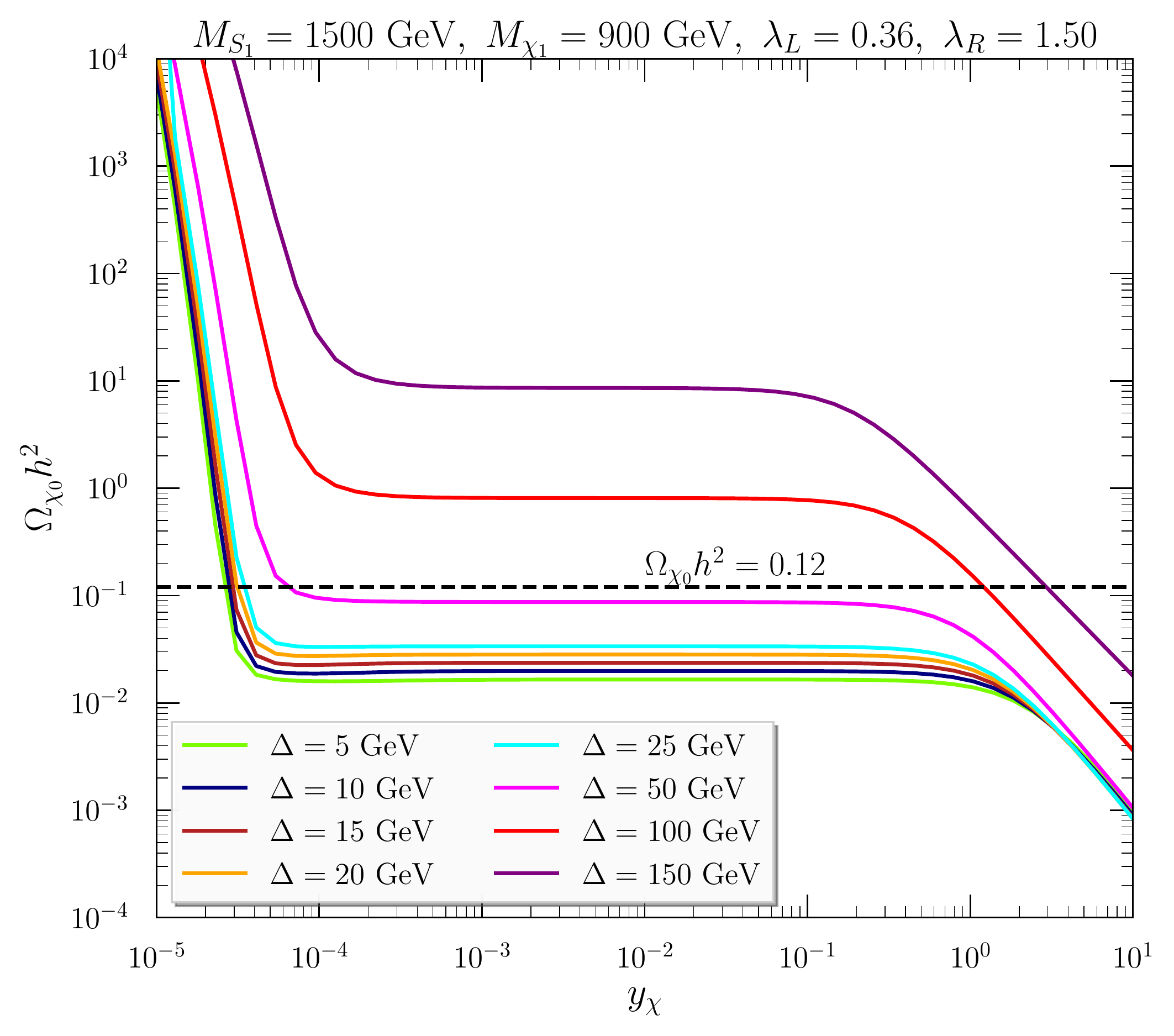}\hfill
    \includegraphics[width=0.49\textwidth]{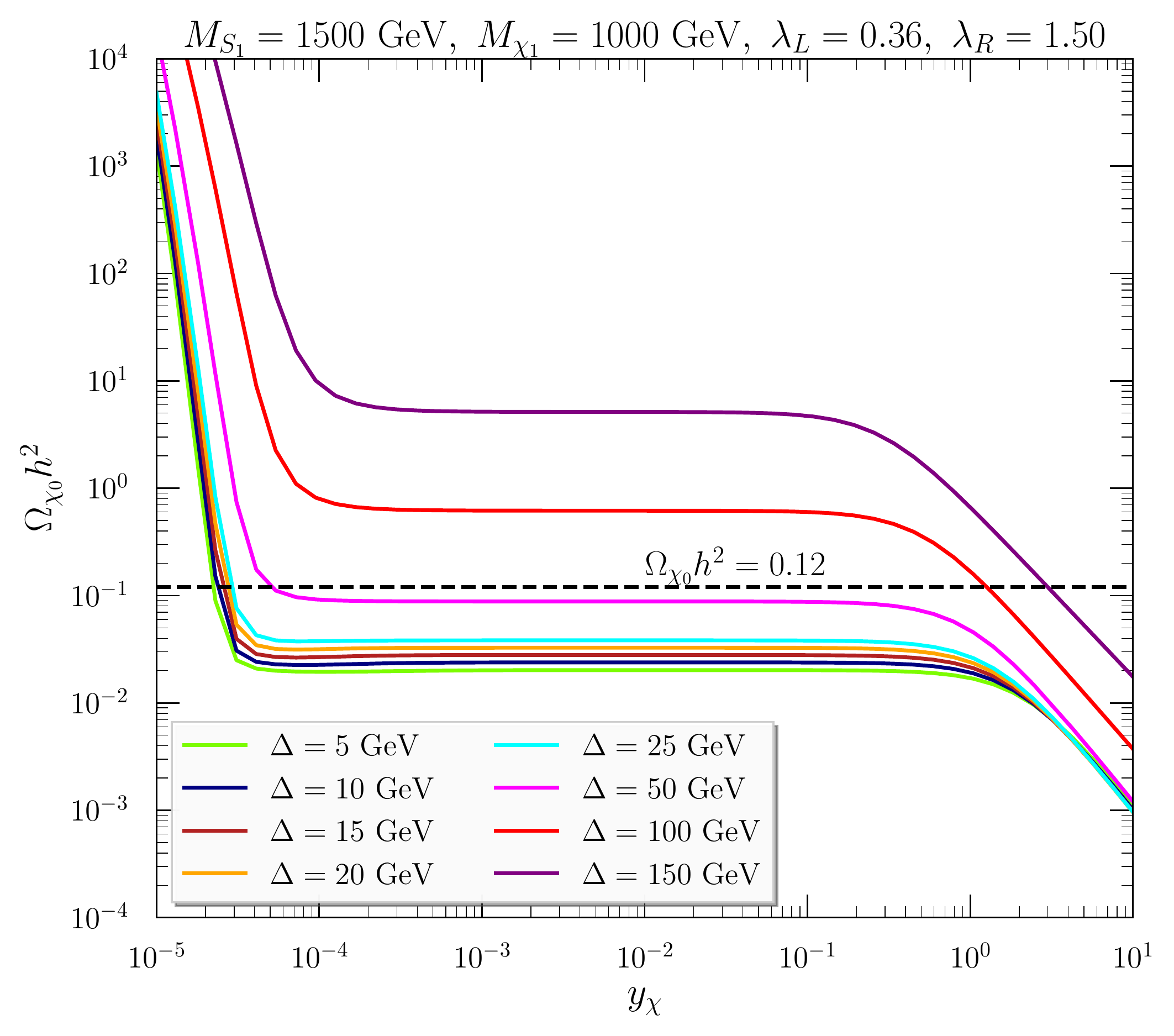}
    \caption{The DM relic abundance as a function of $y_\chi$ in BS2 scenarios for $\mLQ = 1500~{\rm GeV}, \lambda_L = 0.36$ and $\lambda_R = 1.5$. We consider four options for the $\chi_1$ mass consistent with constraints from missing energy searches at the LHC: $M_{\chi_1} = 700~{\rm GeV}$ (upper left), $800~{\rm GeV}$ (upper right), $900~{\rm GeV}$ (lower left) and $1000~{\rm GeV}$ (lower right). The curves in green, navy, sienna, orange, cyan, magenta, red and purple correspond to values of $\Delta$ of 5, 10, 15, 20, 25, 50, 100 and 150~GeV respectively.\label{fig:Omh2:yDM:BS2}}
\end{figure}

We start by investigating the dependence of the relic density on the dark coupling $\ydm$. We present in figures~\ref{fig:Omh2:yDM:BS1} and \ref{fig:Omh2:yDM:BS2} relic density scans for different dark sector parameters within the BS1 and BS2 scenarios, respectively. Both figures exhibit a plateau where DM production is dominated by QCD-induced $\chi_1 \chi_1 \to {\rm SM~SM}$ co-annihilations, hence independent of $y_\chi$. When $y_\chi$ increases, processes such as $\chi_1 \chi_0 \to {\rm SM~SM}$ eventually start to contribute more significantly. Consequently the relic density $\Omega_{\chi_0} h^2$ decreases. The general shape of the curve (a negative slope followed by a plateau and then another negative slope) implies that the interception with the $\Omega h^2$ interval measured by PLANCK~\cite{Aghanim:2018eyx} can happen in the leftmost part  of the curve (CDFO), in its rightmost part (usual thermal freeze-out), or the measurement interval can encompass the plateau. When such a case is realised (like in the upper right panel of figure~\ref{fig:Omh2:yDM:BS2}), the corresponding scenario lies at the border between the standard freeze-out and CDFO regimes. Moreover, the relic density grows with $\Delta$, as the curves shift upwards with increasing $\Delta$ values. As section~\ref{sec:bench} is dedicated to a discussion of the benchmarks, we only mention here that viable BS1 scenarios must be highly compressed due to MET search constraints, so that CDFO would be the preferred regime for points allowed by cosmology. In contrast, MET searches are less severe for BS2 scenarios. We now analyse such a configuration more in detail to assess the complementarity with the DM relic density constraints in the BS2 case.

 \begin{figure}
  \centering
  \includegraphics[width=0.8\linewidth]{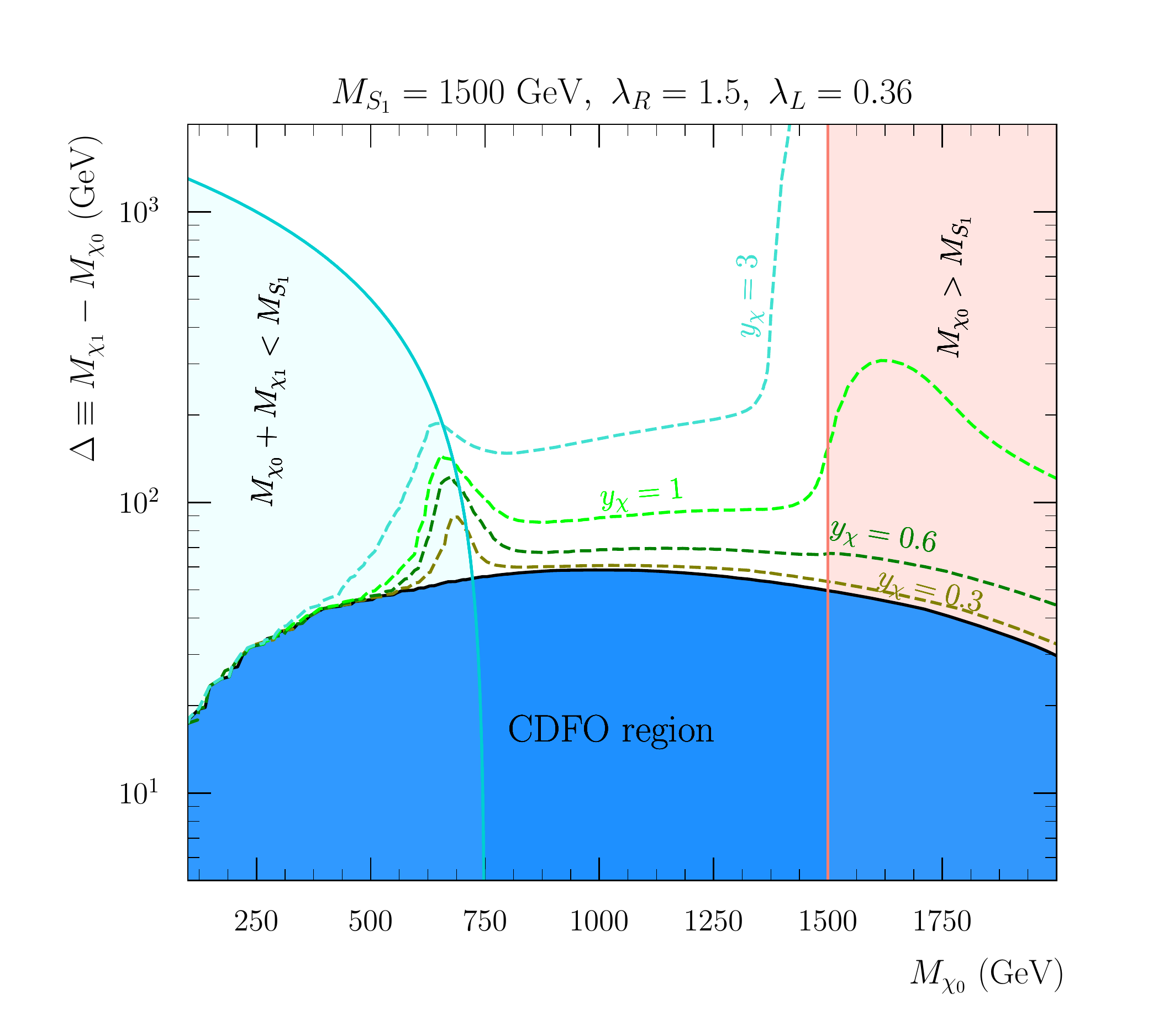}
  \caption{Parameter space satisfying $\Omega h^2 = 0.12$ for the BS2b scenario, shown in the $(\mxZero, \Delta)$ plane. The solid black line denotes the boundary between the standard freeze-out and conversion-driven freeze-out (CDFO) regime, and the thin dashed curves are associated with contours of constant coupling $y_{\chi} = 0.3$~(olive), $0.6$~(green), $1$~(lime) and $3$~(turquoise). The turquoise shaded area corresponds to a parameter space region where the leptoquark dark decays are allowed, while in the shaded salmon area, $\mxZero > \mLQ$.}
\label{fig:RDM:CDFO_WIMP}
\end{figure}

In figure~\ref{fig:RDM:CDFO_WIMP} we display contours of constant $y_{\chi}$ satisfying the correct relic abundance, $\Omega h^2 \simeq 0.12$~\cite{Aghanim:2018eyx}, for the BS2b scenario. All points in the represented $(\mxZero, \Delta)$ plane is allowed by the missing energy and leptoquark searches at the LHC described in section~\ref{sec:lhc}.
The standard DM freeze-out mechanism (initiated by DM self-annihilation processes) is at play above the thick black line, while below the CDFO regime is responsible for generating the relic density. In this last region, DM is under-abundant for sizeable $y_\chi$ couplings, which keep the conversion rate efficient and hence maintain chemical equilibrium in the dark sector. However, much smaller couplings $y_\chi$ in the  $[ 10^{-5}, 10^{-4}]$ range provide viable CDFO solutions, as shown in figures \ref{fig:Omh2:yDM:BS1} and \ref{fig:Omh2:yDM:BS2}. The smallness of the dark coupling yields typically non-prompt $\chi_1$ decays~\cite{Garny:2017rxs}. Moreover, the $S_1$ branching fraction to dark particles is usually negligible, so that invisible $S_1$ decays could not be seen at colliders. The only handle into the dark world at colliders is therefore through $\chi_1$ searches (which is also true for the low $\mxOne$ region with a compressed spectrum). In the standard freeze-out regime (above the thick black line in figure~\ref{fig:RDM:CDFO_WIMP}), co-annihilation is important for small $\Delta$ values, while self-annihilations become important for larger $\Delta$ values and when $M_{\chi_0} > M_{S_1}$. In this case, besides direct leptoquark, missing energy and direct detection searches, information from the resonant leptoquark plus missing energy searches could allow us to establish a firm connection between the flavour anomalies and the dark sector.

Finally, it turns out that the $y_\chi$ contours that we have obtained exhibit two prominent features. The one at $\mxZero\sim M_{S_1}/2$ arises from resonantly enhanced co-annihilations via an $s$-channel leptoquark exchange, while the one at $\mxZero\sim M_{S_1}$ is due to the opening of the $\chi_0\chi_0\to S_1 S_1$ annihilation process. Qualitatively, the phenomenology is very similar to the one of the top-philic parent models studied in~\cite{Baek:2016lnv,Garny:2018icg,Colucci:2018vxz,Cornell:2021crh}, so that for $\mxZero< M_{S_1}/2$ loop-induced DM annihilations into gluons can become important. This configuration is, however, ignored here.

\section{Benchmark scenarios}
\label{sec:bench}
The purpose of this section is to combine all the constraints of diverse origins that we have examined in this work, in order to construct phenomenologically viable and consistent benchmark scenarios relevant for further (theoretical and experimental) studies. Adopted points are presented in the $\{ \mLQ, \mxOne, \mxZero, \lamL, \lamR, \ydm \}$ six-dimensional parameter space of our model, and are chosen so that they comply with current data and could be probed in the future by a combination of the considered searches. This section is not meant to be a comprehensive review of the multiple possibilities, but rather to serve as an illustration of what our simple setup can achieve. This should further motivate the strengthening of both the current collider search program and the direct detection experiments. 

\begin{figure}
    \centering
    \includegraphics[width=.84\textwidth]{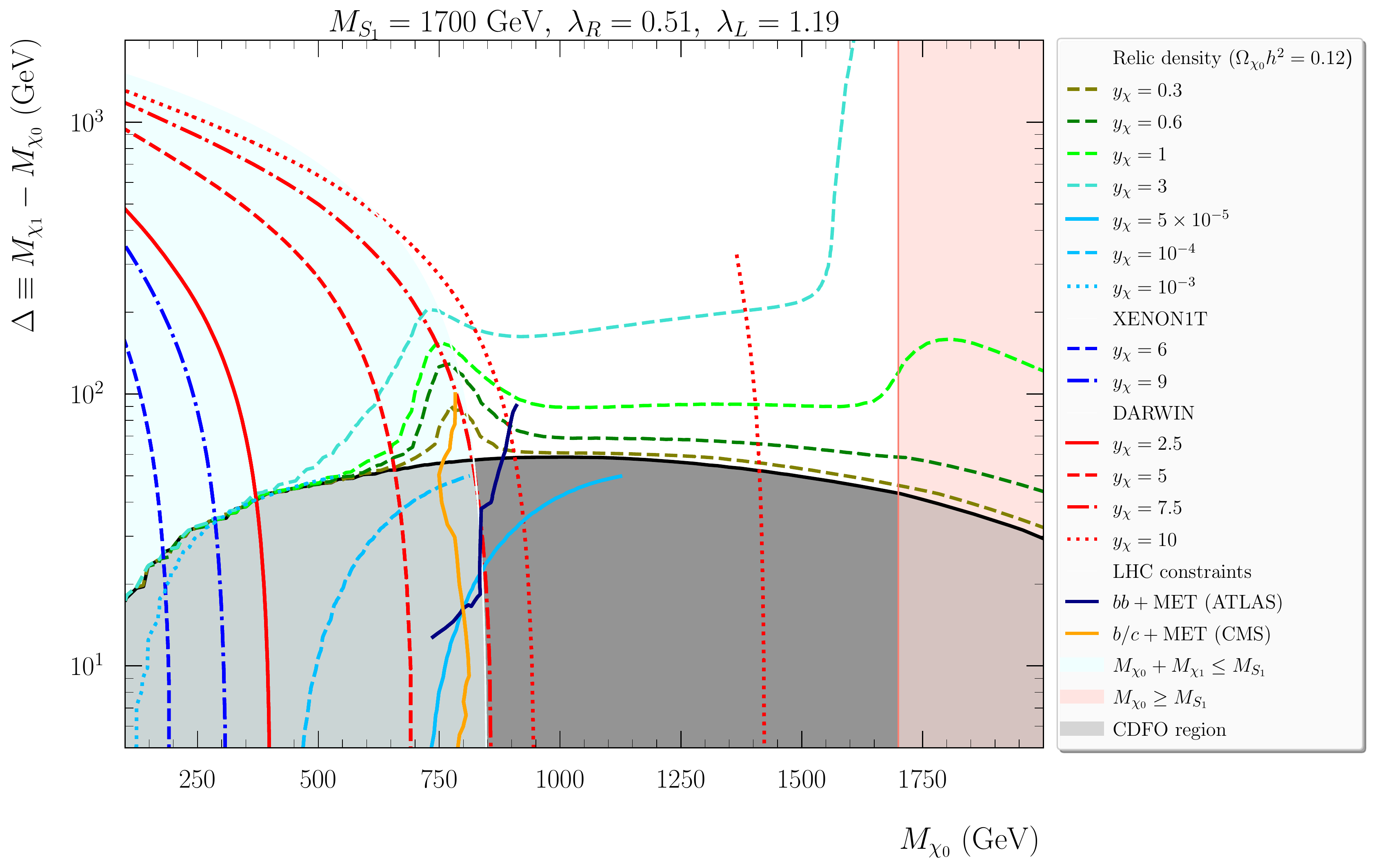}
    \hfill
    \includegraphics[width=.84\textwidth]{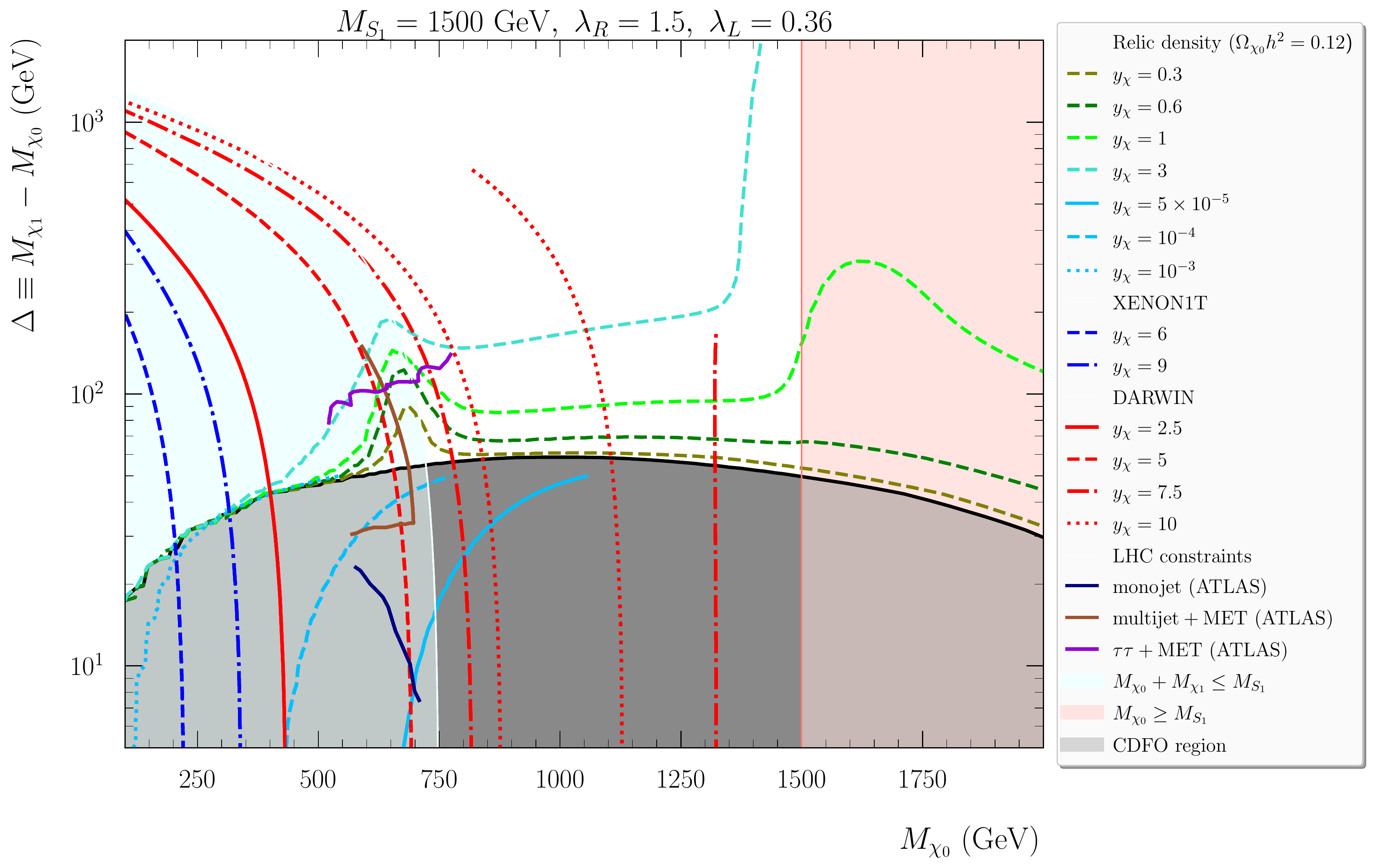}
    \caption{Summary of all existing constraints on BS1c (upper) and BS2b (lower) scenarios, presented in the $(\mxOne, \Delta)$ plane. All points are compatible with the DM relic density as measured by Planck, the $\ydm$ value being inferred from that constraint. We indicate several $\ydm$ isolines, for $y_{\chi} = 0.3$~(olive), $0.6$~(green), $1$~(lime) and $3$~(turquoise). We additionally display the mass combinations excluded by DM direct detection measurements at Xenon1T for a given $\ydm$ value through dashed (for $\ydm=6$) and dash-dotted (for $\ydm=9$) blue lines, and the corresponding expectation of the DARWIN experiment through red lines for $\ydm=2.5$ (solid), 5 (dashed), 7.5 (dash-dotted) and 10 (dotted). LHC constraints exclude the lower $\mxZero$ regime through $b\bar b$ plus missing energy (blue), $b$-jet and $c$-jet plus missing energy (yellow), mono-jet (blue), multi-jet plus missing energy (brown) and di-tau plus missing energy (purple) searches. The light rose area indicates the region where dark leptoquark decays are forbidden.}
    \label{fig:summary_plots}
\end{figure}

We begin with figure~\ref{fig:summary_plots}, in which we show the combined constraints originating from the considered collider searches, DM direct detection and relic density contours in the $(\mxOne, \Delta)$ plane for the BS1c (upper) and BS2b (lower) scenarios. The leptoquark mass has been fixed to 1.7 and 1.5 TeV respectively.

\begin{table}[!t]
\setlength\tabcolsep{8pt}
\resizebox{\textwidth}{!}{\begin{tabular}{lccccccc}
\multicolumn{1}{c}{ Benchmark scenario } & Quantity & BS1d & BS1e & BS2d &  BS2e\\
\midrule
\multirow{6}{*}{Parameters} & $M_{S_1}~({\rm GeV})$  & $1700$ & $1700$ & $1500$ & $1500$ \\
& $M_{\chi_1}~({\rm GeV})$  &  $850$ & $1030$ & $800$ & $800$ \\
& $\Delta~({\rm GeV})$      &  $10$ & $80$ & $200$ & $100$ \\
& $\lambda_L$   &  $1.19$ & $1.19$ & $0.36$ & $0.36$ \\
& $\lambda_R$   &  $0.51$ & $0.51$ & $1.50$ & $1.50$\\
& $\ydm$  &  $10^{-4}$ & $1.0$ & $3.0$ & $0.5$ \\[.6cm]
\multicolumn{7}{l}{{Branching ratios}}  \\
\midrule
\multirow{3}{*}{${\rm BR}(\chi_1 \to \chi_0 X Y)$} & ${\rm BR}(\chi_1 \to \chi_0 c \tau)$ & $29.66\%$ & $15.64\%$ & $94.57\%$ & $94.59\%$ \\
& ${\rm BR}(\chi_1 \to \chi_0 t \tau)$ & $0\%$ & $0\%$ & $0\%$ & $0\%$  \\
& ${\rm BR}(\chi_1 \to \chi_0 b \nu)$ & $70.34\%$ & $84.35\%$ & $4.53\%$ & $5.41\%$ \\
\midrule
\multirow{4}{*}{${\rm BR}(S_1 \to X Y)$} & ${\rm BR}(S_1 \to c \tau)$ & $8.49\%$ & $8.49\%$ & $67.54\%$ & $89.79\%$ \\
& ${\rm BR}(S_1 \to b \nu)$ & $46.25\%$ & $46.25\%$ & $3.89\%$ & $5.17\%$ \\
& ${\rm BR}(S_1 \to t\tau)$ & $45.28\%$ & $45.28\%$ & $3.78\%$ & $5.04\%$ \\
& ${\rm BR}(S_1 \to \chi_1 \chi_0)$ & $\simeq 0\%$ & $\simeq 0\%$ & $24.77\%$ & $0\%$ \\[.6cm]
\multicolumn{7}{l}{{Total widths}} \\
\midrule
\multirow{2}{*}{} & $\Gamma_{\chi_1}~({\rm GeV})$ & $2.18 \times 10^{-20}$ & $1.53 \times 10^{-7}$ & $2.40 \times 10^{-4}$ & $2.56 \times 10^{-7}$ \\
& $\Gamma_{S_1}~({\rm GeV})$ & $1.04 \times 10^{2}$ & $1.04 \times 10^{2}$ & $0.99 \times 10^{2}$ & $0.75 \times 10^{2}$ \\[.6cm]
\multicolumn{7}{l} {{Production cross sections}}  \\
\midrule
\multirow{2}{*}{$\sigma(pp \to \chi_1 X)$~[fb]} & $\chi_1 \chi_1~({\rm LHC})$ & $88.01$ & $24.94$ & $128.79$ & $128.79$  \\
& $\chi_1 \chi_1~({\rm FCC})$ & $31.54 \times 10^{3}$ & $13.13 \times 10^{3}$ & $41.42 \times 10^{3}$ & $41.42 \times 10^{3}$ \\
\midrule
\multirow{6}{*}{$\sigma(p p \to S_1 X)$~[fb]} & $S_1 \tau~({\rm LHC})$ & $0.14$ & $0.14$ & $2.58$ & $2.58$ \\
& $S_1 \nu_\ell~({\rm LHC})$ & $7.1 \times 10^{-2}$ & $7.1 \times 10^{-2}$ & $1.71 \times 10^{-2}$ & $1.71 \times 10^{-2}$ \\
& $S_1 S_1~({\rm LHC})$ & $4.03 \times 10^{-2}$ & $4.03 \times 10^{-2}$ & $0.46$ & $0.46$ \\
& $S_1 \tau~({\rm FCC})$ & $29.69$ & $29.69$ & $429.77$ & $429.77$ \\
& $S_1 \nu_\ell~({\rm FCC})$ & $108.68$ & $108.68$ & $17.38$ & $17.38$ \\
& $S_1 S_1~({\rm FCC})$ & $197.12$ & $197.12$ & $448.71$ & $448.71$ \\[.6cm]
\multicolumn{7}{l}{{Dark Matter}} \\
\midrule
\multirow{2}{*}{$\ydm^{\rm DD}$} & Xenon1T & $15.70$ & $15.71$ & $15.71$ & $15.71$ \\
& DARWIN & $7.22$ & $12.99$ & $5.88$ & $6.39$ \\
\end{tabular}}
\caption{Benchmark scenarios in our model, compatible with an explanation for the $\RD$ anomalies and that satisfy all constraints from cosmology and the LHC. For each point, we provide information on the new state branching ratios, total widths and LO production rates at the LHC, as well as on the current and future bounds from DM direct detection.\label{tab:bench_points}}
\end{table} 

For the BS1 scenario, bounds from the considered missing energy searches at the LHC constrain the leptoquark mass to satisfy $\mLQ>1.6$~TeV and the spectrum to be highly compressed, as shown in the left panel of figure~\ref{fig:RDM:met_bounds}. In figure~\ref{fig:Omh2:yDM:BS1} we have seen that {\it in general}, irrespectively of the particular value of $\mLQ$ and $\Delta$, the relic density constraint is satisfied for $\ydm \sim 10^{-4}$. As anticipated, in our BS1c benchmark scenario the CDFO regime correspondingly takes place in a large part of the parameter space as soon as we impose that the relic density as measured by Planck should be recovered. For illustration, we list in table~\ref{tab:bench_points} a few reference points that are still allowed by data. We provide the values of the six independent model parameters, and we also report the corresponding value of the relic density, the expected DARWIN constraint on $\ydm$ and a rough estimation of the necessary HL-LHC luminosity to detect this point through future leptoquark searches. For the BS1 benchmark slope, the model can thus be tested through $\chi_1$ production and decays at colliders, to which missing energy searches at the LHC are very sensitive. The next handle on it comes from leptoquark searches in mixed visible decay modes. Due to the small value of $\ydm$, direct detection does not further constrain these points, and leptoquark searches in a mixed visible/invisible final state do not provide any meaningful bounds. Depending on the particular value of $\mLQ$ (considering other possible values) we could design scenarios to which HL-LHC searches are sensitive, and others that would rely on future colliders.

For the BS2 case, the actual value of $\Delta$ greatly affects the relic density predictions, as shown in figure~\ref{fig:Omh2:yDM:BS2}. In addition, BS2 benchmarks contrast with those of slope BS1 in which an uncompressed dark spectra is incompatible with an appreciable $S_1S_1$ production rate at the LHC. For a small value of $\Delta = 5 $ GeV we have a very similar situation to the one described for BS1 setups. We will thus only discuss less-compressed spectra in the following. For $\Delta = 50-100$ GeV, the right relic density can be achieved, irrespective of the dark matter mass, for $\ydm \in [10^{-5}-0.2]$ due to the plateau featured by the relic density dependence on $\ydm$. For larger values of $\Delta$ the observed relic can still be recovered, this time with $\ydm \gtrsim 1$. We thus lie in the freeze-out regime. For example, the right panel of figure~\ref{fig:RDM:met_bounds} shows that a configuration in which $\mLQ = 1.5$~ TeV, $\mxZero = 550$~GeV, and $\mxOne = 700$~GeV is barely excluded. Moreover, such a benchmark point can only recover the right relic density with non-perturbative dark couplings $\ydm =10$. Still in contrast with the BS1 situation, BS2 scenarios are also meaningfully reachable through DM direct detection searches. If we increase the leptoquark mass to enhance the dark channel contributions on the considered DM observables, we would reduce at the same time the constraining power of the leptoquark searches at colliders. The phenomenologically most useful handles on the model therefore consist of the missing energy searches in the $\tau \tau$ + MET and mono-jet channels, the leptoquark + MET searches at colliders and DM direct detection. Visible searches for leptoquarks  only come after these, of which the sensitivity to $c \tau$ final states is the least stringent. Corresponding benchmark points which could be used in future analyses are shown in table~\ref{tab:bench_points}. We close this section by noting that a comparison between our scenario and previous work \cite{Arcadi:2021cwg,Guadagnoli:2020tlx,Baker:2021llj} is not a straightforward task. Those works feature a richer particle physics content and they also address a larger set of anomalies, while our minimal model only address $R_{D^{(*)}}$. In general we can only say that our six-dimensional parameter space is more constrained that those constructions featuring a larger dimensional parameter space.

\section{Conclusions and outlook}
\label{sec:conclu}
Among the different explanations for the charged-current flavour anomalies, models with leptoquarks near or at the TeV scale are among the best options. In this work, we have studied the connection between these models and dark matter, which necessarily requires to add new particles and couplings to the theory. In this regard, we have extended the Standard Model with one scalar leptoquark and two dark particles, namely a charged coloured Dirac fermion and a neutral weak singlet Majorana fermion. We have then explored the resulting six-dimensional parameter space of this simplified model vis-a-vis the aforementioned flavour anomalies, the dark matter relic density,  direct detection prospects, and the LHC constraints. 

When all these constraints are simultaneously taken into account we are left with two possible regimes for dark matter: the traditional freeze-out mechanism and the conversion-driven freeze-out (co-scattering) one. In the first regime, the dark sector couples to the leptoquark mediator with a similar strength to that required to reconcile measurements and predictions for the $\RD$ anomalies. In the second regime, the leptoquark couples faintly to the dark sector, which implies that the leptoquark dark branching fraction is negligible. The associated phenomenology at the LHC therefore consists of seemingly disconnected ``flavour-anomaly-inspired'' leptoquark signals and traditional missing energy + X signatures. Depending on the level of compression of the dark sector spectrum, it may be possible to resolve the leptons and jets originating from the $S_1 \to \chi_1 \chi_0 \to \ell q \chi_0 \chi_0$ decay chain, which would then allow for the establishment of a connection between dark matter and the flavour anomalies. In the case where the leptoquark branching fraction into the dark sector is non-negligible, then the strongest indication of the leptoquark-dark matter connection is through a leptoquark pair production signal in which one of the leptoquarks decays into the dark sector and the other one into hard leptons and jets. Only one available analysis, from the CMS collaboration, addresses such a search, assuming the decay of leptoquarks into second generation fermions. We encourage both collaborations to incorporate in their program searches for  the broad range of signatures corresponding to this crucial final state, and in particular to include the decay of leptoquarks into third generation fermions. Such searches would indeed enrich the current dark matter program of the LHC experiments.

At several points we have pointed out the existence of two important gaps in the campaign to optimally cover the leptoquark parameter space. As the solutions to the $\RD$ anomalies involve large couplings connecting second generation quarks with third generation leptons, we firstly advocate to study the $c \tau c \tau$ final state at colliders, and if possible and depending on the $c$-tagging efficiencies, also include the corresponding $c \tau t \tau$ and $b \nu c \tau$ mixed final states. Secondly, it is important to also include searches for the production of a single leptoquark decaying largely into $c \tau$ ({\it i.e.}\ a search targeting the $c \tau \tau$ final state).

Our results have shown that the exploration of common solution to the $\RD$ anomalies and the dark matter puzzle is an interesting research avenue. Based on the results of this work, we have suggested few benchmark scenarios consistent with all the constraints and amenable to possible discovery in the future. We plan to expand the preliminary results presented here into a comprehensive study of the model in the context of not only the current constraints, but also by considering future projections at the HL-LHC, future colliders and the coverage of future DM direct detection experiments.

\subsection*{Acknowledgements}
We would like to thank Jordan Bernigaud, David Marzocca and Marco Nardecchia for their collaboration in the early stages of this work, and we would like to express our gratitude to the organisers of the Les Houches ``Physics at TeV Colliders'' 2019 Session for setting up an interesting workshop and providing an ideal atmosphere for scientific exchange. We would like to thank Jonathan Kriewald for useful comments on the manuscript.
BF and AJ would like to thank Abdollah Mohammadi for kindly providing cutflow tables for some benchmark scenarios and for his assistance throughout the implementaion and the validation of the CMS leptoquark plus missing energy analysis in {\sc MadAnalysis}~5. The work of AJ is supported in part by a KIAS Individual Grant No. QP084401 via the Quantum Universe Center at Korea Institute for Advanced Study and by the National Research Foundation of Korea, Grant No.~NRF-2019R1A2C1009419.
The work of AL was supported by the S\~ao Paulo Research Foundation (FAPESP), project
2015/20570-1. 
JH acknowledges support from the DFG via the Collaborative Research Center TRR 257 and the F.R.S.-FNRS as a Charg\'e de recherche.
The work of AP and GB was funded by the RFBR and CNRS project number 20-52-15005. The work of AP was also supported in part by an AAP-USMB grant and by the Interdisciplinary Scientific and Educational School of Moscow University for Fundamental and Applied Space  Research. 
The work of DS is based upon work supported by the National Science Foundation under Grant No. PHY-1915147.
JZ is supported by the {\it Generalitat Valenciana} (Spain) through the {\it plan GenT} program (CIDEGENT/2019/068), by the Spanish Government (Agencia Estatal de 
Investigaci\'on) and ERDF funds from European Commission (MCIN/AEI/10.13039/501100011033, Grant No. PID2020-114473GB-I00). All Feynman diagrams shown in this document have been produced using \textsc{FeynArts} version 3.9~\cite{Hahn:2000kx}.

\appendix
\section{\fr~ model}
\label{app:model}

In this appendix we collect all relevant information about the \fr\ implementation of the model introduced in section~\ref{sec:lagrangian}. Both the \fr\ model file and the corresponding UFO libraries are publicly available on \url{https://feynrules.irmp.ucl.ac.be/wiki/LQDM}, which also includes \mg\ parameter cards for the benchmark points considered in this work\footnote{An implementation of a more general model for scalar LQs can be found in \cite{Crivellin:2021ejk}. We must note that this implementation does not target leptoquark models with dark matter candidates.}.

\begin{table}[t]
\renewcommand{\arraystretch}{1.4}
\setlength\tabcolsep{12pt}
\centering\resizebox{\textwidth}{!}{
\begin{tabular}{c c c c c c}
  Field & Spin & Representation & Self-conjugate & \fr\ name & PDG code\\
  \hline\hline
  $S_1$    & 0   & $({\bf 3}, {\bf 1})_{-1/3}$ & no  & {\tt LQ} & 42\\
  $\chi_0$ & 1/2 & $({\bf 1}, {\bf 1})_0$      & yes & {\tt chi0} & 5000522\\
  $\chi_1$ & 1/2 & $({\bf 3}, {\bf 1})_{-1/3}$ & no  & {\tt chi1} & 5000521\\
\end{tabular}}
\caption{New particles supplementing the Standard Model, given
  together with their representation under $SU(3)_c\times SU(2)_L \times U(1)_Y$. We additionally indicate whether the particles are self-conjugate, we introduce their name in the \fr\ implementation and the associated PDG identifier. As the PDG code 42 is the official PDG value for a leptoquark~\cite{Tanabashi:2018oca} , there are no issues when running parton showering and/or hadronisation code.\label{tab:RDM:fields}}
\begin{tabular}{c c c c}
  Coupling & \fr\ name & Les Houches block name\\
  \hline\hline
  $(\lambda_{\sss L})_{ij}$ & {\tt lamL} & {\tt LQLAML}\\
  $(\lambda_{\sss R})_{ij}$ & {\tt lamR} & {\tt LQLAMR}\\
  $\ydm$ & {\tt yDM} & {\tt DMINPUTS}\\
\end{tabular}
\caption{New physics couplings beyond the SM, given together with the associated \fr\ symbol and the Les Houches block name that can be refered to when using the model with a high-energy physics numerical tool.}
\label{tab:RDM:params}
\end{table}

The field content of the new physics sector of our simplified model is summarised in table~\ref{tab:RDM:fields}, in which we also show the corresponding representation under the gauge and Poincar\'e groups, the potential Majorana nature of the different particles, the adopted symbol in the \fr\ implementation and the Particle Data Group (PDG) identifier that has been chosen for each particle. The new physics coupling parameters are collected in table~\ref{tab:RDM:params} that additionally includes the name used in the \fr\ model and the Les Houches blocks~\cite{Skands:2003cj} in which the numerical values of the different parameters can be changed by the user when running tools like \mg~\cite{Alwall:2014hca} or \micromegas~\cite{Belanger:2014vza}. The \fr\ implementation finally includes the Lagrangian~\eqref{eq:RDM:lag}, and has been validated and carefully checked alongside the guidelines sketched in \cite{Christensen:2009jx} (Hermiticity, cross sections for basic processes, {\it etc.}).

\section{\texorpdfstring{$t$}{t}-channel lepton exchange contributions to leptoquark pair production}
\label{app:tchannel}
As mentioned in section~\ref{sec:pheno}, the last diagram in the third row of figure~\ref{fig:collider_diagrams} is often neglected in any phenomenological or experimental analysis. The naive scaling of this diagram indeed goes as $\lamR^2$, which is (naively) assumed to be subdominant relative to the QCD-induced contributions (that scale as the strong coupling $\alpha_s$). However, the considered solutions to the $\RD$ anomalies sometimes involve ${\cal O}(1)$ values for $\lamR$, so that this diagram could have a substantial impact on a cross section~\cite{Borschensky:2020hot,Borschensky:2021hbo}. Deciding whether this contribution could be neglected is very non-generic, and the issue must be addressed on a case-by-case basis, for any individual benchmark. This task is performed in the present appendix.

In the new physics setup examined in this work, we consider $\lamR$ leptoquark couplings to second-generation quarks and third-generation leptons. Therefore, $t$-channel lepton exchange contributions to leptoquark pair production can only be induced by a charm-anticharm initial state. The corresponding total rate $\sigma\equiv \sigma (pp\to S_1 \overline{S}_1)$ can thus be written as
\be\bsp
   \sigma =&\ \sigma(gg\to S_1 \overline{S}_1) 
       \ +\ \sum_{q=u,d,s} \sigma(q\bar{q} \to S_1 \overline{S}_1) 
       \ +\ \sigma(c\bar{c} \to S_1 \overline{S}_1) \\
  \equiv &\ \sigma_1 + \kappa_1 \Big(1 + \kappa_2 \lambda_{R}^4 + \kappa_3 \lambda_{R}^2\Big).
 \esp\label{eq:sigma_with_interference}\ee
In the first line of the above expression, we have explicitly singled out the $c\bar c$ contributions, and ignored any PDF-suppressed $b\bar b$ contribution. In its second line, we have introduced a semi-analytical form for the cross section that is well suited for numerical estimates. All contributions stemming from initial states different from the $c\bar c$ one are collected into the $\sigma_1$ coefficient, whilst the $c\bar c$ component has been rewritten as a polynomial in $\lamR$ involving three numerical coefficients $\kappa_1$, $\kappa_2$ and $\kappa_3$.

All the free parameters $\sigma_1$ and $\kappa_i$ (with $i=1,2,3$) depend on the leptoquark mass $\mLQ$, the parton distribution functions and the collider centre-of-mass energy. We assess the impact of the $t$-channel lepton exchange diagrams by fitting them to the numerical results obtained with \mg, that is used to convolve LO matrix elements with the \texttt{NNPDF30\_lo\_as\_0118} PDF set~\cite{Ball:2014uwa}. Considering four representative values of $\mLQ = 500$, 1000, 1500 and 2000~GeV, we obtain
\begin{eqnarray}
\sigma = \begin{cases}
		321.54  + 0.34 \!\times\! \bigg(1 - 1.88 \lambda_R^2 + 5.06 \lambda_R^4 \bigg)~{\rm fb} & {\rm for}\ M_{S_1} \!=\! 500~{\rm GeV}, \\
		3.48 + 1.28\!\times\! \bigg(1 - 3.19 \lambda_R^2 + 9.42 \lambda_R^4 \bigg) \times 10^{-3}~{\rm fb} & {\rm for}\ M_{S_1} \!=\! 1000~{\rm GeV}, \\ 
        0.13 + 0.03 \!\times\! \bigg(1 - 3.74 \lambda_R^2 + 9.78 \lambda_R^4 \bigg) \times 10^{-3}~{\rm fb} & {\rm for}\ M_{S_1} \!=\! 1500~{\rm GeV}, \\ 
        7.64 \!\times\! 10^{-3} + 6.66 \!\times\! \bigg(1 - 3.66 \lambda_R^2 + 9.94 \lambda_R^4 \bigg) \!\times\! 10^{-7}~{\rm fb} & {\rm for}\ M_{S_1}\!=\! 2000~{\rm GeV}.
		\end{cases} \hspace{1cm}
		\label{eq:xsc:tch}
\end{eqnarray}
We can immediately see that $\sigma_1$ is about three orders of magnitude larger than the $\kappa_1$ prefactor. Moreover, for moderate values of $\lamR$ there is a partial cancellation between the quadratic and the quartic terms associated with the $c\bar c$ contributions to the cross section, as they come with opposite signs. Hence $t$-channel exchange diagrams are only relevant for very large values of $\lamR$. 

\begin{figure}[!t]
    \centering
    \includegraphics[width=1.01\textwidth]{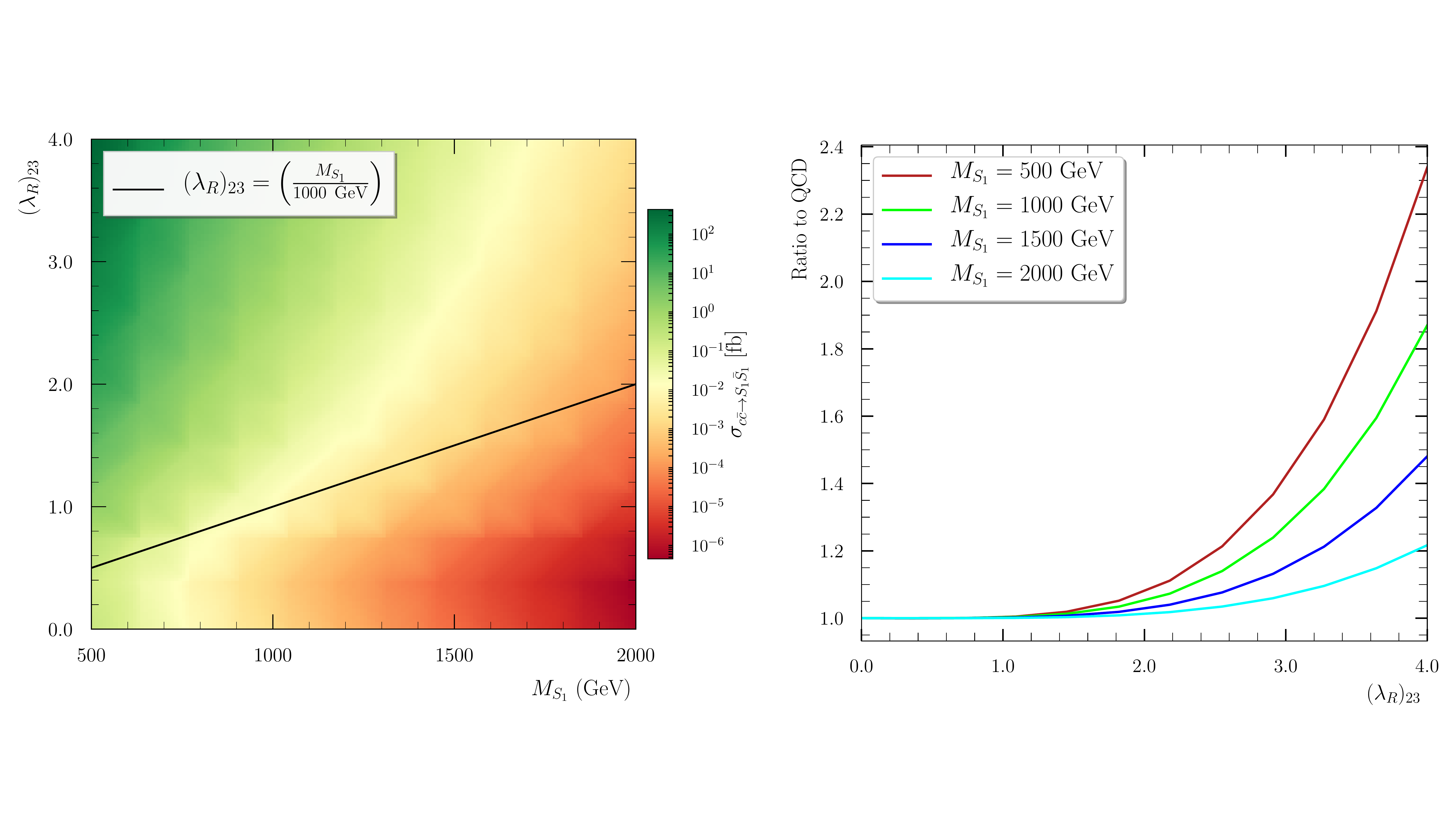}
    \vspace{-2cm}
    \caption{{\it Left}: Contribution to the leptoquark pair production cross section at the LHC that originates from $c\bar{c}$ scattering. The results are shown as a function of $M_{S_1}$ and $(\lambda_R)_{23}$, and the black line corresponds to the BS2 benchmark slope introduced in the context of the $R_D$ anomalies. {\it Right}: Ratio of the total cross section (including lepton $t$-channel exchanges) to the pure QCD contribution, shown as a function of $(\lambda_R)_{23}$ for $M_{S_1} = 500~{\rm GeV}$ (red), $1000~{\rm GeV}$ (lime), $1500~{\rm GeV}$ (blue) and $2000~{\rm GeV}$ (cyan).}
    \label{fig:xs:tch}
\end{figure}

We show results corresponding to a larger scan, together with the best fit point for the $\RD$ anomalies BS2, in the left panel of figure~\ref{fig:xs:tch}. In the right panel of this figure, we compute the ratio of $\sigma$ over the pure QCD-induced LO rate ({\it i.e.}\ the rate that is obtained without including any $t$-channel lepton exchange contribution), and we show it as a function of $\lamR$ for the four different masses employed in equation~\eqref{eq:xsc:tch}. We can safely conclude from those results that for the benchmark points defined in table~\ref{tab:RDM:benchmark_points}, the corrections that originate from $t$-channel lepton exchange lie at the sub-percent level for $\mLQ=1$ TeV. For $\mLQ= 1.6$ TeV (which for BS2 would imply $\lamR=1.6$), the overall impact increases to about 2 \%, which is well below the typical size of the theory errors associated with our predictions.

\section{Reinterpretation of MET searches}
\label{app:met_searches}
In this appendix we provide details of the reinterpretation procedures that enabled us to obtain the results displayed in figure~\ref{fig:RDM:met_bounds}. Although most MET searches are interpreted within supersymmetric scenarios, several of them can be reinterpreted for the process $p p \to  \chi_1\chi_1$, with $\chi_1 \to L q \chi_0$, where $L = e,\mu,\tau$ or $\nu$ and $q = b,c$ or $t$. Furthermore, results for simplified models describing sbottom pair-production and decay ($p p \to \tilde{b}\tilde{b}$, $\tilde{b} \to b \tilde{\chi}_1^0$) can, within a good approximation, be directly applied to the channel $p p \to  \chi_1\chi_1$, $\chi_1 \to \nu\ b \chi_0$, since the signal efficiencies will be very similar.

\subsection{CMS $b/c$ + MET}

The CMS search~\cite{CMS:2017kil} for $b$ or $c$-jets and missing energy (CMS-SUS-16-032)
can be sensitive to the scenarios discussed in this work, since it targets many of the final states generated by $\chi_1$ production and decay. The search has signal regions dedicated to compressed scenarios, in which secondary vertices are used to discriminate between signal and background. Such signal regions can in particular be sensitive to $b$-jets with very small $p_T$. As we have seen, the compressed regions are of primordial importance for us.

The search was implemented using {\sc CheckMATE~2} and was validated for both the compressed and non-compressed signal regions. The validation aimed to reproduce the official CMS exclusion curves obtained for the compressed stop ($p p \to \tilde{t} \tilde{t}$, $\tilde{t} \to c \tilde{\chi}_1^0$) and sbottom ($p p \to \tilde{b} \tilde{b}$, $\tilde{b} \to b \tilde{\chi}_1^0$) supersymmetric simplified models. These topologies are also relevant to validate the implementation of the $b$-tagging efficiencies provided as auxiliary material. Finally, the CMS collaboration has provided covariance matrices to allow for the statistical combination of the relevant signal regions using the simplified likelihood framework~\cite{Collaboration:2242860}. All the limits shown in this work were computed using these covariance matrices, following the  SModelS~\cite{Ambrogi:2018jqj} implementation. Further details of the implementation can be found in \url{https://github.com/andlessa/RDM/tree/master/myCheckMate3Files/validation}.

\begin{figure}
    \centering
    \includegraphics[width=0.47 \textwidth]{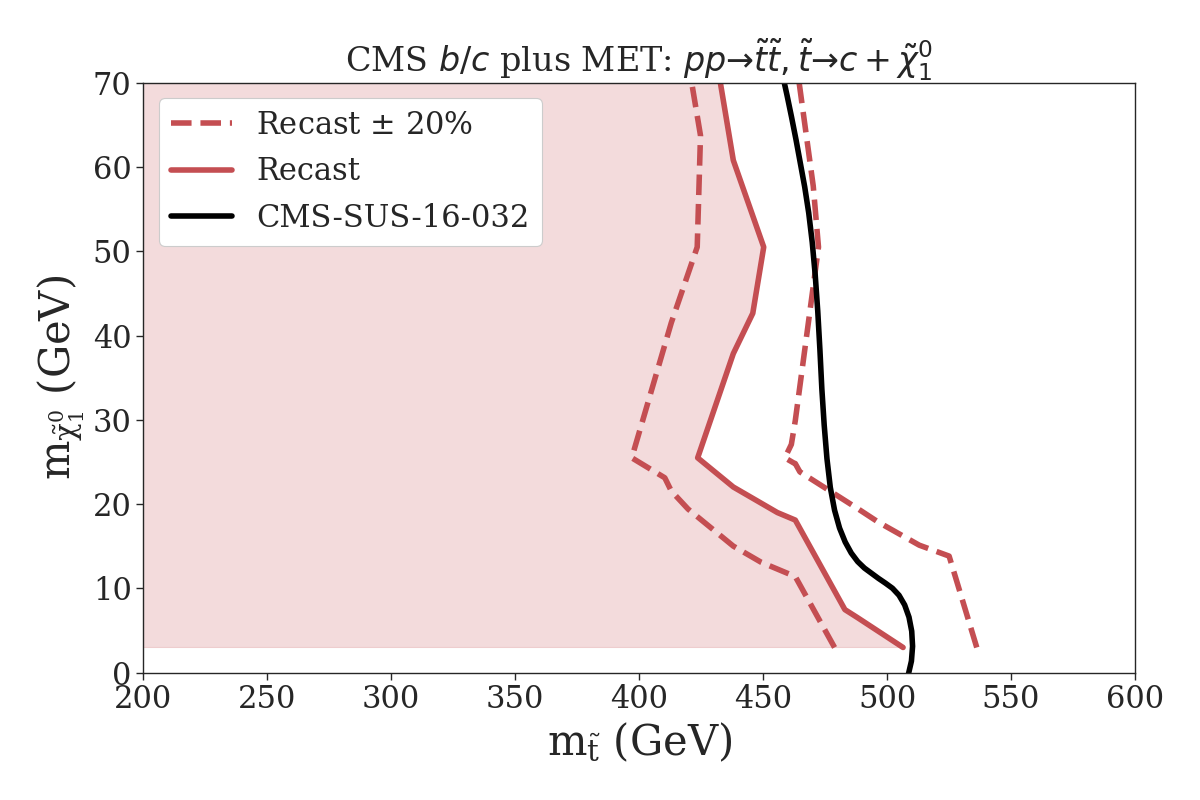}
    \includegraphics[width=0.47 \textwidth]{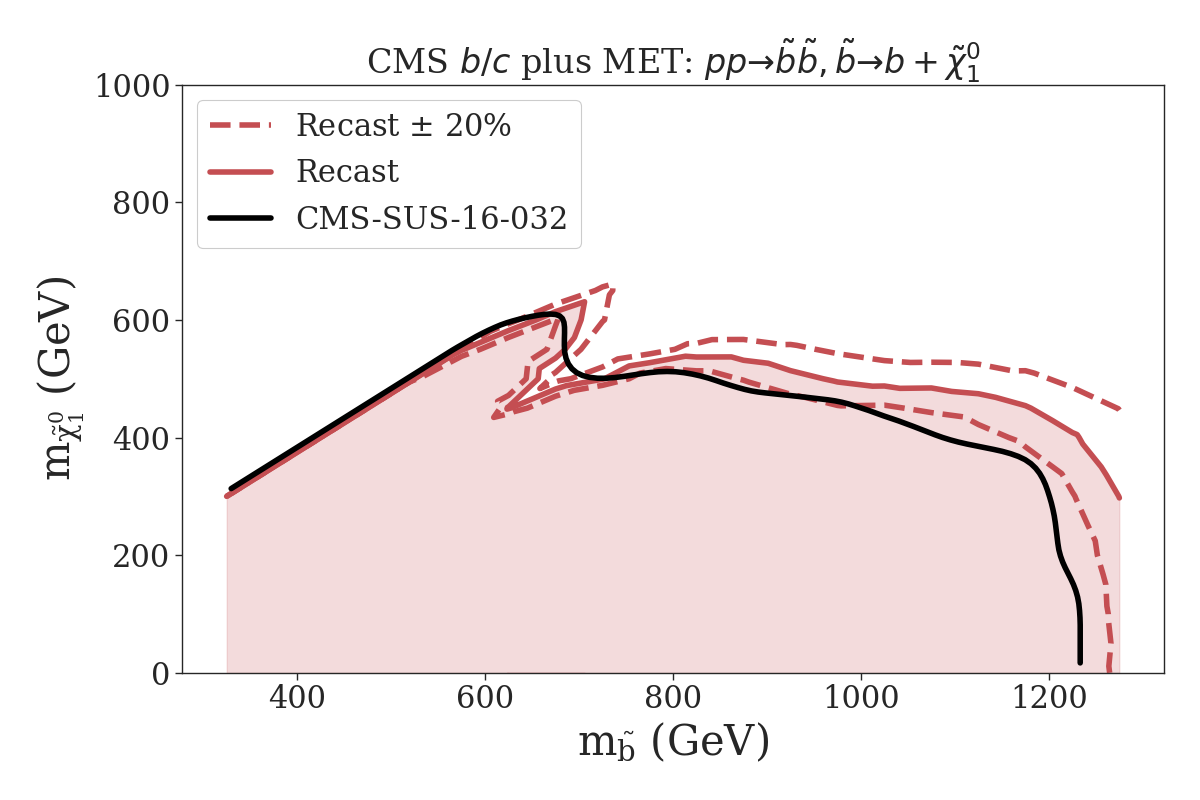}
    \caption{Validation of the CMS search for missing energy in association with $b-$ or $c-$ jets~\cite{CMS:2017kil} implemented using {\sc CheckMATE~2}. The solid black line shows the official exclusion curve, while the solid red curve shows the one obtained through recasting. The left panel shows the results obtained for the compressed stop simplified model, while the right panel displays the results obtained in the sbottom case.}
    \label{fig:val_CMS_bcMET}
\end{figure}

We display in figure~\ref{fig:val_CMS_bcMET} the validation figures obtained in the compressed stop scenario (left panel), and for the sbottom case (right panel). The solid red curves correspond to the exclusion obtained with our {\sc CheckMATE~2} implementation, while the solid black curve corresponds to the official CMS curve. Curves corresponding to a 20\% variation in the signal are also shown as dashed lines. As it can be seen, the curves agree with the official exclusion within a 20\% uncertainty. Furthermore, the region with a compressed sbottom (upper left corner in the right panel), which is the most relevant for the results shown in this work, is well described by our recasting.

\subsection{ATLAS $bb$+MET}

In addition to the CMS search for $b$-jets plus MET, we have also considered the corresponding ATLAS search (ATLAS-SUSY-2018-34~\cite{ATLAS:2021yij}) that relies on a luminosity of 139~fb$^{-1}$ of data. For this search, however, we do not rely on a detailed recasting of the event selection, but simply apply the upper limits obtained for the sbottom simplified model ($p p \to \tilde{b} \tilde{b}$, $\tilde{b} \to b \tilde{\chi}_1^0$) to the channel 
 $p p \to  \chi_1 \chi_1$, $\chi_1 \to \nu\ b \chi_0$, as discussed before. Since no auxiliary material was available until the completion of this work, upper limits on sbottom pair-production cross sections were digitised and included in a private version of the {\sc SModelS}~\cite{Ambrogi:2018ujg} database.  {\sc SModelS} was then used to compute the effective signal cross section, $\sigma(p p \to \chi_1 \chi_1) \times BR(\chi_1 \to b \nu \chi_0)^2$, which was next compared against the official upper limits implemented in the database.
 
\begin{figure}
 	\centering
 	\includegraphics[width=0.47 \textwidth]{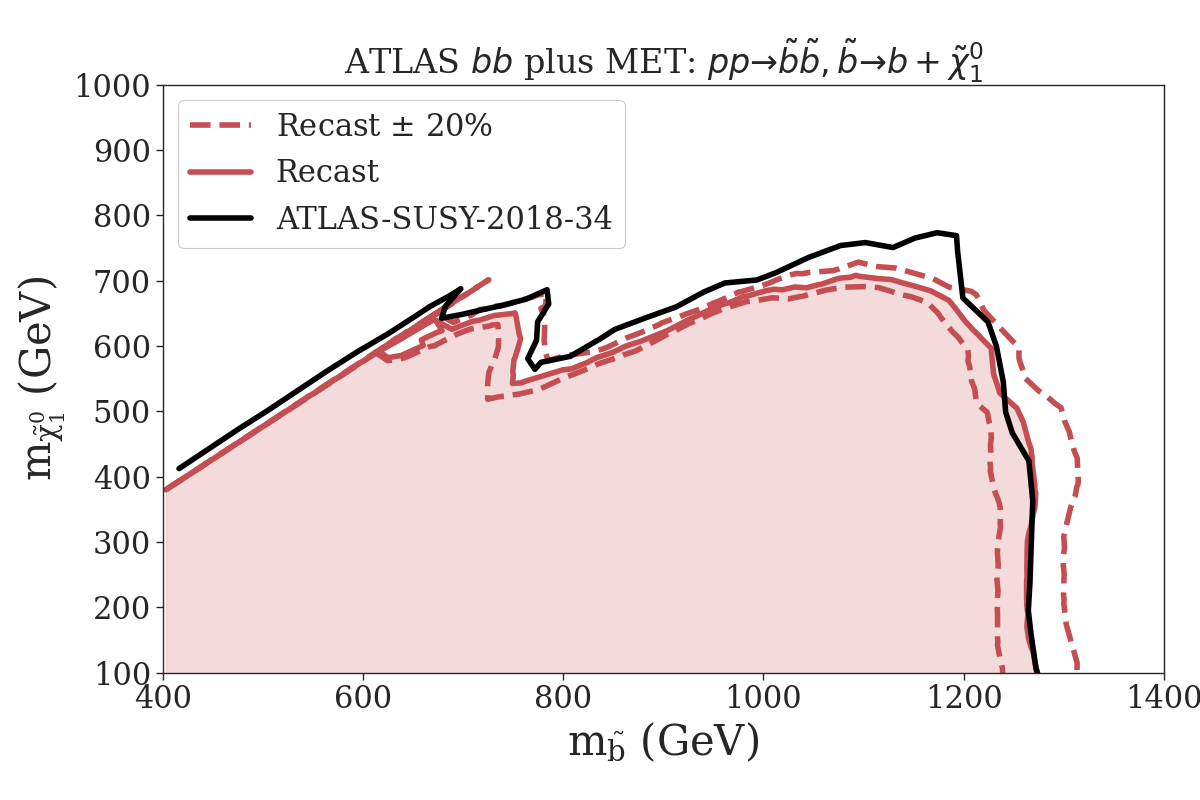}
 	\caption{Validation of the ATLAS search for $b$-jets plus MET~\cite{ATLAS:2021yij} as implemented in the {\sc SModelS} database. The solid black line shows the official exclusion curve for the sbottom simplified model considered by the ATLAS collaboration, while the solid red curve shows the corresponding curve obtained through recasting.}
 	\label{fig:val_ATLAS_bbMET}
 \end{figure}

In order to verify that the analysis has been properly implemented in the {\sc SModelS} database, we compare in figure~\ref{fig:val_ATLAS_bbMET} the official ATLAS exclusion curve for the sbottom scenario with the curve obtained applying {\sc SModelS} to the same simplified model. As we can see the results agree well.

\subsection{ATLAS mono-jet}

Since $\chi_1$ can be nearly mass degenerate with the dark matter candidate ($\chi_0$), mono-jet searches are relevant for constraining this region of parameter space. We have implemented in {\sc CheckMATE~2} the ATLAS mono-jet search~\cite{ATLAS:2021yij} with 139~fb$^{-1}$ of luminosity (ATLAS-EXOT-2016-06), which targets a hard jet and missing energy. The search considers multiple bins in MET divided into inclusive ($\mbox{MET}_{\rm min} < \mbox{MET}$) and exclusive  ($\mbox{MET}_{\rm min} < \mbox{MET} < \mbox{MET}_{\rm max}$) signal regions. Some of the simplified models considered by the ATLAS collaboration as targers for this search are the compressed sbottom ($p p \to \tilde{b} \tilde{b}$, $\tilde{b} \to b \tilde{\chi}_1^0$) and compressed stop ($p p \to \tilde{t} \tilde{t}$, $\tilde{t} \to c \tilde{\chi}_1^0$) ones.

\begin{figure}
    \centering
    \includegraphics[width=0.47 \textwidth]{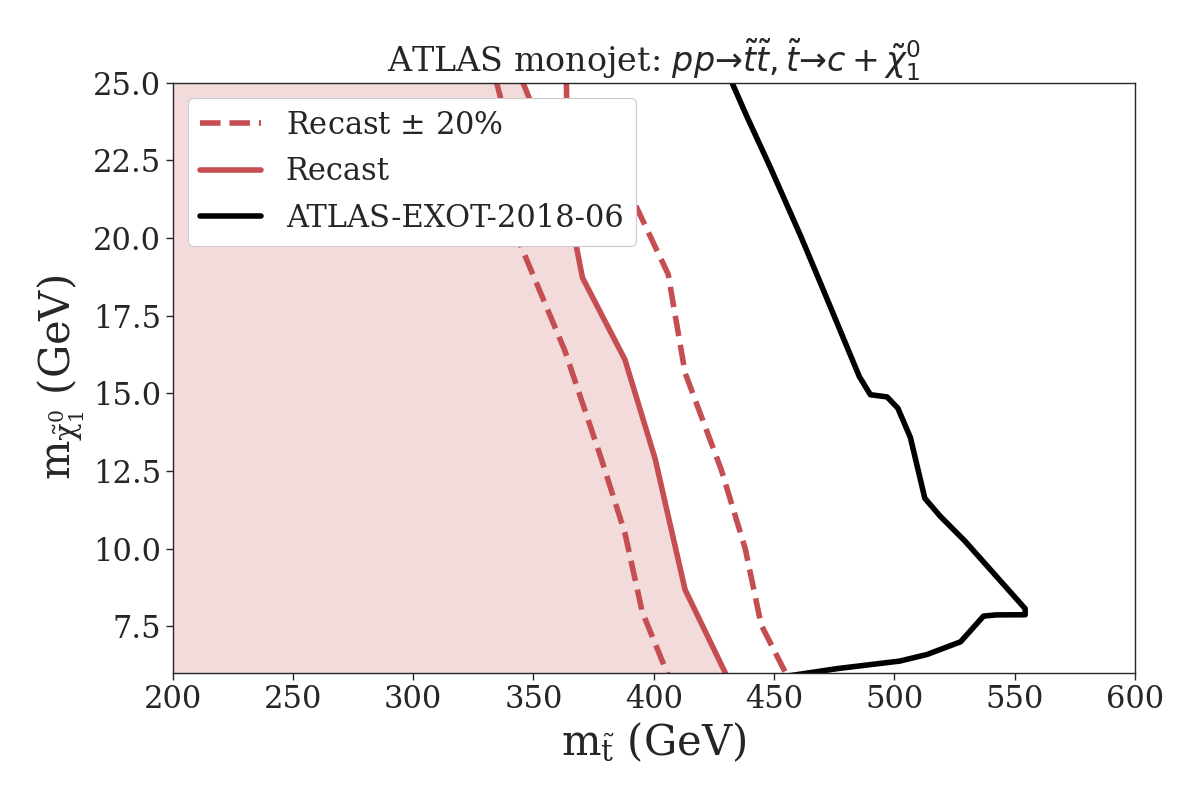}
    \includegraphics[width=0.47 \textwidth]{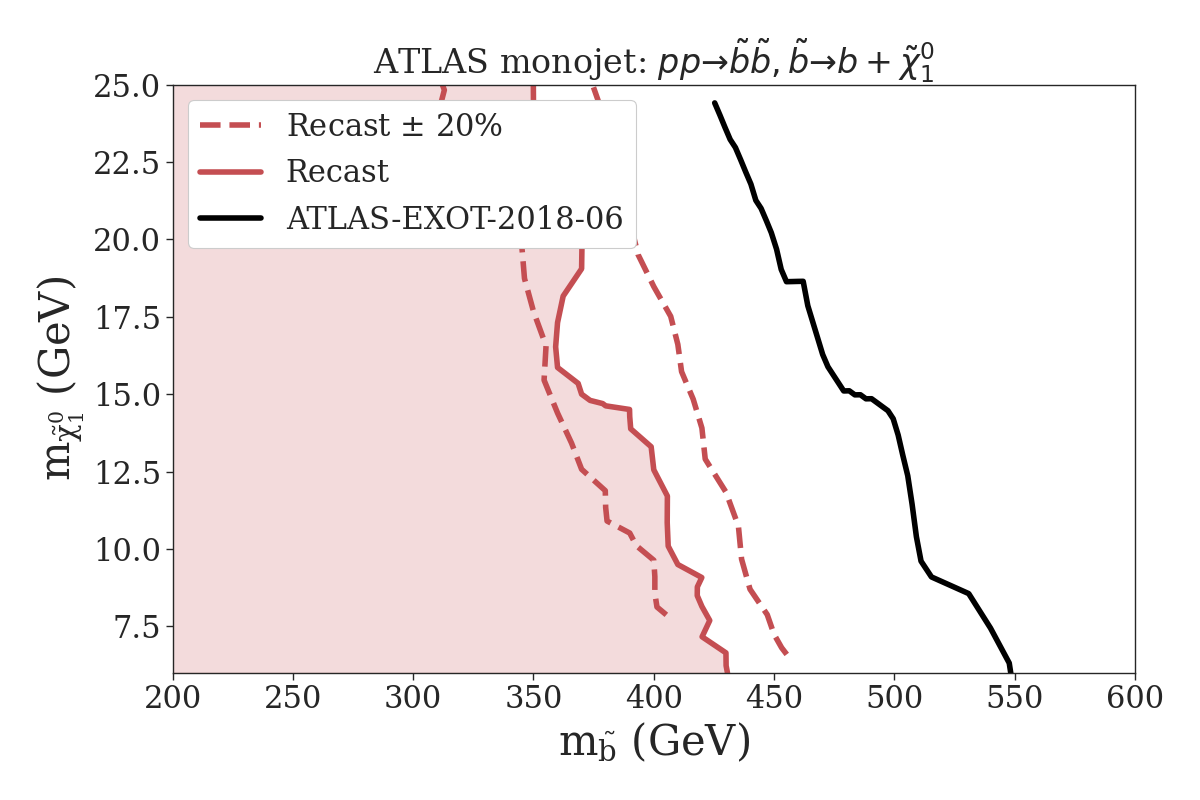}
    \caption{Validation of the mono-jet ATLAS search~\cite{ATLAS:2021yij} implemented using {\sc CheckMATE~2}. The solid black line shows the official exclusion curve, while the solid red curve shows the one obtained through recasting. The left panel shows the results obtained for the compressed stop simplified model, while the right panel displays the results for the compressed sbottom scenario.}
    \label{fig:val_ATLAS_monojet}
\end{figure}

In order to validate the recasting, we have scanned the parameter spaces of these simplified models and computed the exclusion curves using our {\sc CheckMATE~2} implementation. The results are shown in figure~\ref{fig:val_ATLAS_monojet}, where the solid black curve displays the official exclusion obtained by the ATLAS collaboration. The recasting results are found to be somewhat conservative, leading to a smaller excluded region when compared to the official results. However, the ATLAS exclusion profits from a fit of all exclusive signal regions, which can not be done in the recasting approach. Therefore it is not surprising that the recasting provides weaker limits than the official results. In addition, we have verified that those results are compatible with those that would be obtained when employing the implementation of the most recent run~2 CMS mono-jet analysis in {\sc MadAnalysis}~5. Such an implementation, with the corresponding validation, has been officially provided by the CMS collaboration. We refer the interested reader to refs.~\cite{CMS:2021far,IRF7ZL_2021} for details.

\subsection{CMS soft-lepton(s) plus MET}

This analysis targeted a search of new physics in events with two oppositely charged soft leptons that can be either electrons or muons in addition to missing energy \cite{CMS:2018kag}. The search has been performed using data collected at $\sqrt{s} = 13~{\rm TeV}$ and an integrated luminosity of $\mathcal{L} = 39.5~{\rm fb}^{-1}$. The CMS collaboration has interpreted this search in supersymmetric models whereas the charginos and neutralinos with nearly zero mass-splitting are pair produced. In the model considered in this work, this final-state signature can arise from $\chi_1$ pair production, followed by the decay $\chi_1 \to c\tau\ \chi_0 \to c\ \ell \bar{\nu}_\ell \nu_\tau\ \chi_0$. We have used \textsc{MadAnalysis}~5 which contains a validated implementation of this search. We encourage the reader to examine section 19 of \cite{Brooijmans:2020yij} and section 5.3 of \cite{Araz:2020lnp} for details about the analysis and its validation, and reference \cite{DVN/YA8E9V_2020} for the corresponding source codes.

\subsection{ATLAS multi-jet plus MET}
For this study we employ the \textsc{MadAnalysis}~5 implementation of the ATLAS-CONF-2019-040 note~\cite{ATLAS-CONF-2019-040} already used in the preliminary results presented in~\cite{Brooijmans:2020yij}. The ATLAS collaboration has published a newer version of this analysis~\cite{ATLAS:2020syg}, however since this study has not yet been implemented and validated, we opt for using its older version. We remark that for the stop/sbottom and neutralino case, the updated version of the study does not alter qualitatively the results, which furthers supports the use of the older yet validated study. 

\subsection{ATLAS $\tau \tau$+MET}

Most of the analyses discussed so far target jets plus MET or first and second generation leptons plus MET. However, in some of the scenarios discussed in this work, $\chi_1$ has a sizeable branching ratio to a $\tau c$ final state. Therefore the ATLAS search for hadronic taus plus MET (ATLAS-SUSY-2018-04) at 139~fb$^{-1}$~\cite{Aad:2019byo} can be relevant to test such scenarios. This search considers two signal regions: a first region targets harder taus (high mass) and a second region other softer taus (low mass). It is important to point out that the ATLAS search vetoes $b$-jets, so that a fraction of the $\chi_1 \to \tau c$ signal will be lost due to the mistagging of $c$-jets as $b$-jets. We have assumed the mistagging efficiency already implemented in {\sc CheckMATE~2}, based on~\cite{ATLAS:2015dex}.

\begin{figure}
    \centering
    \includegraphics[width=0.95 \textwidth]{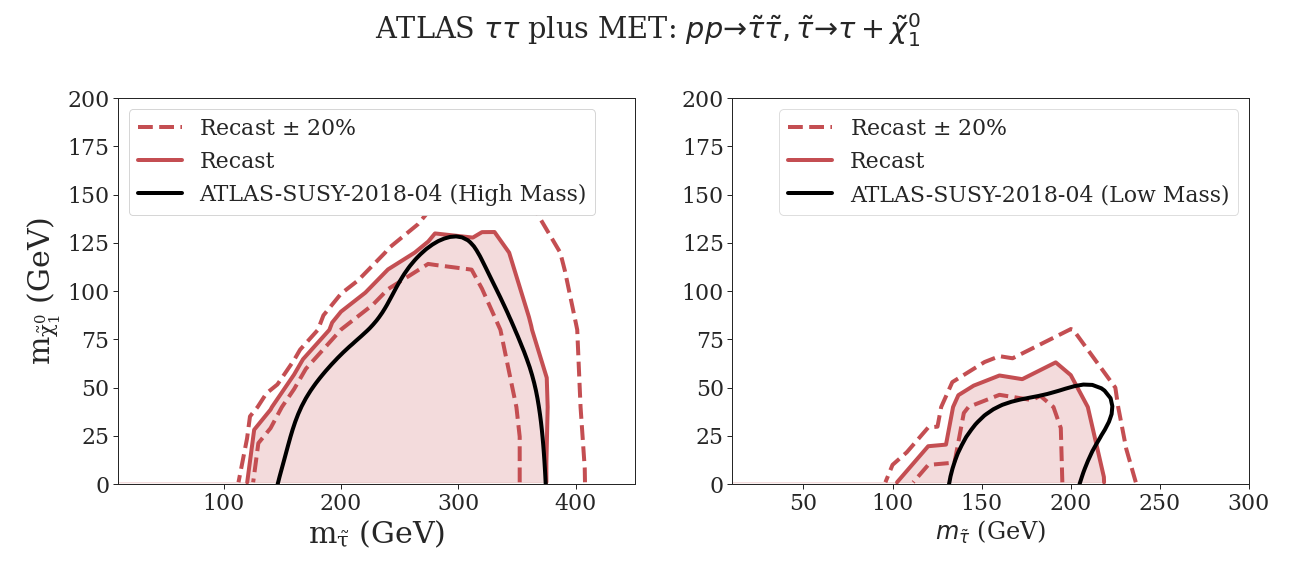}
    \caption{Validation of the hadronic tau plus MET ATLAS search~\cite{Aad:2019byo} implemented using {\sc CheckMATE~2}. The solid black line shows the official exclusion curve, while the solid red curve shows the one obtained through recasting. The left panel shows the results obtained for the high mass signal region, while the right panel displays the results for the low mass one.}
    \label{fig:val_ATLAS_tataMET}
\end{figure}

The simplified model considered by the ATLAS collaboration consists of stau pair production followed by $\tilde{\tau} \to \tau \tilde{\chi}_1^0$. Due to presence of $c$-jets in our signal it is not possible to directly apply the stau efficiencies (or upper limits) to our scenario. Hence a full recasting is necessary in order to reinterpret the results. We have used the {\sc CheckMATE~2} tool to implement and the validation was done for the stau simplified model in the low mass and high mass signal regions\footnote{During the completion of this work, a corresponding {\sc MadAnalysis}~5 implementation independently appeared~\cite{Fuks:2021wpe,Lim:2021tzk,UN3NND_2020}. It has not been used in this work.}. The ATLAS collaboration has provided separate exclusion curves for each signal region and these are compared to the ones obtained through recasting, as shown in figure~\ref{fig:val_ATLAS_tataMET}. As shown, the recasting results agree well with the exclusion curves, given the expected recasting uncertainties.

\section{Resonant leptoquark plus MET production at the LHC and the HL-LHC}
\label{app:lqmet_HL-LHC}
\subsection{CMS search at the LHC in the muon channel: details of the selection}
In this subsection, we discuss details of the event selection of the CMS search~\cite{Sirunyan:2018xtm} and its implementation~\cite{Fuks:2020xxz, Fuks:2021wpe} in {\sc MadAnalysis}~5~\cite{Conte:2012fm,Conte:2014zja,Dumont:2014tja,Conte:2018vmg} . 

The event selection comprises two steps: a preselection and a signal region definition. Preselected events are required to feature at least one isolated high-$p_T$ muon with $p_T > 60$~GeV and $|\eta| < 2.4$, dubbed as \texttt{SignalMuon}. They should next contain at least one high-$p_T$ jet with $p_T > 100$~GeV and $|\eta| < 2.4$, such a jet (coined a \texttt{SignalJet}) being isolated by $\Delta R > 0.5$ from the leading muon candidate. Vetoes are then required in order to reduce the contamination from $t\bar{t}, Z+$~jets and $W+$~jets backgrounds: events are vetoed if they contain $b$-tagged jets with $p_T > 30$~GeV and $|\eta| < 2.4$, electrons with $p_T > 15$~GeV and $|\eta| < 2.4$, or hadronically-decaying tau leptons with $p_T > 20$~GeV and $|\eta| < 2.3$. The selection then vetoes  events featuring a second muon with an electric charge opposite to the one of the leading muon, if the invariant mass ($m_{\mu\mu}$) of the reconstructed muon pair is compatible the decay of a $Z$-boson candidate, \emph{i.e.}~if $|m_{\mu \mu} - M_Z| < 10$~GeV ($M_Z = 91.2$~GeV being the $Z$-boson mass). Furthermore, the missing transverse energy is required to be larger than $100$~GeV,  and the missing transverse momentum, mainly originating from the decay of the second leptoquark in the signal case, is required to be well separated in azimuth from both the leading muon and the leading jet,
  \be
  |\phi_{\mathrm{miss}} - \phi_{\mu}| > 0.5\qquad\text{and}\qquad  |\phi_{\mathrm{miss}} - \phi_{\mathrm{jet}}| > 0.5.
  \ee
Finally, the transverse mass of the system constituted of the leading muon and the missing momentum is requried to be larger than $500$~GeV.

\begin{table}[t]
\setlength\tabcolsep{16pt}
{\begin{tabular}{@{}cccc@{}} \toprule
Cut &  $\mLQ = 500$~GeV & $\mLQ = 1000$~GeV & $\mLQ = 1500$~GeV  \\
\toprule
Initial events & $300032~(100\%)$ & $300032~(100\%)$ & $300032~(100\%)$ \\
\texttt{SignalMuon} & $29592~(9.86\%)$ & $40270~(13.42\%)$ & $44906~(14.97\%)$ \\
\texttt{SignalJet} & $28178~(95.22\%)$ & $39654~(98.47\%)$ & $44444~(98.97\%)$ \\
$b$-\texttt{Veto}  & $25910~(91.95\%)$ & $35892~(90.52\%)$ & $40292~(90.65\%)$ \\
$\tau_h$-\texttt{Veto} & $17098~(65.99\%)$ & $22153~(61.72\%)$ & $24036~(59.65\%)$ \\
$e$-\texttt{Veto}  & $14973~(87.57\%)$ & $19151~(86.45\%)$ & $20594~(85.67\%)$ \\
\texttt{ZMassWindow} & $14824~(99.01\%)$ & $19041~(99.42\%)$ & $20519~(99.64\%)$ \\
\MET-\texttt{threshold} & $11775~(79.43\%)$ & $17896~(93.98\%)$ & $19913~(97.05\%)$ \\
$\Delta\phi(\textrm{jet},~ $\MET) ~$> 0.5$ & $11603~(98.54\%)$ & $17562~(98.13\%)$ & $19523~(98.04\%)$ \\
$\Delta\phi(\mu,~$\MET) ~$>0.5$ & $6434~(55.54\%)$ & $9991~(56.88\%)$ & $10870~(55.67\%)$ \\
$m_T > 500$~GeV & $370~(5.75\%)$ & $1798~(17.99\%)$ & $3428~(31.53\%)$ \\
\bottomrule
\end{tabular}
\caption{Cutflow of the CMS selection~\cite{Sirunyan:2018xtm} for $\mLQ = 500$, $1000$ and $1500$~GeV. The numbers inside the parentheses correspond to the cut efficiency ($\epsilon$) defined as $\epsilon = n_i/n_{i-1}$, with $n_k$ being the number of events surviving the $k^{\rm th}$ selection. Results are shown for $M_{\chi_0} = 100$~GeV and $\DX = 0.1$.}
 \label{tab:efficiency-LQmuc}}
\end{table}

\begin{figure}[!t]
\centering
\includegraphics[width=0.495\linewidth]{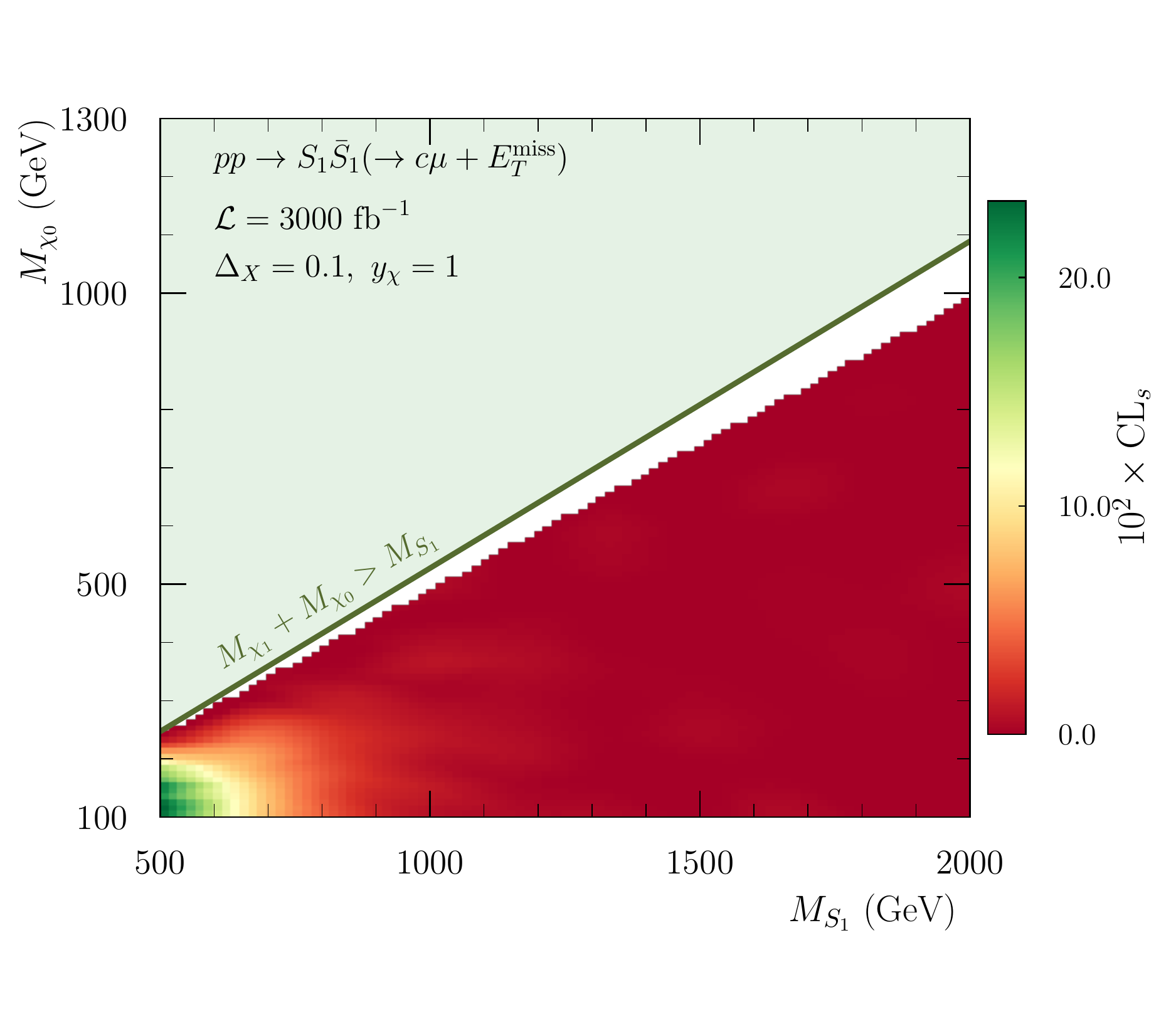}
\hfill
\includegraphics[width=0.495\linewidth]{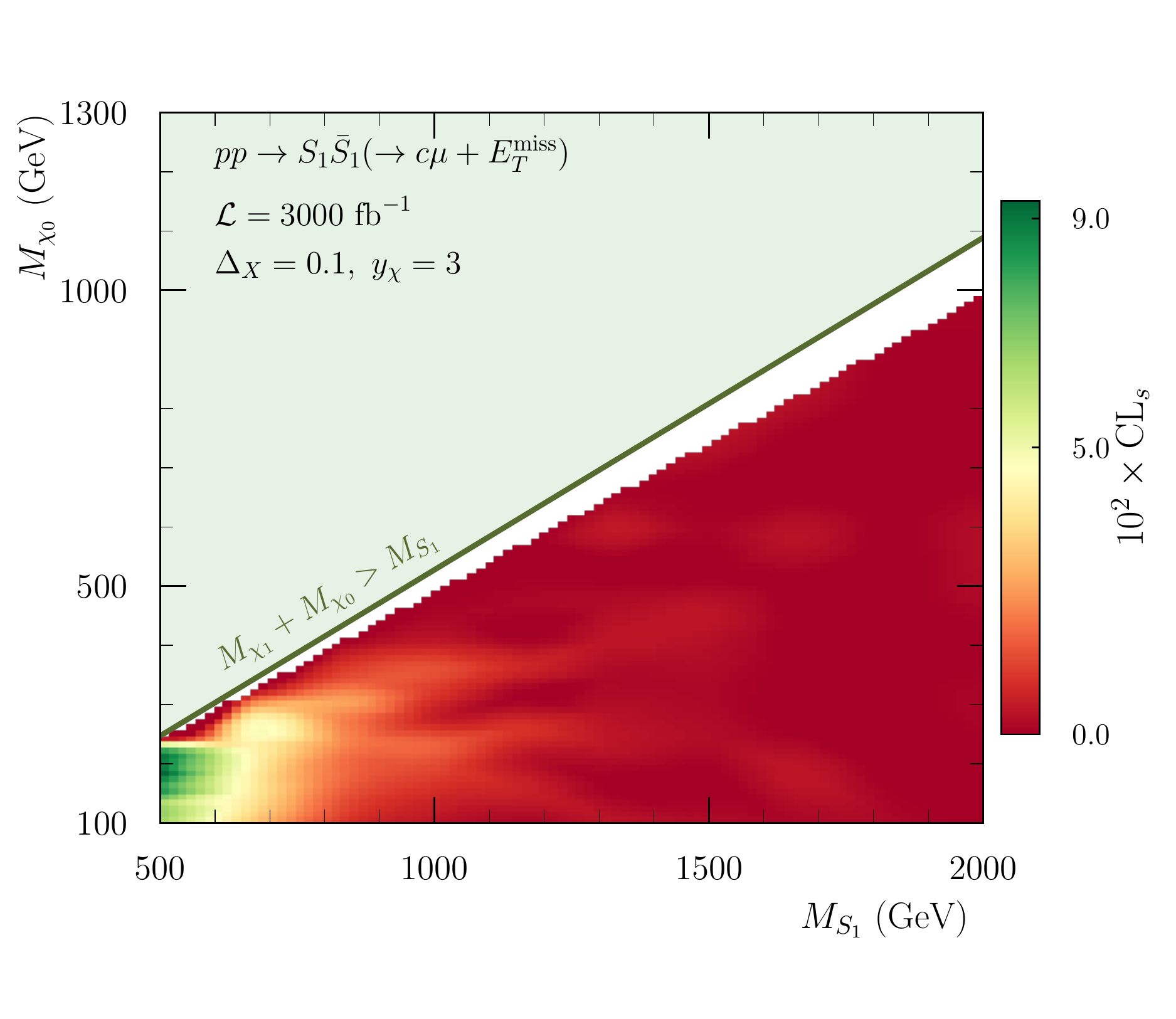}
\vfill
\includegraphics[width=0.495\linewidth]{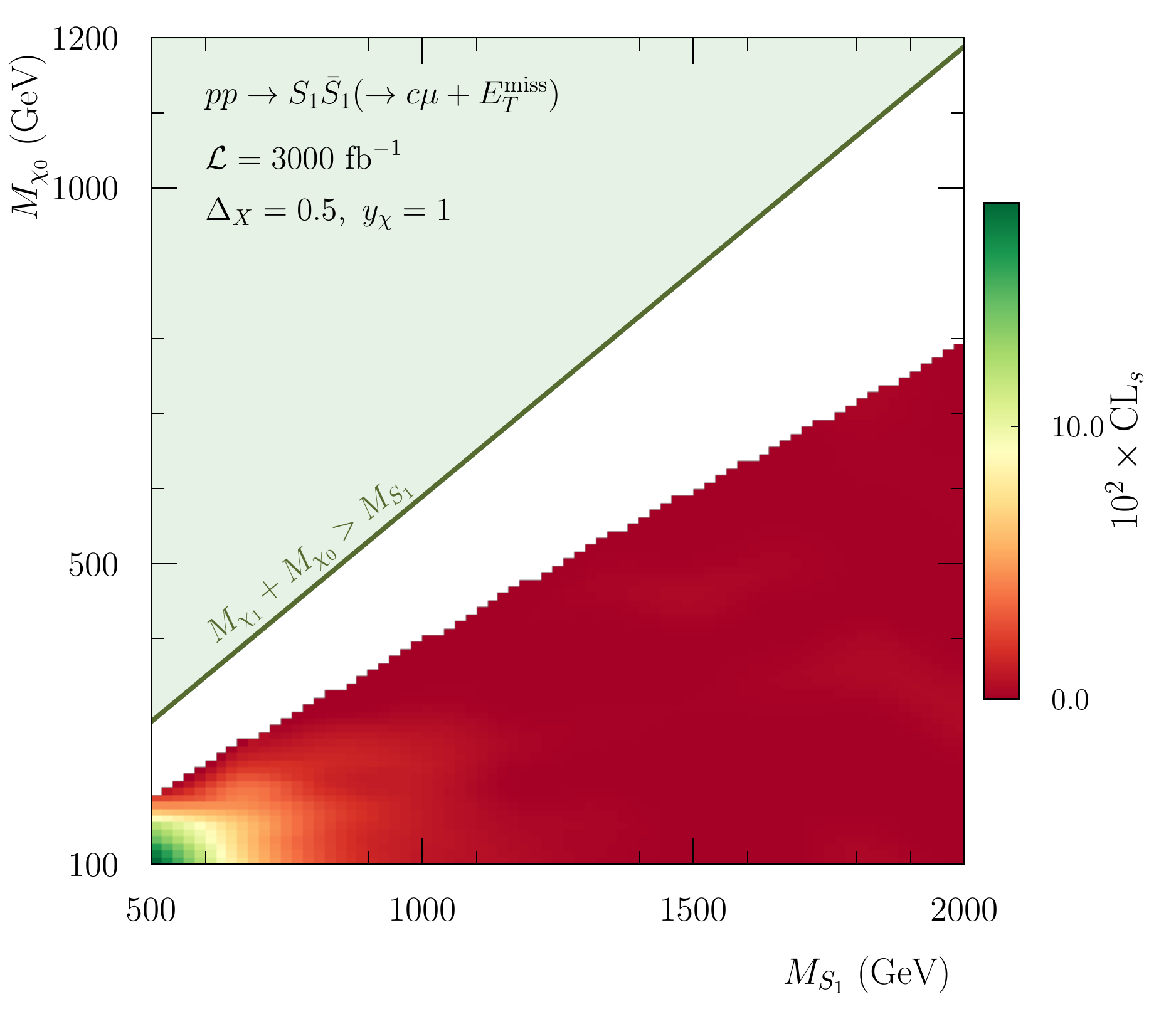}
\hfill
\includegraphics[width=0.495\linewidth]{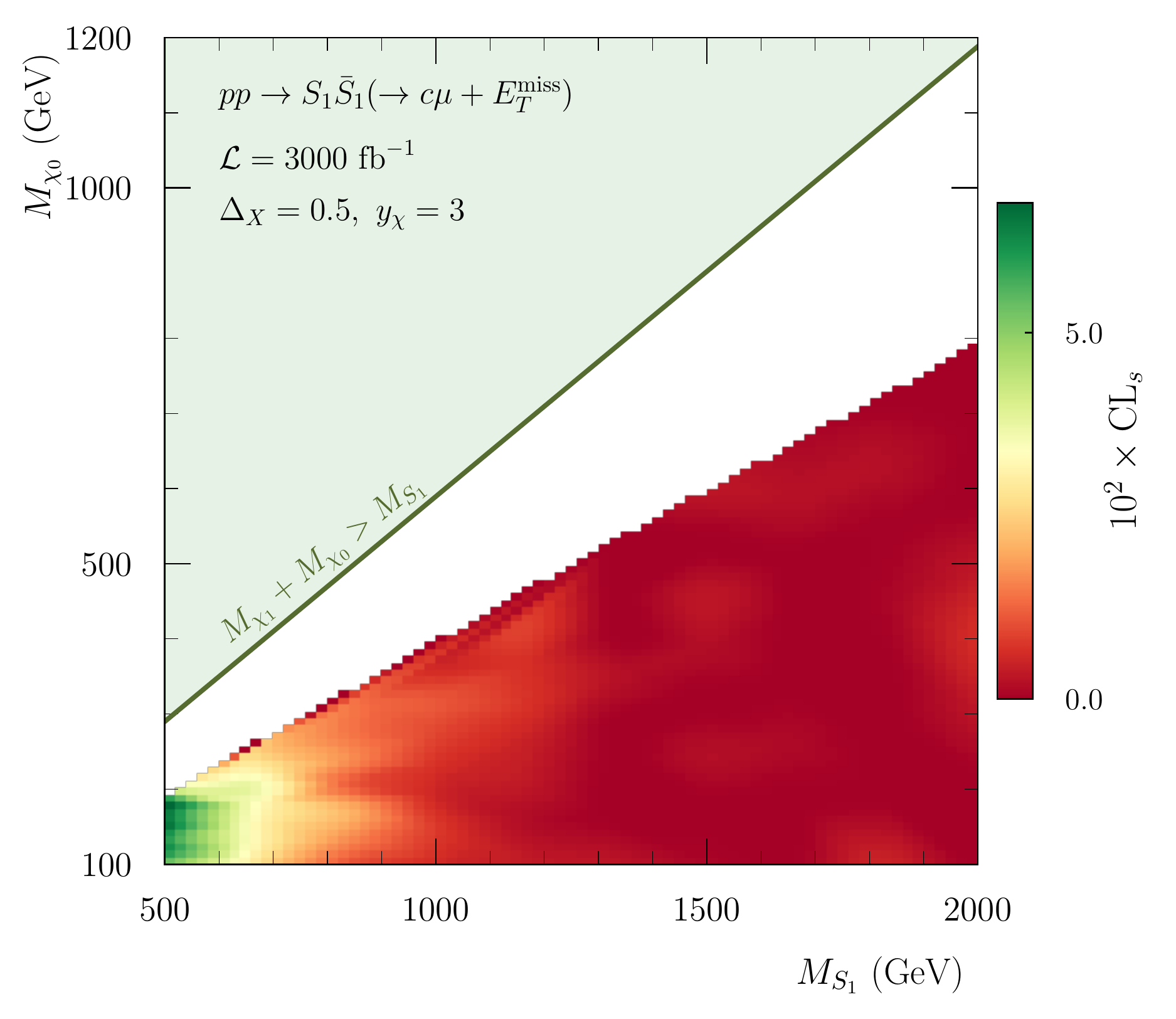}
\caption{Same as figure~\ref{fig:CLs-resonantLHC} but for a luminosity $\mathcal{L} = 3000$ fb$^{-1}$.}
\label{fig:CLs-resonantHLLHC}
\end{figure}

Detailed cutflow charts are provided in table~\ref{tab:efficiency-LQmuc} for leptoquark masses of $\mLQ = 500$~GeV, 1~TeV and 1.5~TeV. This shows that the typical efficiency of the considered CMS run 2 analysis in our scenarios is rather weak, reaching at most 1\%--2\% in the considered mass setups. As a second illustration of this poor constraining power, we display in figure~\ref{fig:CLs-resonantHLLHC} HL-LHC exclusions for the BP1, BP2, BP3 and BP4 benchmarks defined in section~\ref{sec:lqmet}. These are obtained by extrapolating the number of events for the signal and the background using the method explained in \cite{Araz:2019otb}.

Even at the HL-LHC, the CMS analysis~\cite{Sirunyan:2018xtm} is not sensitive to leptoquark plus dark matter scenarios providing an explanation for the $R_D$ anomalies. As already mentioned in section~\ref{sec:lqmet}, this is not surprising as tau-enriched final states are more relevant than muon-enriched ones in our benchmarks. This motivate the design of a similar analysis targeting the tau channel. Such an analysis is not (yet?) existing in the ATLAS and CMS experimental programs, so that we dedicate the next section to roughly assess its potential relevance in present and future data.
   
\subsection{Resonant leptoquark plus MET: prospects at the future HL-LHC}
In this section, we discuss a simple search strategy which can be designed specifically to test our 
scenarios. It relies on a final state including a hadronically decaying tau lepton $\tau_h$, instead of a high-$p_T$ muon as in the CMS analysis~\cite{Sirunyan:2018xtm}. The associated signature consists of at least one $\tau_h$, one jet, and a large amount of missing transverse momentum. The process under consideration indeed reads
 \begin{eqnarray}
 p p \to S_1 S_1 \to c \tau\ \chi_1 \chi_0 \to c\tau\ \chi_0 \tau c \chi_0,
 \end{eqnarray}
a generic illustrative Feynman diagram being shown in figure~\ref{fig:FD-LQMET}. As the second leptoquark decay leads to a large amount of missing transverse energy in addition to a soft $\tau$ and a soft charm jet, the requirement of having at least two hard $\tau_h$ and two hard jets in the final state would kill a significant amount of signal events. The corresponding signal efficiency is indeed typically of about $10\%$--$12\%$ for $\tau_h$ and jet $p_T$ thresholds of 20 and 25~GeV respectively, all objects being additionally imposed to be central ($|\eta|<2.4$). Such a selection is thus ignored in the design of our analysis strategy.

Accounting for the presence of barely reconstructed soft objects, the dominant background originates from $W/Z+$jets, $t\bar{t}$, di-boson, and single top production. Background simulations rely on the same Monte Carlo setup as in section~\ref{sec:lhc} with two exceptions. $W$-boson and top-quark decays are handled with {\sc MadSpin}~\cite{Artoisenet:2012st} and \textsc{MadWidth}~\cite{Alwall:2014bza}, and we use the \texttt{CMS} parametrisation shipped with {\sc Delphes}~3 to simulate the response of a typical LHC detector, bar a few modifications. Our $b$-tagging efficiency is parametrised by~\cite{CMS:2012rta}
\begin{eqnarray}
   \mathcal{E}_{b|b}(p_T) = 0.85 \tanh(2.5 \times 10^{-3} p_T)\left(\frac{25}{1+ 0.063 p_T}\right),
\end{eqnarray}
with corresponding mistagging probabilities of a light and a charm jet as a $b$-jet given by
\begin{eqnarray}
   \mathcal{E}_{j|b}(p_T) &=& 0.01 + 3.8 \times 10^{-5} p_T, \nonumber \\
   \mathcal{E}_{c|b}(p_T) &=& 0.25 \tanh(1.8 \times 10^{-2} p_T)\left(\frac{1}{1+ 1.3 \times 10^{-3} p_T}\right).
\end{eqnarray}
This implies a maximum $b$-tagging efficiency of $70\%$ for a mistagging probability of a charm (lighter) jet of about 20\% ($1\%$--$8\%$). In addition, our parametrisation relies on a tau-tagging efficiency of $85\%$~\cite{ATLAS:2017mpa}, that is associated with a mistagging rate of a light jet as a hadronic tau of about $1\%$. We have finally added the PDG code of the dark matter state in our model ($\chi_0$) to the list of particles leaving no energy in both the electromagnetic (\texttt{ECal}) and hadronic (\texttt{HCal}) calorimeters.
   
 \begin{table}
\setlength\tabcolsep{10pt}
\begin{tabular}{lccc}
\toprule
\multicolumn{1}{c} { Process } & Cross section $[\mathrm{pb}]$ & Generated events & $\omega_{i}$ \\
\toprule
$p p \rightarrow W^{\pm}\left(\rightarrow \tau^{\pm} \nu\right)+$ jets & $4.18 \times 10^{4}$ & $144.09 \times 10^6$ & $2.9 \times 10^{-4}$ \\
$p p \rightarrow Z(\rightarrow \nu \nu)+$ jets & $7.22 \times 10^{3}$ & $47.91 \times 10^6$ & $1.5 \times 10^{-4}$ \\
$p p \rightarrow Z / \gamma^{*}\left(\rightarrow \tau^{+} \tau^{-}\right)+$ jets & $2.33 \times 10^{3}$ & $24.62 \times 10^6$ & $9.5 \times 10^{-5}$ \\
$p p \rightarrow t \bar{t} \rightarrow b \nu \ell b j j$ & $2.09 \times 10^2$ & $1.56 \times 10^6$ & $1.3 \times 10^{-4}$ \\
$p p \rightarrow V V$ & $1.02 \times 10^2$ & $1.02 \times 10^6$ & $1.0 \times 10^{-4}$ \\
$p p \rightarrow t q+t W$ & $0.74 \times 10^2$ & $0.76 \times 10^6$ & $9.7 \times 10^{-5}$ \\
\bottomrule
\end{tabular}
\caption{Cross sections, number of generated events and event weights for the different background processes leading to a $c \tau$ plus MET signature. Single top quark production $(t q+t W)$ includes decays into a single-lepton final state.\label{tab:backgrounds}}
\end{table}   

In our simulations, leading-order hard-scattering $W/Z$+jets events include up to two extra partons, the merging of the related exclusive samples employing the \textsc{Mlm} scheme~\cite{Mangano:2006rw,Alwall:2008qv} with a merging scale $Q_0 = 30$~GeV (or equivalently \texttt{xqcut} $= 20~$GeV in the \textsc{MadGraph5\_aMC@NLO} language). We enforce the weak boson decays to be leptonic, and off-shell and interference effects are accounted for in $Z+$~jets production (for which we allow for decays in both charged leptons and neutrinos). Similarly, the top quark is imposed to decay leptonically in its single-production mode, and semi-leptonically in its pair-production mode. Details on the background cross sections (as returned by {\sc MadGraph5\_aMC@NLO}) and on the number of generated events are provided in table~\ref{tab:backgrounds}.

\begin{figure}
\centering
\includegraphics[width=0.495\textwidth]{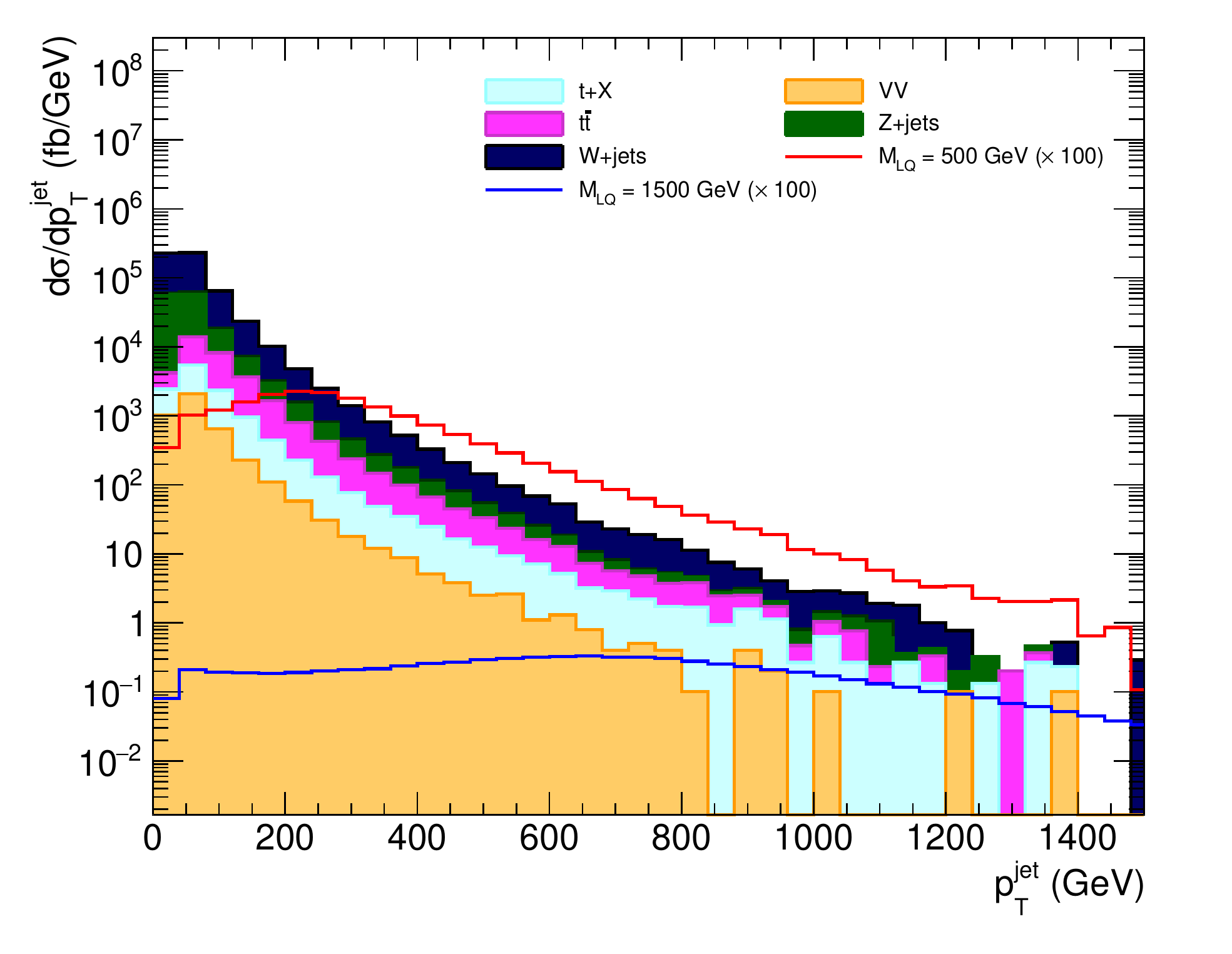}
\hfill
\includegraphics[width=0.495\textwidth]{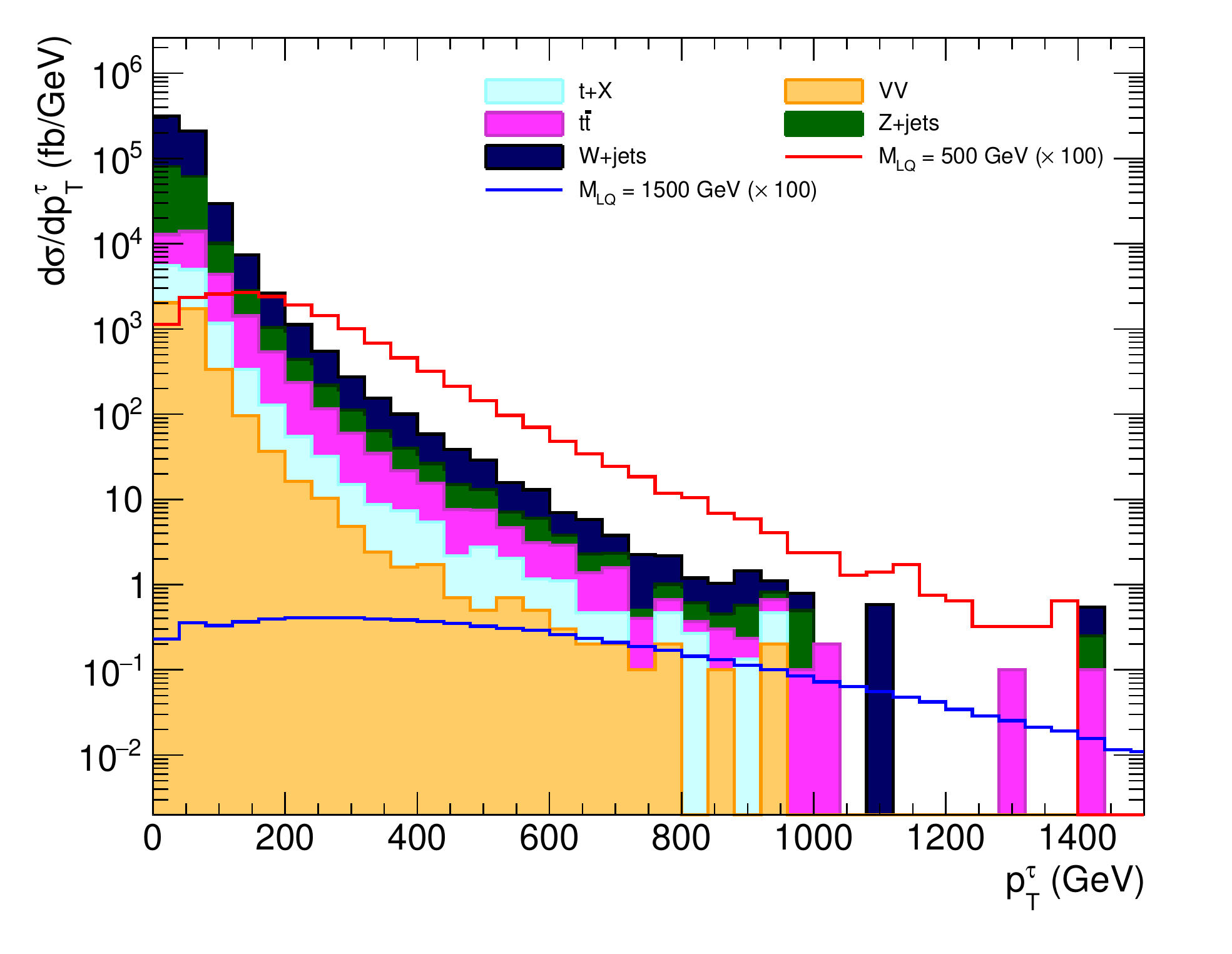}
\vfill
\includegraphics[width=0.495\textwidth]{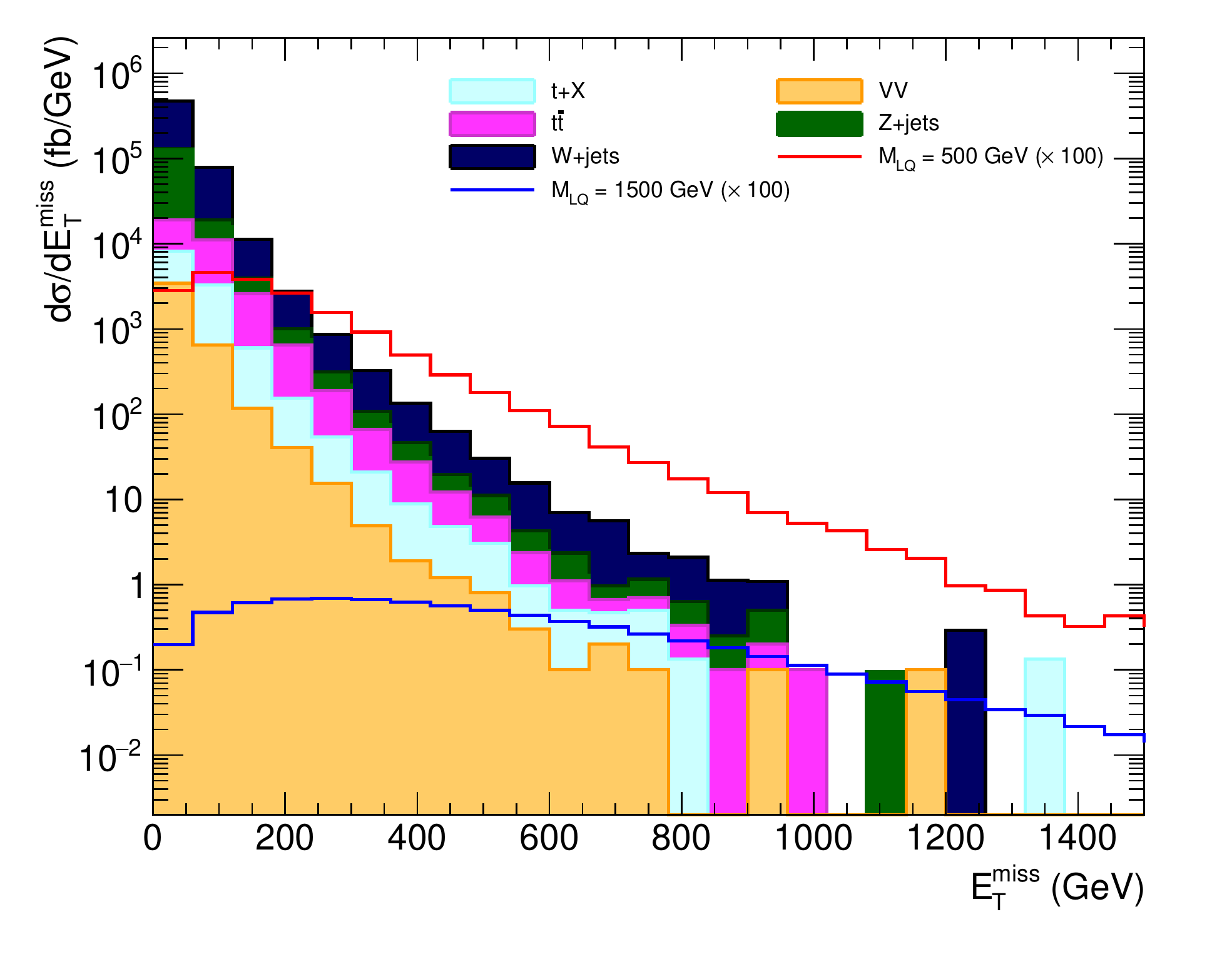}
\hfill
\includegraphics[width=0.495\textwidth]{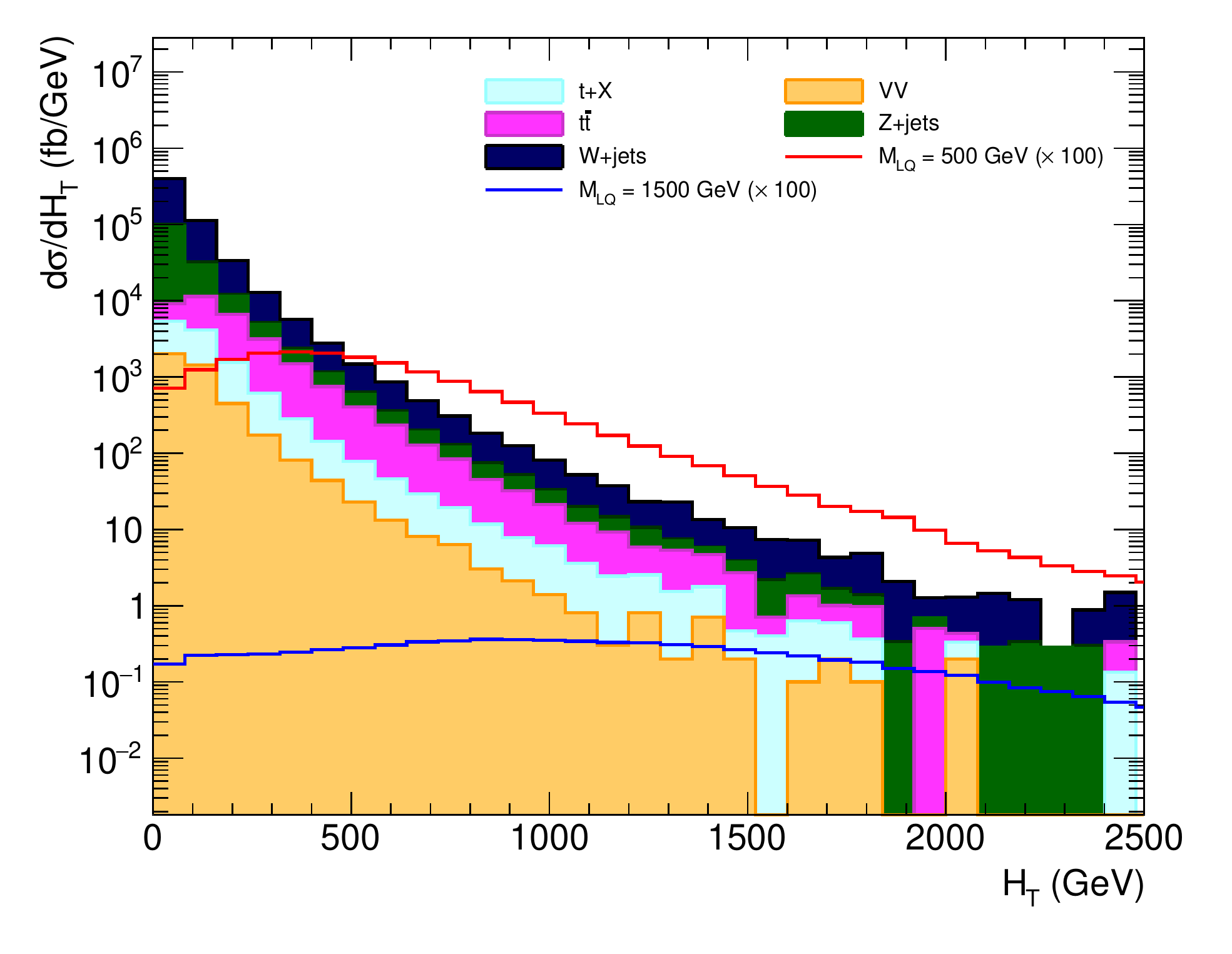}
\vfill
\includegraphics[width=0.495\textwidth]{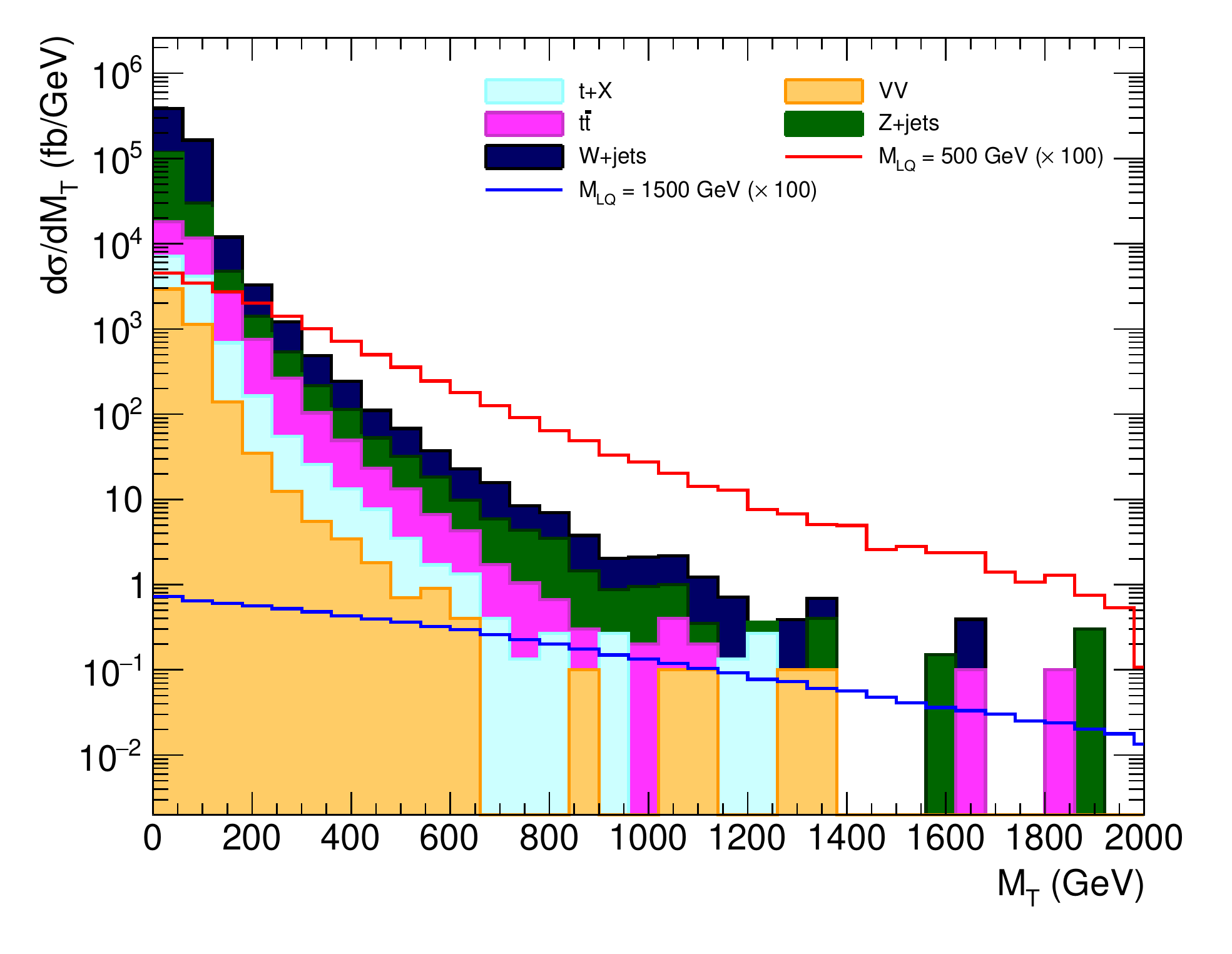}
\hfill
\includegraphics[width=0.495\textwidth]{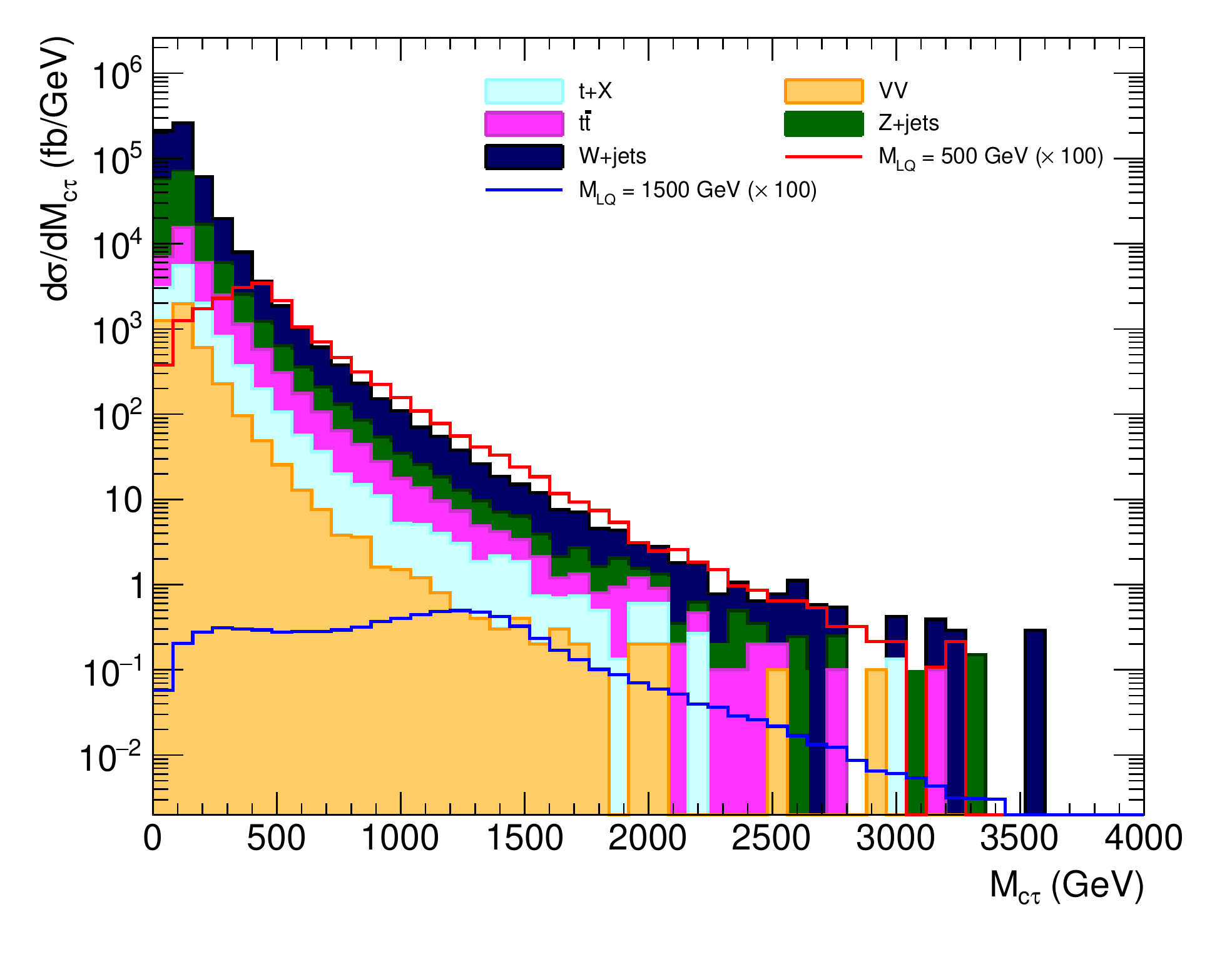}
\caption{Differential cross sections for the leading jet $p_T$ (top left), the leading tau $p_T$ (top right), the missing transverse energy (middle left), the scalar sum $H_T$ of all jet transverse momenta (middle right), the transverse mass $M_T$ of the leading tau lepton and the missing momentum (bottom left), and the invariant mass of the leading tau/leading jet system (bottom right). The distributions are shown for the di-boson ($VV$, orange), single top (cyan), $t\bar{t}$ (magenta), $Z+$jets (dark green), and $W+$jets (navy) backgrounds, and for signal scenarios in which $M_{S_1} = 500$~GeV (red) and $1500$~GeV (blue), $M_{\chi_0} = 100$~GeV and $\Delta_X = 0.5$.}
\label{fig:distributions}
\end{figure}

Several differential distributions for low-level and high-level observables are shown in figure~\ref{fig:distributions} to guide our analysis strategy. Events are selected if they contain at least one hadronically decaying tau lepton with $p_T > 20$~GeV and $|\eta| < 2.4$, in addition to at least one jet with $p_T > 25$~GeV and $|\eta|<2.4$. Furthermore, the leading hadronic tau is required to have a transverse momentum larger than $60~$GeV and to be isolated by $\Delta R > 0.5$ from the leading jet candidate. On the other hand, the leading jet is required to have a $p_T > 100~$GeV, and events are vetoed if they contain a charged lepton (electron or muon) with $p_T > 15~$GeV and $|\eta^\ell| < 2.5$. To reduce the contamination from the $t\bar{t}$ and single top backgrounds, a veto on events containing at least one $b$-tagged jet with $p_T > 30~$GeV and $|\eta| < 2.4$ is enforced, and we moreover veto events containing a second hadronic tau if the invariant mass $m_{\tau_h \tau_h}$ of the reconstructed tau pair is compatible with a $Z$-boson decay, \emph{i.e.}~if $|m_{\tau_h \tau_h} - M_Z| < 10$~GeV. We then require that the missing transverse energy is larger than 100~GeV, and that the missing momentum is well separated in azimuth from both the leading jet and the leading tau, $|\phi_{\rm miss} - \phi_{\rm jet, tau}| > 0.5$.

The efficiency for the signal process after the above selection lies in the $6.5\%$--$16.8\%$ range for leptoquark masses in the $500$--$2000~$GeV mass window. On the other hand, for a specific leptoquark mass, the signal efficiency slightly decreases when the mass of the $\chi_0$ state increases. For instance, for $\mLQ = 500~$GeV, the efficiency decreases from $9.7\%$ for $M_{\chi_0} = 100~$GeV to $6.5\%$ for $M_{\chi_0} = 240~$GeV. For the various backgrounds, we provide a detailed cutflow in table~\ref{tab:efficiency-LQtauc}.

\begin{figure}[!h]
\centering
\includegraphics[width=0.42\linewidth]{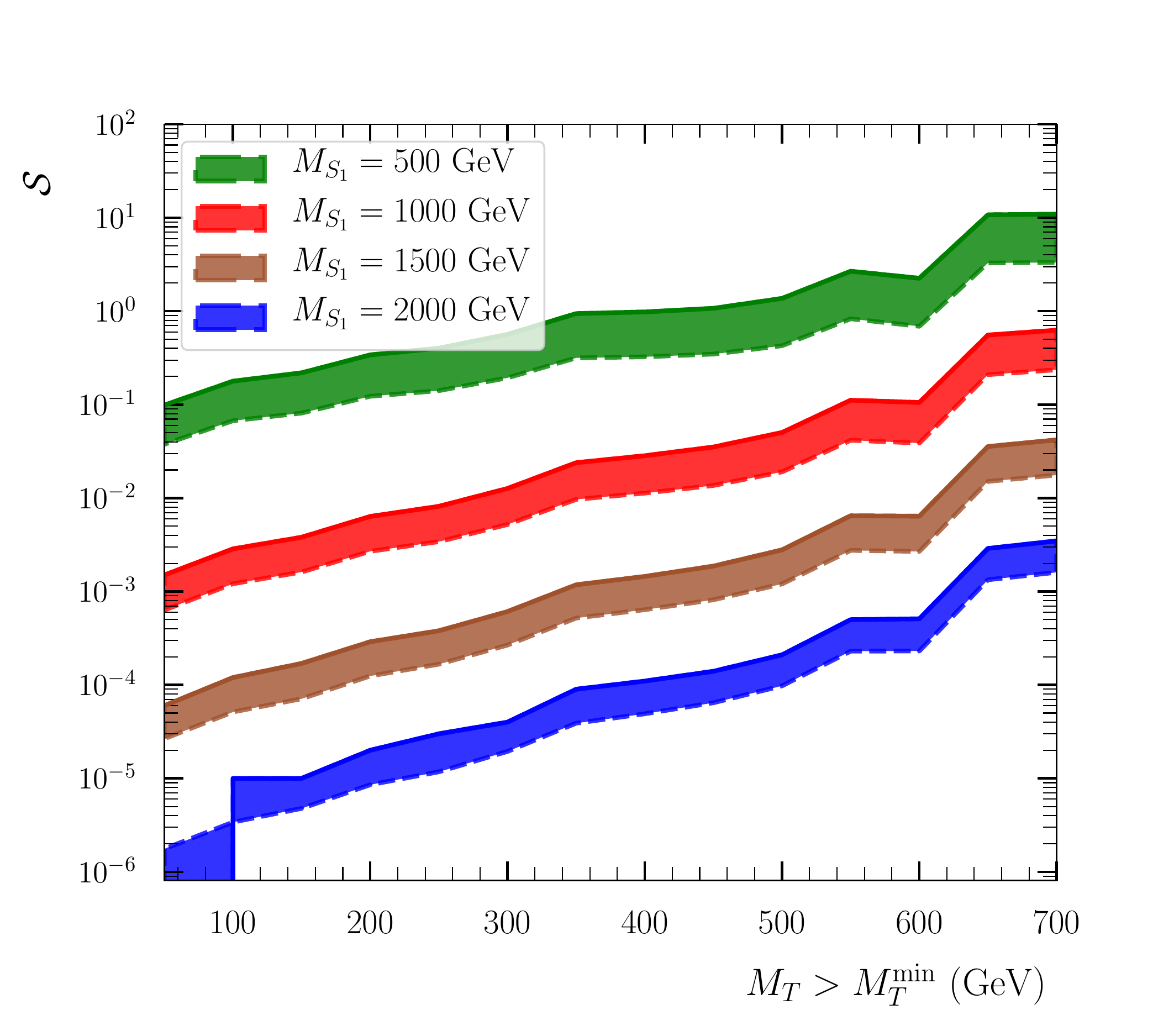}\hfill
\includegraphics[width=0.42\linewidth]{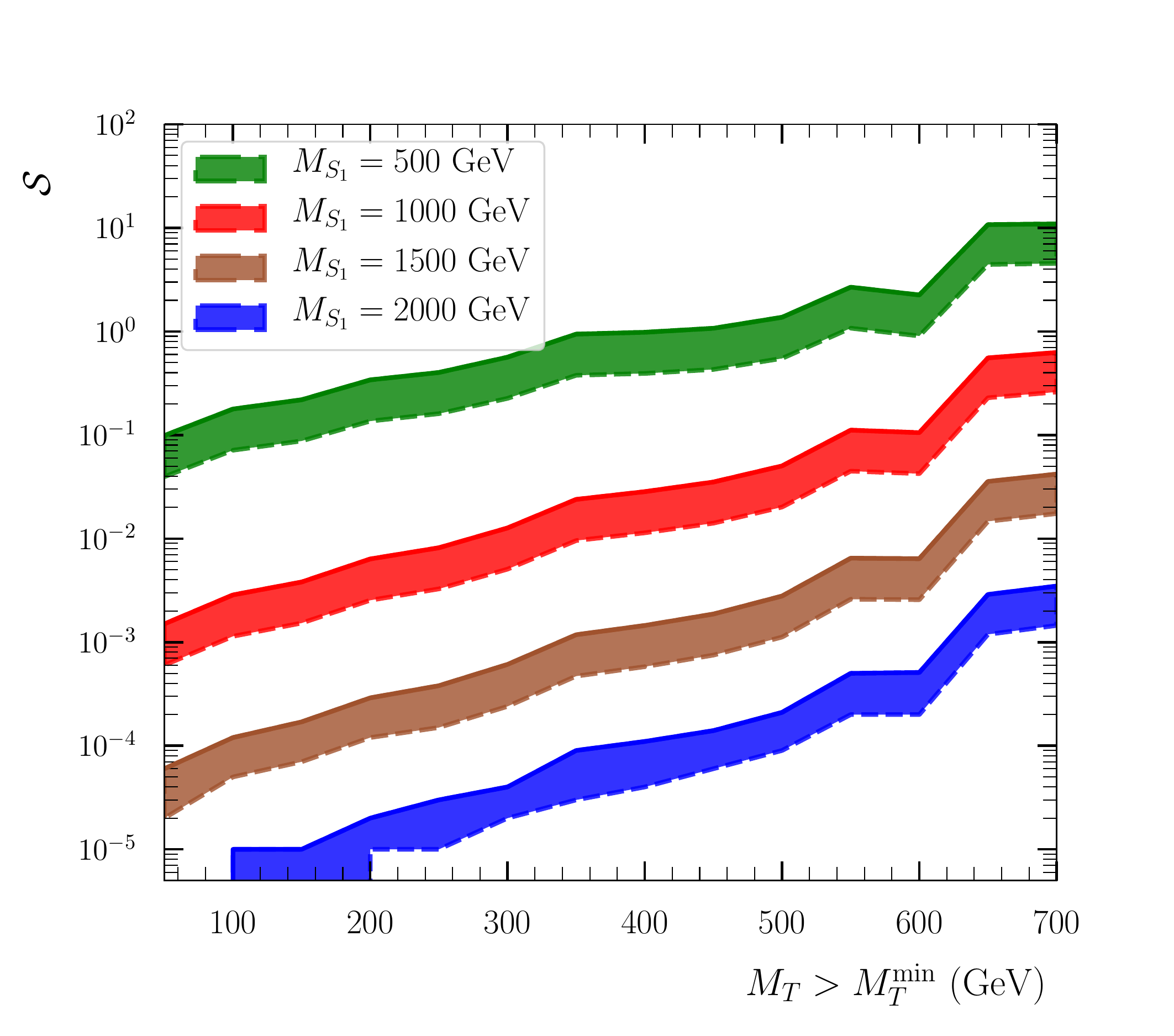}
\caption{Signal significance as a function of the cut on $M_T$ for $\Delta_X = 0.5$ (left) and $0.1$ (right), for  $\mLQ = 500~$GeV (green), $1~$TeV (red), $1.5~$TeV (brown), and $2~$TeV (blue), and for $M_{\chi_0} = 100~$GeV. We present results for a $2\%$ (solid) and $5\%$ (dashed) systematical uncertainty on the background. \label{fig:Z-MT-bins}} \vspace{0.3cm}
\includegraphics[width=0.425\linewidth]{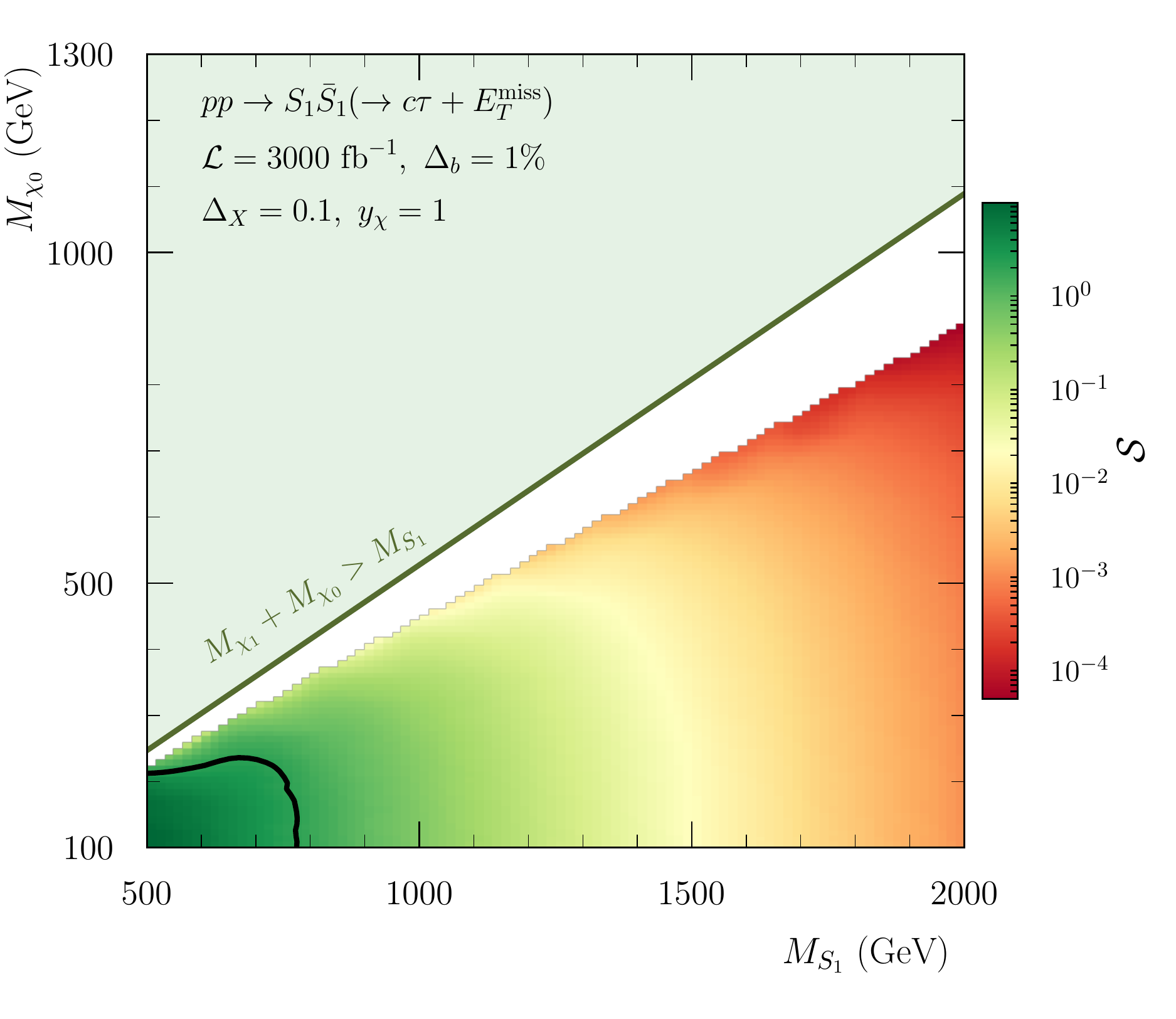}\hfill
\includegraphics[width=0.425\linewidth]{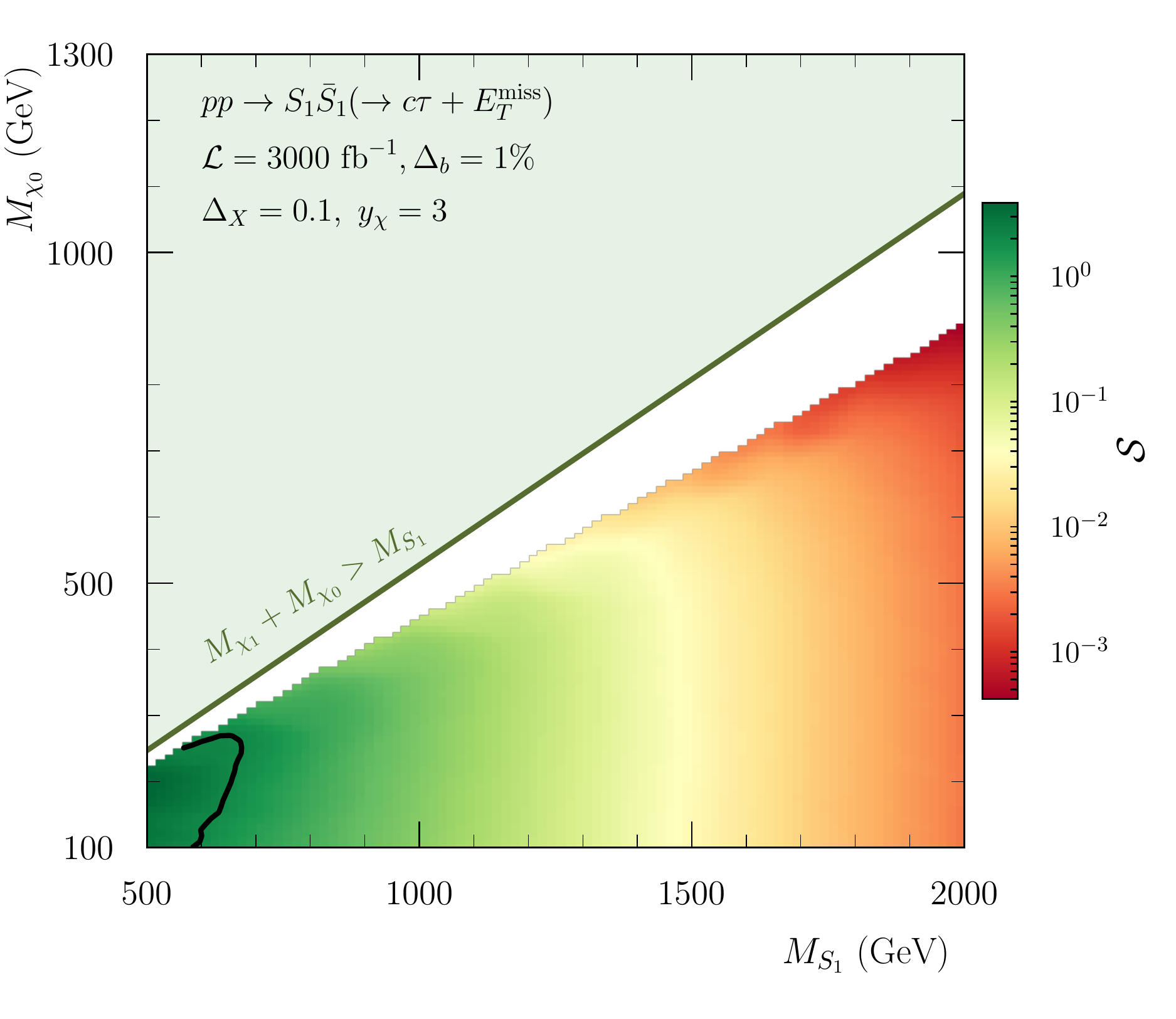}
\includegraphics[width=0.425\linewidth]{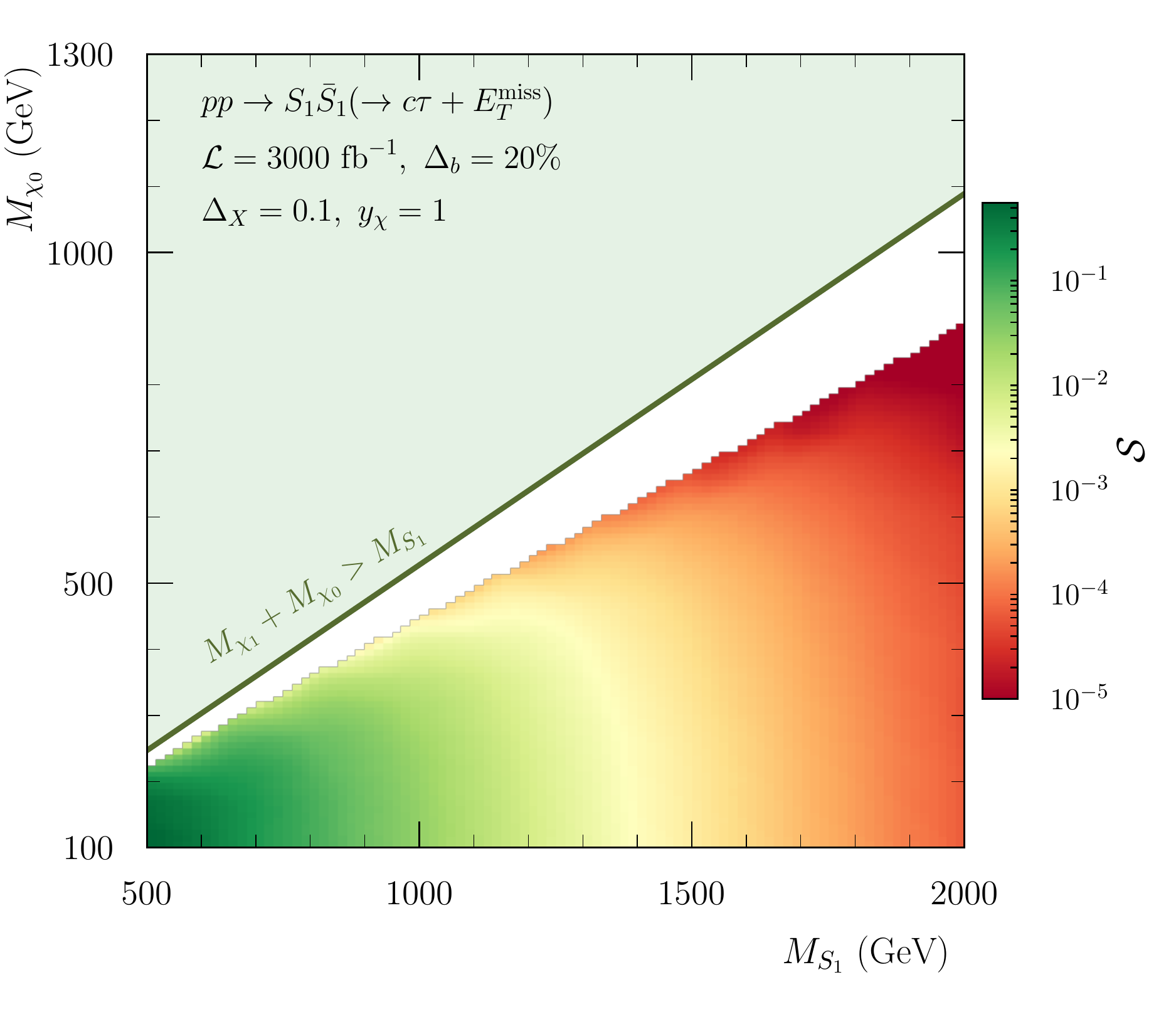}\hfill
\includegraphics[width=0.425\linewidth]{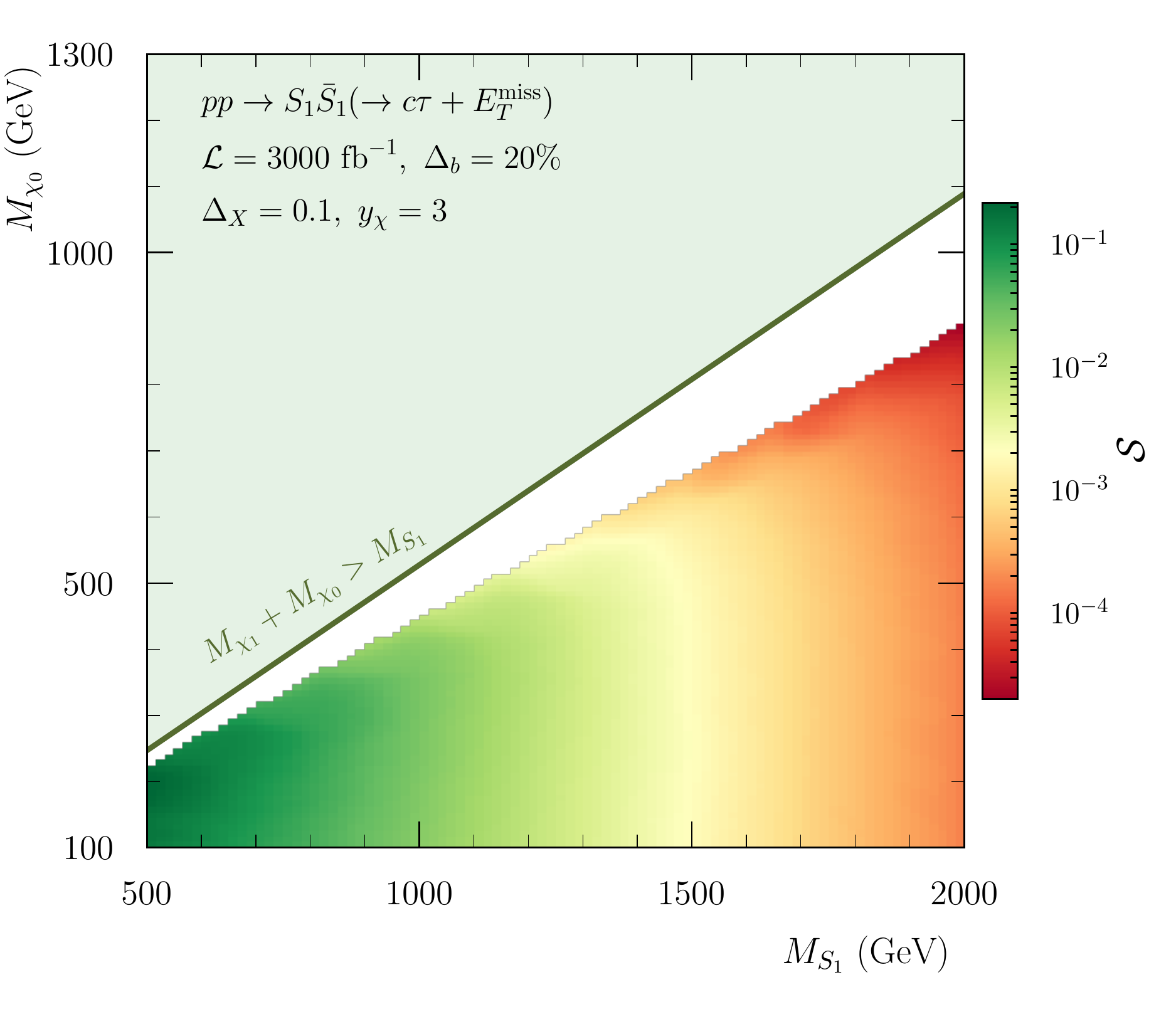}
\caption{Exclusion contours in the $(M_{S_1}, M_{\chi_0})$ mass plane for $y_\chi=1$ (left) and $3$ (right) in the proposed search for leptoquark + dark matter in $3000~{\rm fb}^{-1}$ of LHC collisions at $13$~TeV. We assume a $1\%$ (top) and $20\%$ (bottom) uncertainty on the background. The solid black lines correspond to a 95\% confidence exclusion contour, and the dark green line defines the kinematical boundary above which the dark $S_1$ decay is forbidden. All results are shown for $\Delta_X = 0.1$. \label{fig:Z-LQMET-DX01}}
\end{figure}

\begin{figure}
\centering
\includegraphics[width=0.425\linewidth]{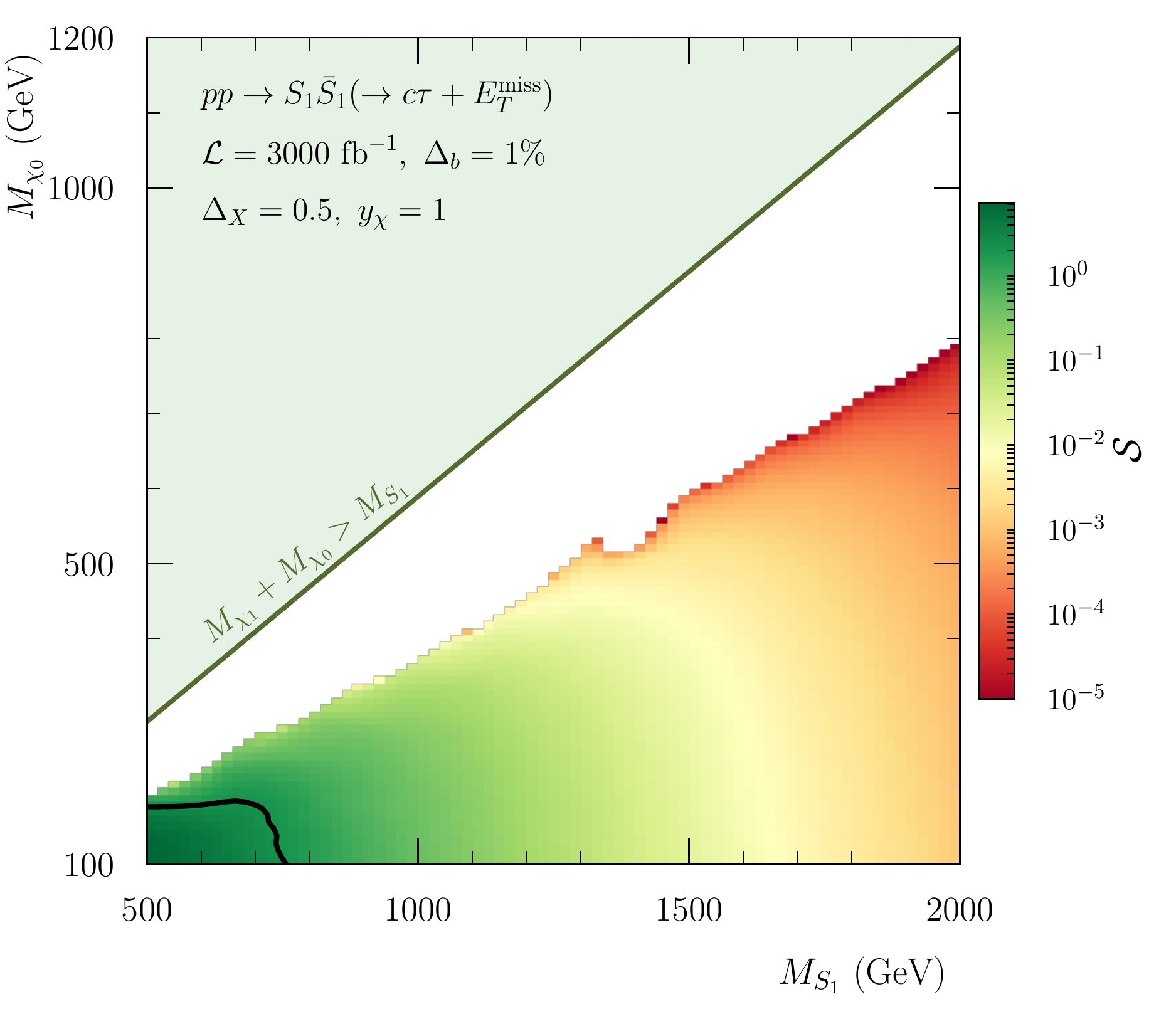}\hfill
\includegraphics[width=0.425\linewidth]{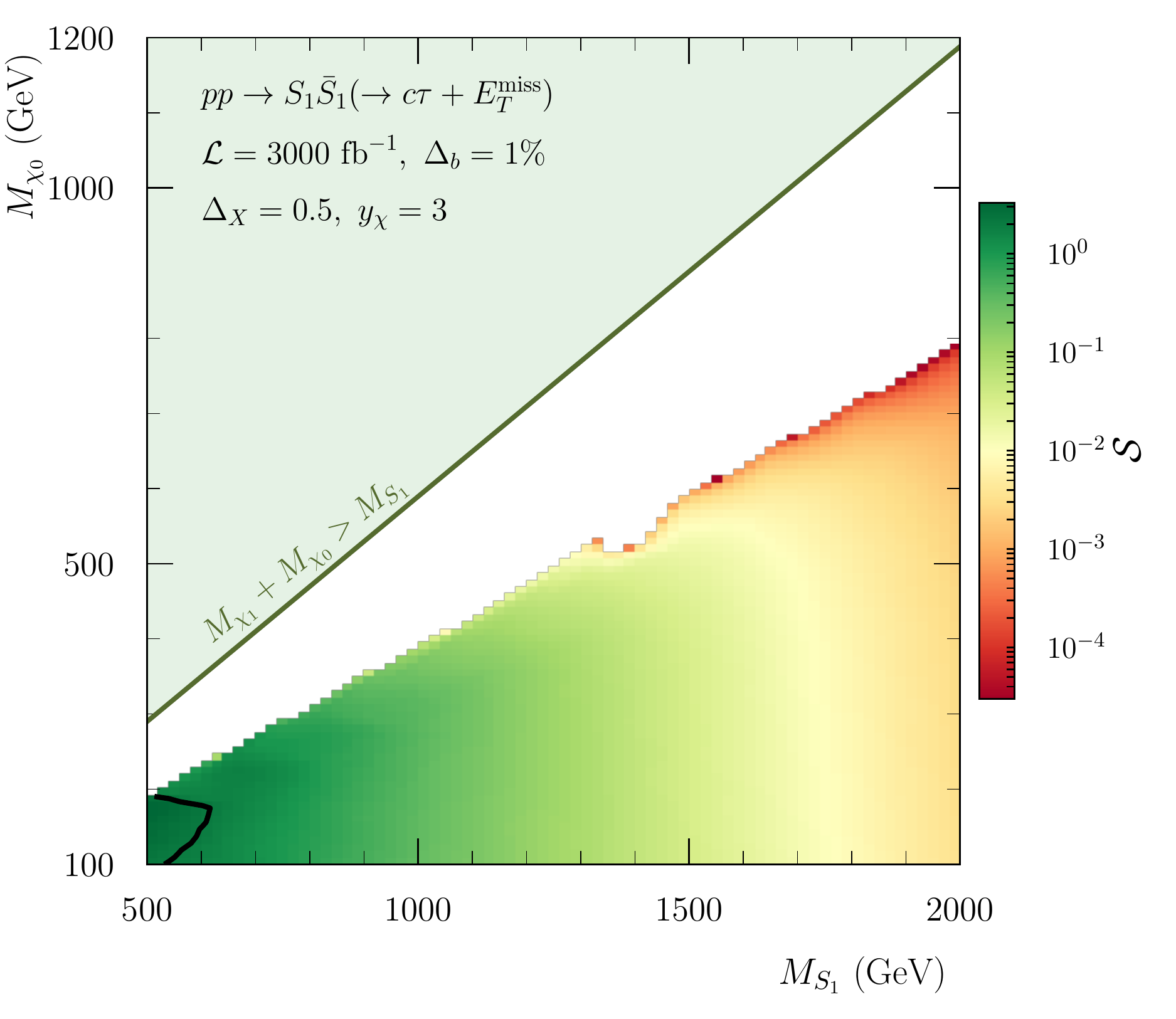}
\includegraphics[width=0.425\linewidth]{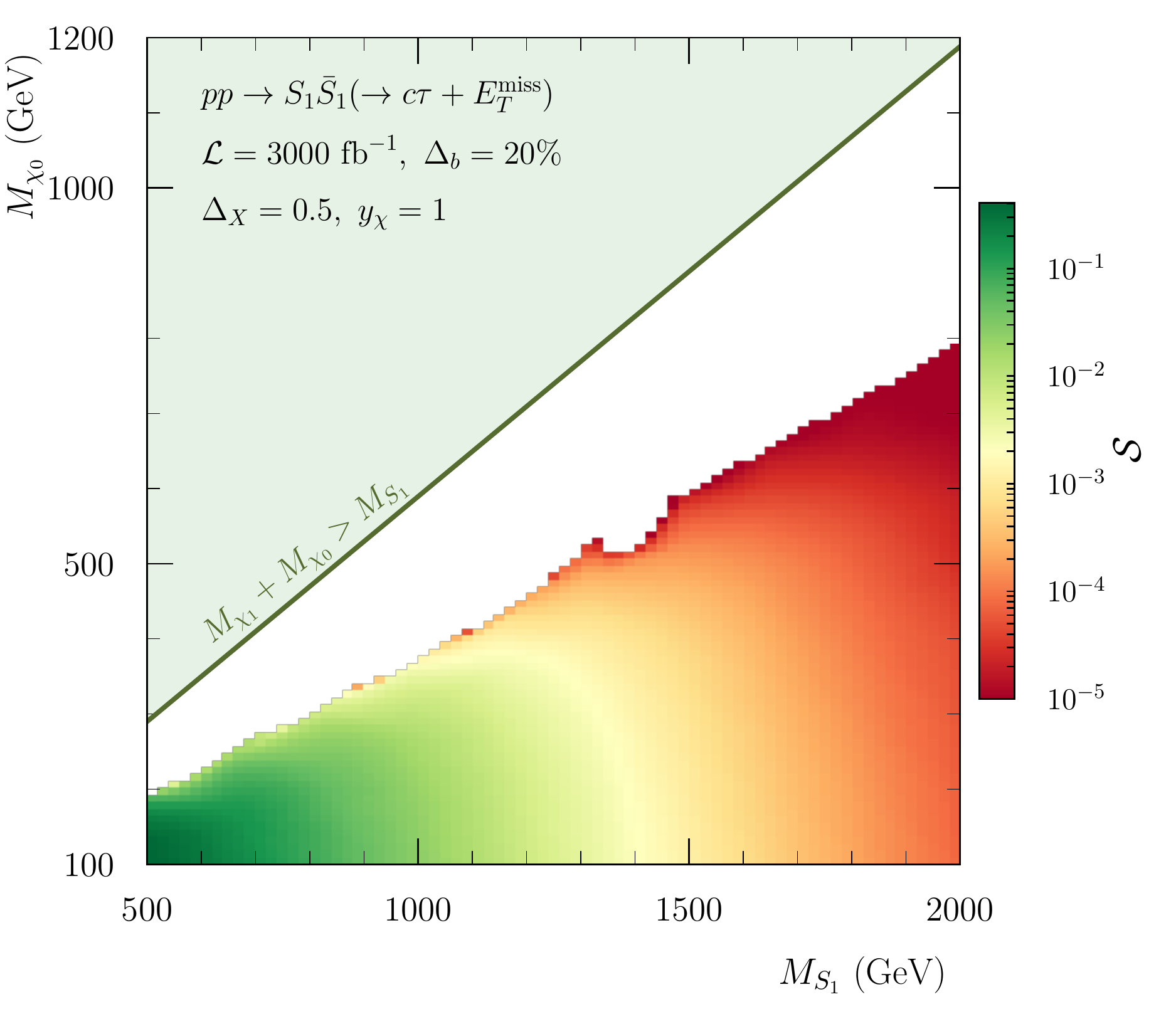}\hfill
\includegraphics[width=0.425\linewidth]{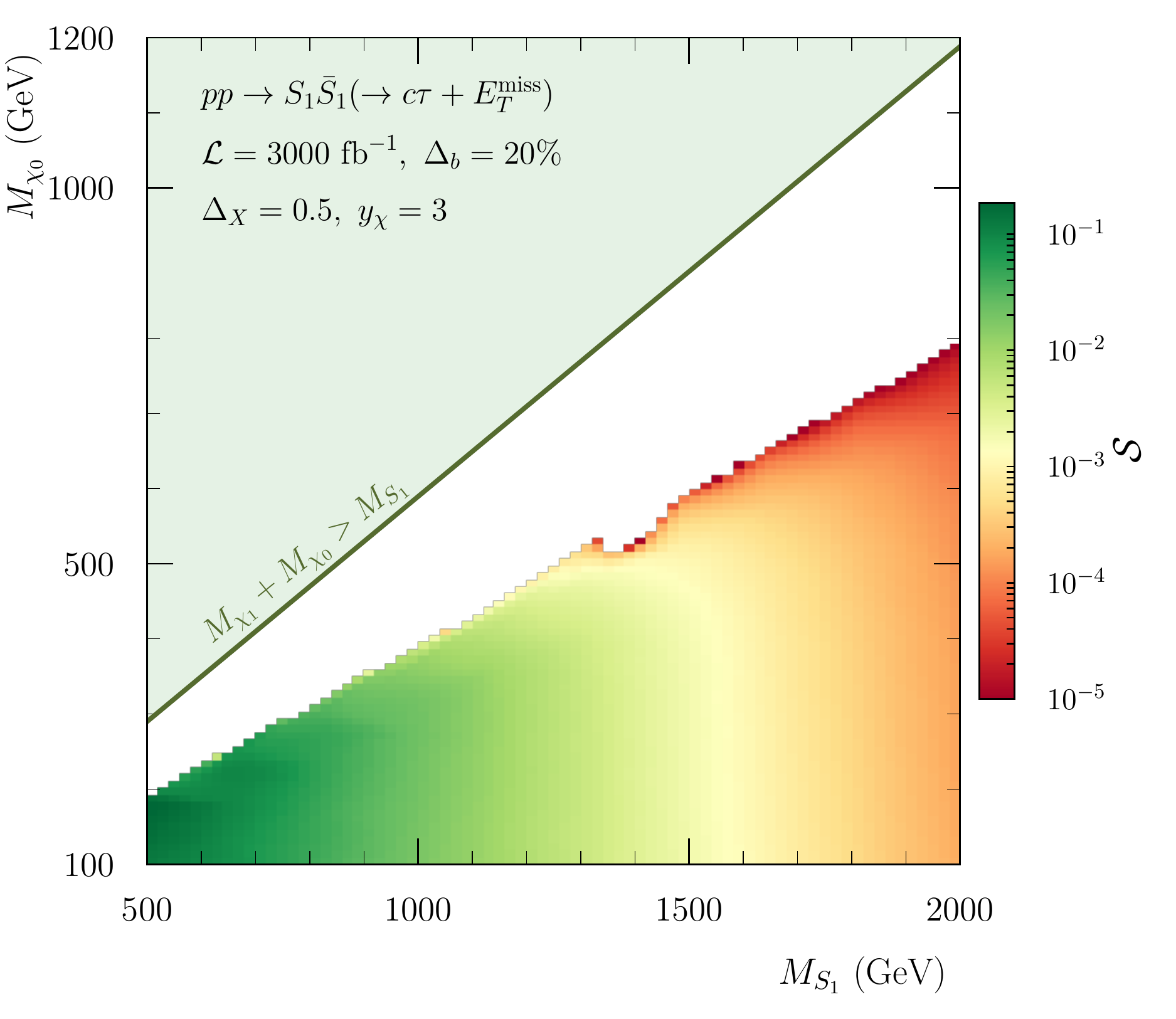}
\caption{Same as figure~\ref{fig:Z-LQMET-DX01} but for $\Delta_X = 0.5$.}
\label{fig:Z-LQMET-DX05}
\end{figure}

We finally define a signal region by using, as in the CMS analysis in the muon channel, the transverse mass $M_T$ of the system comprising the leading tau and the missing momentum. The obtained signal significance as a function of the cut on this variable is displayed in figure~\ref{fig:Z-MT-bins} for various leptoquark masses and dark mass splittings $\Delta_X$. We observe that the cut $M_T > 700~$GeV maximises the signal significance for all mass values, so that we make use of such a cut as a final analysis selection. In the results presented in this section, we estimate the analysis significance by relying on an approximation valid for $N_b \gg N_s$ ({\it i.e.} as in our case)~\cite{Cowan:2010js},
\begin{eqnarray}
\mathcal{S} = \frac{N_s}{\sqrt{N_b + \delta_b^2}},
\label{eq:Z}
\end{eqnarray}
where $N_s$, $N_b$ are the number of signal and background events populating the signal region. Moreover, $\delta_b = x N_b$ allows us to set the uncertainty on the background estimate by fixing $x$. The errors on the background yields may stem from theory (missing higher-order corrections) or from more experimental sources like the jet energy resolution and the jet energy scale. The estimate of these uncertainties is beyond the scope of this work that only consists of preliminary steps to complete the dark matter and LHC search program at the LHC and HL-LHC. We therefore assume two extreme cases in which $x = 1\%$ and $20\%$. The corresponding significance is shown in figure~\ref{fig:Z-LQMET-DX01} for $\Delta_X = 0.1$ and in figure~\ref{fig:Z-LQMET-DX05} for $\Delta_X = 0.5$.

\begin{landscape}
\begin{table}
\setlength\tabcolsep{8pt}
\begin{center}
\resizebox{1.45\textwidth}{!}{\begin{tabular}{ccccccc} 
                   &         $W$+ jets   &       $Z(\to \nu\bar{\nu})+ jets$  & $Z/\gamma^*(\to \tau\tau)$ + jets & $t\bar{t}$ &     Di-boson & $t+X$ \\ 
\toprule
Initial events   & $125.4 \times 10^9$ &   $21.7 \times 10^9$ & $6.9 \times 10^9$ &      $626.7 \times 10^6$   &  $307.5 \times 10^6$  &   $230.1\times 10^6$ \\
&   $(100\pm0)\%$    &    $(100\pm0)\%$ & $(100\pm0)\%$ & $(100\pm0)\%$ & $(100\pm0)\%$ & $(100\pm0)\%$ \\ 
\toprule
 $\tau_h$ with $p_T^\tau > 60$ GeV  &     $192.3 \times 10^6$ &  $9.1 \times 10^6$ &           $47.7 \times 10^6$ &    $33.7 \times 10^6$ &  $3.1 \times 10^6$ &  $5.3 \times 10^6$ \\ 
& $(0.15  \pm 0.00033)\%$   & $(0.042\pm0.00029)\%$  & $(0.68\pm0.00166)\%$ &   $(5.38 \pm 0.01808)\%$  &  $(0.99\pm0.00983)\%$ & $(2.28 \pm 0.017135)\%$ \\ \toprule
 Jet with $p_T > 100$ GeV  &  $58.4 \times 10^6$ & $2.6 \times 10^6$ & $13.1 \times 10^6$ & $13.9 \times 10^6$ &  $1.0 \times 10^6$ &         $1.9 \times 10^6$ \\                                        & $(0.04 \pm 0.00017)\%$ & $(0.0118 \pm 0.00016)\%$  &  $(0.19 \pm 0.00087)\%$ &  $(2.21 \pm 0.011781)\%$ & $(0.34 \pm 0.00572)\%$ & $(0.85 \pm 0.01056)\%$ \\ \toprule
Lepton veto     &    $54.1 \times 10^6$ &  $2.6 \times 10^6$  &  $11.0 \times 10^6$ &         $10.9 \times 10^6$ &  $948.0 \times 10^3$ &  $1.4 \times 10^6$ \\
  & $(0.04 \pm 0.00017)\%$ & $(0.01188 \pm 0.000156)\%$  &  $(0.16 \pm 0.00079)\%$  &  $(1.74 \pm 0.010472)\%$ & $(0.31 \pm 0.00548)\%$ & $(0.62 \pm 0.00903)\%$ \\ \toprule 
$b$-jet veto   &      $50.9 \times 10^6$ &   $2.3 \times 10^6$  &   $10.4 \times 10^6$ &          $2.2 \times 10^6$    &   $808.7 \times 10^3$   &  $485.4 \times 10^3$ \\
& $(0.040 \pm 0.00016)\%$ & $(0.01060 \pm 0.000148)\%$   &  $(0.15 \pm 0.00077)\%$  &  $(0.35 \pm 0.004771)\%$ & $(0.26 \pm 0.00507)\%$ & $(0.21 \pm 0.00526)\%$ \\ \toprule
$Z$-mass window    &  $50.8 \times 10^6$ &  $2.3 \times 10^3$ &  $9.8 \times 10^6$     & $2.2 \times 10^2$   &  $802.1 \times 10^3$    &   $479.1 \times 10^3$ \\ 
& $(0.0405 \pm 0.00017)\%$ & $(0.01059\pm 0.000148)\%$  & $(0.14 \pm 0.00075)\%$ &  $(0.35 \pm 0.004744)\%$ & $(0.26 \pm 0.00505)\%$ & $(0.208\pm0.00523)\%$ \\ \toprule
$E_T^{\rm miss} > 100$ GeV   &   $10.9 \times 10^6$    &   $991.8 \times 10^3$ &            $939.1 \times 10^3$ &  $750.2 \times 10^3$ &  $166.9 \times 10^3$   &  $150\times 10^3$ \\
& $(0.00875 \pm 0.00008)\%$ & $(0.00457 \pm 0.00009)\%$  &  $(0.013 \pm 0.00023)\%$ &  $(0.12 \pm 0.002769)\%$ &  $(0.054\pm0.00230)\%$ & $(0.065 \pm 0.00292)\%$ \\ \toprule
$|\phi_{\rm miss} - \phi_{\tau_h}| > 0.5$ & $3.8 \times 10^6$ &  $892.3 \times 10^3$ & $181.9 \times 10^3$    &  $382.7 \times 10^3$    &  $56.9 \times 10^3$   &  $83.0 \times 10^3$ \\
&  $(0.00304 \pm 0.00004)\%$ & $(0.00411 \pm 0.00009)\%$ & $(0.0026 \pm0.00010)\%$  & $(0.06 \pm 0.001978)\%$ & $(0.018 \pm 0.00134)\%$ & $(0.036 \pm 0.00217)\%$ \\ \toprule
$|\phi_{\rm miss} - \phi_{\rm jet}| > 0.5$ & $3.6 \times 10^6$  &  $842.2 \times 10^3$      &  $164.6 \times 10^3$    &    $367.1 \times 10^3$    &   $54.2 \times 10^3$  &    $80.3 \times 10^3$ \\ 
& $(0.00290 \pm 0.00004)\%$ & $(0.00388 \pm 0.00009)\%$ & $(0.0023 \pm 0.00009)\%$ & $(0.058 \pm 0.001937)\%$ & $(0.017 \pm 0.00131)\%$ & $(0.034 \pm 0.00214)\%$
\end{tabular}}
\end{center}
\caption{Detailed cutflow including each selection stage of our analysis targeting the $c\tau$ plus MET final state. The numbers in the second rows correspond to the efficiencies defined by $\epsilon = n_i / n_0$, with $n_i$ being the number of events surviving the $i^{\rm th}$ cut. The uncertainty on the efficiencies are estimated by assuming that the yields $N_i$ follow a binomial distribution~\cite{Paterno:2004cb}.}
\label{tab:efficiency-LQtauc}
\end{table}
\end{landscape}

 It is found that our analysis can probe leptoquark masses ranging up to $800$ GeV for $\Delta_X = 0.1$ and $y_\chi = 1$, and if we assume a $1\%$ uncertainty on the background yields. For larger $\Delta_X = 0.5$ values, the sensitivity gets milder. For more significant  uncertainties on the background ($x=20\%$), the sensitivity decreases significantly and our analysis becomes insensitive to the scenarios considered. A straightforward modification of the (existing) $c \mu$ analysis does therefore not perform well for the $\tau c$ final state and improvements are in order. A more refined selection relying on other kinematics variables could hence be beneficial, or on the usage of supervised machine learning techniques. We leave this for a future study.

\section{Dark matter direct detection at the one-loop level}
\label{app:dd_calculation}

The computation of the DM direct detection rate in our model follows closely the method outlined in~\cite{Mohan:2019zrk}. For the sake of brevity, we only present our main results and refer the reader to the original reference for further details. The spin-independent dark matter-nucleon scattering cross section is given by
\begin{equation}\label{eq:sigmaSIgeneral}
\sigma_{\rm SI}^N = \frac{4}{\pi} \left( \frac{\mxZero m_N}{\mxZero + m_N} \right)^2 \left| f_N \right|^2,
\end{equation}
where $m_N$ (with $N = \lbrace n,p \rbrace$) is the nucleon mass and $f_N$ are form factors which need to be computed. In our model, the leading contributions to these form factors stem from $\chi_0$ scattering off gluons. These can be described, in the low-energy limit, through the effective Lagrangian
\be
  {\cal L}^{\rm eff}_{g} = 
  f_G \bar{\chi}_0\chi_0 ~{\cal O}_{g}^{(0)}
  +i \frac{g^{(1)}_G} {m_{\chi}}\ \bar{\chi}_0 \big(\partial^{\mu}\gamma^{\nu} + \partial^{\nu}\gamma^{\mu}\big) \chi_0 \  {\cal O}_{g,\mu\nu}^{(2)}
  + \frac{g^{(2)}_G}{m_{\chi}^2}\ \bar{\chi}_0(i\partial^{\mu}) (i\partial^{\nu})\chi_0 \ {\cal O}_{g,\mu\nu}^{(2)} \ .
\label{eq:DDEFT}\ee
In this expression,
\begin{eqnarray}
  {\cal O}_g^{(0)} = G^A_{\mu \nu} G^{A \mu \nu} \qquad\text{and}\qquad
  {\cal O}^{(2)\mu\nu}_g = -G^{A \mu\lambda} G^{A \nu}_{\phantom{A \nu} \lambda} + \frac14 g^{\mu\nu} (G^A_{\alpha\beta})^2\ ,
\end{eqnarray}
where $G_{\mu\nu}^A$ is the gluon field strength tensor. The associated Wilson coefficients $f_G$, $g_G^{(1)}$ and $g_G^{(2)}$  are, in turn, related to the form factors $f_N$ through
\begin{equation}\label{eq:hadronicME}
\frac{f_N}{m_N} =
-\frac{8\pi}{9\alpha_s}f_{T_G} f_G 
+\frac{3}{4} G(2)\left(g^{(1)}_G
+g^{(2)}_G\right) \ ,
\end{equation}
where $f_{T_G}$ and $G(2)$ are  hadronic matrix elements~\cite{Mohan:2019zrk, Hill:2014yxa}. In the above analytical results, we recall that we have ignored any quark contributions as these are subdominant for our model and choice of parameters.

In order to compute the Wilson coefficients $f_G$, $g_G^{(1)}$ and $g_G^{(2)}$, we have used \textsc{FeynArts}~\cite{Hahn:2000kx}, \textsc{FormCalc}~\cite{Hahn:2016ebn} and \textsc{PackageX}~\cite{Patel:2015tea} to calculate the $\chi_0 g$ scattering matrix element at one-loop (leading order). We have then matched the results to the effective interactions~\eqref{eq:DDEFT}. In practice, this is achieved by applying appropriately chosen projectors on the full amplitude, as detailed in~\cite{Mohan:2019zrk}. The resulting expressions read
\begin{align}
\label{eq:DDfG}
\nonumber f_G=&\,\frac{ \alpha _s \,\,\ydm^2}{96\,\pi \, \lambda^4 \mxOne \mLQ^2
   \left((\mxZero-\mxOne)^2-\mLQ^2\right)}
 \bigg[\lambda^2 \bigg(-\mxZero^4 
   \left(\mxOne^2+2 \mLQ^2\right)\\
\nonumber &+
  \mxZero^3 \left(\mxOne \mLQ^2-2\mxOne^3\right)+2\mxZero^2 \left(\mxOne^4+2 \mLQ^4\right)+ \mxZero^5 \mxOne+
   6 \mxOne^4 \mLQ^2\\
\nonumber&-
  { \mxZero} \left(5 \mxOne^3\mLQ^2-\mxOne^5 + 2 \mxOne \mLQ^4\right) -3\mxOne^2 \mLQ^4-2 \mLQ^6-\mxOne^6\bigg) \\
  &-12\mxOne^2 \mLQ^4M_{\chi_0}^2\Lambda\left(\mxZero^2-\mxZero \mxOne+\mxOne^2-\mLQ^2\right) \bigg]\ ,
\end{align}

\begin{align}
g_{G}^{(1)} =& -\frac{\alpha_s \ydm^2}{48 \pi  M_{\chi_0}^2 \lambda ^3} \Bigg[\Bigg(\left(M_{\chi_0}^2-M_{\chi_1}^2+M_{S_1}^2\right) \Lambda 
\nonumber \\
& \times \left(3 M_{\chi_0}^4-2 M_{\chi_0}^2 \left(M_{\chi_1}^2+2 M_{S_1}^2\right)+\left(M_{\chi_1}^2-M_{S_1}^2\right)^2\right) 
\nonumber\\\label{eq:DDG1}
&+\left(3 M_{\chi_0}^2-M_{\chi_1}^2+M_{S_1}^2\right) \lambda \Bigg)
+M_{\chi_0}^2 \lambda^2
+\lambda^3\log (\frac{M_{\chi_1}}{M_{S_1}}) \Bigg] \ ,\qquad
\end{align}

\begin{align}
g_{G}^{(2)}=& \frac{\alpha_s \ydm^2}{24 \pi  M_{\chi_0}^3 \lambda ^3}\Bigg[
2 M_{\chi_0}^2 \left((M_{\chi_0}+M_{\chi_1})^2-M_{S_1}^2\right) \Lambda \Bigg(M_{\chi_0}^8-4 M_{\chi_0}^6 M_{S_1}^2 
\nonumber \\
&+4 M_{\chi_0}^5 M_{\chi_1} \left(M_{S_1}^2-M_{\chi_1}^2\right)+6 M_{\chi_0}^4 M_{S_1}^4+6 M_{\chi_0}^3 \left(M_{\chi_1}^5-M_{\chi_1} M_{S_1}^4\right)
\nonumber \\
&
-2 M_{\chi_0}^2 \left(M_{\chi_1}^6-3 M_{\chi_1}^2 M_{S_1}^4+2 M_{S_1}^6\right)-2 M_{\chi_0} M_{\chi_1} \left(M_{\chi_1}^2-M_{S_1}^2\right)^3
\nonumber \\
&+\left(M_{\chi_1}^2-M_{S_1}^2\right)^4\Bigg)+M_{\chi_0}^2 (M_{\chi_0}+M_{\chi_1}-M_{S_1}) (M_{\chi_0}+M_{\chi_1}+M_{S_1}) \lambda 
\nonumber \\ 
&\times\Bigg(3 M_{\chi_0}^4-M_{\chi_0}^2 \left(M_{\chi_1}^2+5 M_{S_1}^2\right)+4 M_{\chi_0} M_{\chi_1} \left(M_{S_1}^2-M_{\chi_1}^2\right)
\nonumber \\ 
\label{eq:DDG2}&+2 \left(M_{\chi_1}^2-M_{S_1}^2\right)^2\Bigg)-2 \log \left(\frac{M_{\chi_1}}{M_{S_1}}\right) \lambda ^3
\Bigg]
\ ,
\end{align}
where we have made use of the definitions 
\begin{eqnarray}
\Lambda
&= &
\frac{\lambda}{M_{\chi_0}^2} \log \left(\frac{\lambda
   -\mxZero^2+\mxOne^2+\mLQ^2}{2 \mxOne \mLQ}\right)\nonumber, \\
\lambda&=&\sqrt{M_{\chi_0}^4-2 M_{\chi_0}^2 (M_{\chi_1}^2+M_{S_1}^2)+(M_{\chi_1}^2-M_{S_1}^2)^2}\ .
\end{eqnarray}	
By combining eqs.~\eqref{eq:sigmaSIgeneral}, \eqref{eq:hadronicME}, ~\eqref{eq:DDfG}, \eqref{eq:DDG1} and ~\eqref{eq:DDG2} we can finally compute predictions for the spin-independent DM scattering cross section in our model.

\section{Implementation of conversion-driven freeze-out in \micromegas}
\label{app:cdfo_micro}

To implement the CDFO mechanism in \micromegas, we rely on the Boltzmann equations describing the thermal evolution of the abundances of both $\chi_0$ and $\chi_1$, the two particles in the dark sector of our model. These abundances are defined by 
\begin{eqnarray}
  Y_{\chi_i} &=& \frac{n_{\chi_i}}{s(T)},
\end{eqnarray}
where $s(T)$ is the entropy density originating from the SM particle contributions, $n_{\chi_0}$ is the $\chi_0$ number density and $n_{\chi_1} $ is the total number density of $\chi_1$ and  $\bar{\chi}_1$.  The coupled set of Boltzmann equations then reads
\be\bsp
\frac{\diff Y_{\chi_i}}{ \diff T}=&\    
 \frac{1}{3H}\frac{\diff  s(T)}{\diff T}  \left[  
\langle v\sigma_{\chi_i\chi_i 00}  \rangle \Big( Y_{\chi_i}^2 -{Y_{\chi_i}^{\eq}}^2\Big)  
+\langle v\sigma_{\chi_i\chi_i  \chi_j\chi_j }  \rangle \Bigg( Y_{\chi_i}^2 -\bigg(Y_{\chi_j}\frac{
Y_{\chi_i}^{\eq}}{Y_{\chi_j}^{\eq}}\bigg)^2 \Bigg) \right. \\
&\ + \left. \langle v\sigma_{\chi_j\chi_i 0 0} \rangle ( Y_{\chi_i}Y_{\chi_j} -
Y_{\chi_i}^{\eq}Y_{\chi_j}^{\eq}) + \frac{\Gamma_{\chi_i\to\chi_j}}{s} \bigg( Y_{\chi_i} -
Y_{\chi_j}\frac{Y_{\chi_i}^{\eq}}{Y_{\chi_j}^{\eq}}\bigg)\right],
\esp\label{eq:Y}\ee
where $\chi_j=\chi_{1-i}$, $Y_k^{\eq}= n_k^{\eq}/s$, and $n_k^{\eq}$ is the equilibrium number density. In addition, $\langle v\sigma\rangle$ denote thermally-averaged cross sections that can be explicitly written  as   
\begin{eqnarray}
 \langle v\sigma_{\chi_0\chi_0 00} \rangle &=&  \frac{Tg_{\chi_0}^2}{8 \pi^4( n_{\chi_0}^{\eq})^2 }  \int \sqrt{s} p^2(s) K_1\!\left(\frac{\sqrt{s}}{T}\right)
 \sigma_{\chi_0,\chi_0\rightarrow S_1,\bar{S}_1}(s) \,\diff s,
\\
 \langle v\sigma_{\chi_1\chi_1 00} \rangle &=&  \frac{2Tg_{\chi_1}^2}{8 \pi^4 ( n_{\chi_1}^{\eq})^2 } \sum_{{\rm SM}} \int \sqrt{s} p^2(s) K_1\!\left(\frac{\sqrt{s}}{T}\right)
\sigma_{\chi_1,{\bar\chi_1}\rightarrow {\rm SM}}(s) \,\diff s, \label{eq:chi1ann}
\\
 \langle v\sigma_{\chi_0\chi_1 00} \rangle &=&  \frac{2Tg_{\chi_0}g_{\chi_1}} {8 \pi^4  n_{\chi_0}^{\eq}
n_{\chi_1}^{\eq}  } \sum_{{\rm SM}}   \int \sqrt{s} p^2(s) K_1\!\left(\frac{\sqrt{s}}{T}\right)
\sigma_{\chi_0,\chi_1\rightarrow {\rm SM}}(s) \,\diff s,
\\
 \langle v\sigma_{\chi_0\chi_0 \chi_1\chi_1} \rangle &=&  \frac{T g_{\chi_0}^2}{8 \pi^4 ( n_{\chi_0}^{\eq})^2 }  \int \sqrt{s} p^2 (s) K_1\!\left(\frac{\sqrt{s}}{T}\right)
 \sigma_{\chi_0,\chi_0\rightarrow \chi_1,{\bar\chi_1}}(s) ds ,
\\
\langle v\sigma_{\chi_1\chi_1 \chi_0\chi_0} \rangle &=& \langle v\sigma_{\chi_0\chi_0
\chi_1\chi_1} \rangle \left(\frac{n_{\chi_0}^{\eq}}{n_{\chi_1}^{\eq}}\right)^2,
\end{eqnarray}
where $g_{\chi_0}=2, \;\; g_{\chi_1}=6$  are the numbers of spin/colour states
encompassed in the $\chi_0$ and $\chi_1$ fields respectively. Moreover, $s$ stands for the usual Mandelstam variable and $p$ is the momentum of the incoming particles in the centre-of-mass frame. In the above expressions a ``0'' indicates a particle that is even under the discrete  $\mathcal{Z}_2$ symmetry that stabilises dark matter. All SM particles are naturally included here, and in our particular case also $S_1$.

In eq.~\eqref{eq:Y}, the $\Gamma_{\chi_1\to\chi_0}$ quantity contains two contributions, the decay term which dominates at low temperatures and guarantees the final disappearance of $\chi_1$, and the scattering term which is important at large temperatures,
\be\bsp
   \Gamma_{\chi_1\to\chi_0} =&\ \frac{ K_1\!\left(\frac{M_{\chi_1}}{T}\right)}{ K_2\!\left(\frac{M_{\chi_1}}{T}\right)} \Gamma_{\chi_1}
  \!+\! \frac{2 T g_{\chi_1}}{8 \pi^4 n_{\chi_1}^{\eq}} \sum_{{\rm SM}} g_{\rm SM} \int \sqrt{s} p^2(s) K_1\left(\frac{\sqrt{s}}{T}\right)
\sigma_{\chi_1,{\rm SM}\rightarrow \chi_0,{\rm SM'}}(s) \,\diff s, \!\!\\
 \Gamma_{\chi_0\to\chi_1} =&\  \Gamma_{\chi_1\to\chi_0}\frac{n_{\chi_1}^{\eq}} {n_{\chi_0}^{\eq}}.\label{eq:convrate}
\esp\ee
Here, $\Gamma_{\chi_1}$ is the $\chi_1$ decay width, and $\sigma_{\chi_1,{\rm SM}\rightarrow \chi_0,{\rm SM'}}(s)$ denotes the cross section associated with the $\chi_1{\rm SM}\rightarrow \chi_0{\rm SM'}$ scattering process via a $t$-channel leptoquark exchange. The sum runs over all possible channels characterized by the SM particle in the initial state of the scattering process, and $g_{\rm SM}$ denotes the number of degrees of freedom of this particle.

For couplings $y_\chi\lesssim 10^{-4}$ required for CDFO, annihilation of the dark sector into the SM is dominated by pair-annihilation of $\chi_1$, eq.~\eqref{eq:chi1ann}, while conversions within the dark sector are dominated by the two terms in eq.~\eqref{eq:convrate}. All other terms are found to be numerically irrelevant but are taken into account in our numerical computations. 

\bibliographystyle{JHEP}
\bibliography{RDM}

\end{document}